\documentclass[numbook, envcountsect, envcountsame, envcountreset, runningheads, smallextended]{svjour3}

\makeatletter
\def\makeheadbox{}
\date{}
\gdef\@date{}
\renewcommand{\and}{\unskip ,\ }
\makeatother

\usepackage{mathptmx}
\usepackage{graphicx} 
\usepackage{latexsym,amsfonts,amssymb,amsmath}
\usepackage{hyperref}
\usepackage{nameref}
\usepackage{etoolbox}
\usepackage{color}

\makeatletter
      \spnewtheorem*{Proof}{}{\itshape*\bfseries*}{\rmfamily}
      \def\Proof@Opargbegintheorem #1#2#3#4{#4\trivlist \item
      {#3#2\@thmcounterend \ }}
\AtBeginEnvironment{Proof}{\let\@Opargbegintheorem\Proof@Opargbegintheorem}
      \makeatother

\appto\ProcessRunnHead{\markboth{\truncate{220pt}{\authrun}\csuse{@author}}{\truncate{220pt}{\titrun}{\csuse{@title}}}}
\def\truncate#1#2{\ifdim#1<\wd#2\relax \rlap{\fboxsep=1pt\fboxrule1pt\color{red}\fbox{\hbox to#1{\strut\hss}}}\fi}
\smartqed 

\usepackage{accents}
\usepackage{mathtools}
\usepackage{bm}

\usepackage{mathrsfs}
\usepackage{dirtytalk}
\usepackage{shuffle}
\usepackage{tensor}
\usepackage{diagbox}

\usepackage[ruled]{algorithm2e}

\usepackage{listings}
\usepackage{alphabeta}
\newcommand{\deltaup}{\text{\delta}}

\newcommand{\bbN}{\mathbb{N}}
\newcommand{\N}{\mathbb{N}}
\newcommand{\E}{\mathbb{E}}
\newcommand{\R}{\mathbb{R}}
\newcommand{\G}{\mathcal{G}}
\renewcommand{\d}{\dif}
\newcommand{\eps}{\varepsilon}
\newcommand{\indep}{\perp\hspace{-0.6em}\perp}
\newcommand{\half}{\frac{1}{2}}
\newcommand{\halff}{\frac{1}{4}}
\newcommand{\bbP}{\mathbb{P}}
\newcommand{\bbE}{\mathbb{E}}
\newcommand{\bbR}{\mathbb{R}}

\newcommand{\Vv}{\mathcal{V}}
\newcommand{\Ww}{\mathcal{W}}
\newcommand{\dif}{\mathrm{d}}

\newcommand{\jo}{\mathrm{I}}
\newcommand{\jj}{\mathrm{II}}
\newcommand{\jjj}{\mathrm{III}}

\newcommand{\bfX}{\bm{X}}
\newcommand{\bfY}{\bm{Y}}

\newcommand{\bfW}{\bm{W}}

\newcommand{\barX}{\overline{\bfX}{}}

\renewcommand{\a}{\alpha}
\renewcommand{\b}{\beta}
\renewcommand{\c}{\gamma}
\newcommand{\pf}{\mathfrak{p}}

\newcommand{\cev}[1]{\reflectbox{\ensuremath{\vec{\reflectbox{\ensuremath{#1}}}}}}

\def\subclassname{{\bfseries Mathematics Subject Classification
}\enspace}
\def\subclass#1{\par\addvspace\medskipamount{\rightskip=0pt plus1cm
\def\and{\ifhmode\unskip\nobreak\fi\ $\cdot$
}\noindent\subclassname\ignorespaces#1\par}}
\def\JELname{{\bfseries JEL Classification}\enspace}
\def\JEL#1{\par\addvspace\medskipamount{\rightskip=0pt plus1cm
\def\and{\ifhmode\unskip\nobreak\fi\ $\cdot$
}\noindent\JELname\ignorespaces#1\par}}

\begin{document}

\journalname{}

\title{Rough differential equations for volatility\thanks{EF is supported by EPSRC Programme Grant EP/S026347/1.
IG and OB acknowledge financial support from the EPSRC grant EP/T032146/1.
AJ is supported by the EPSRC grants EP/W032643/1 and EP/T032146/1.
}
}


\author{Ofelia Bonesini
\and
Emilio Ferrucci\and
\quad\quad\quad\quad Ioannis Gasteratos
\and
Antoine Jacquier}

\authorrunning{O. Bonesini, E. Ferrucci, 
I. Gasteratos, A. Jacquier} 

\institute{O. Bonesini \at
              Department of Mathematics, London School of Economics and Political Sciences \\
              \email{o.bonesini@lse.ac.uk}
           \and
           E. Ferrucci \at
              Mathematical Institute, University of Oxford\\
\email{emilio.rossiferrucci@maths.ox.ac.uk}
           \and
           I. Gasteratos \at
              Institute of Mathematics, TU Berlin \\
              \email{i.gasteratos@tu-berlin.de}  
           \and
           A. Jacquier \at
              Department of Mathematics, Imperial College London \\
        \email{a.jacquier@imperial.ac.uk}  
}
\maketitle
\begin{abstract}
    We introduce a canonical way of performing the joint lift of a Brownian motion~$W$ and a low-regularity adapted stochastic rough path~$\bfX$, extending~\cite{DOR15}. Applying this construction to the case where $\bfX$ is the canonical lift of a one-dimensional fractional Brownian motion (possibly correlated with~$W$) completes the partial rough path of~\cite{FT22}. 
    We use this to model rough volatility with the versatile toolkit of rough differential equations (RDEs), namely by taking the price and volatility processes to be the solution to a single RDE. 
    We argue that our framework is already of interest when $W$ and $X$ are independent, as correlation between the price and volatility can be introduced in the dynamics. The lead-lag scheme of~\cite{FHL16} is extended to our fractional setting as an approximation theory for the rough path in the correlated case. Continuity of the solution map transforms it into a numerical scheme for RDEs. We numerically test this framework and use it to calibrate a simple new rough volatility model to market data.
    \end{abstract}
    \keywords{Rough paths \and rough volatility \and rough differential equations \and Wong-Zakai approximations \and lead-lag approximations \and stochastic Volterra equations}
\subclass{60L20 \and 60L90 \and 60G22\and 65C30 \and 91G20 \and 91G60 }

\setcounter{tocdepth}{3}
\tableofcontents

\section{Introduction}\label{sec:Intro}
\label{intro}
Stochastic volatility models describe the dynamics of asset prices subject to randomness and whose instantaneous variance also evolves randomly in time. 
They are typically expressed by a pair of stock price and  variance processes governed by a system of It\^o stochastic differential equations (SDEs) and driven by correlated Brownian motions;
there, the log-price is a semimartingale while its instantaneous variance is a Markovian diffusion process. 
While these models---among which 
the Heston model~\cite{heston1993closed}, the Bergomi model~\cite{bergomi2005smile} and the SABR model~\cite{hagan2002managing} are the most popular---capture many market behaviours, 
they however suffer from notorious shortcomings (hard to fit short-dated smiles, many parameters to fit,\ldots). In sharp contrast, 
the rough volatility paradigm abandons the Markovian assumptions of the variance process, 
replacing the latter by \emph{rough} volatility, driven by 
a process of lower H\"older regularity, 
such as fractional Brownian motion (fBm) with Hurst parameter $H<\half$, for example in the Rough Bergomi model~\cite{BFG16} or in~\cite{alos2007short}, or, more generally, by a stochastic Volterra process with singular kernels, as is the case for the rough Heston model~\cite{el2019characteristic}
and affine rough models~\cite{abi2019affine}.

Such models are not new per se, and strong inspiration came from the long-memory models suggested in~\cite{comte1996long}
and the literature on stochastic Volterra processes \cite{coutin2001stochastic,decreusefond1999stochastic}.
They were however brought to light and developed in the $H<\half$ case
in~\cite{alos2007short} and~\cite{gatheral2018volatility}, 
the latter coining the term \emph{rough volatility},
and have since gained large popularity due to their remarkable fit to market data and their ability to capture---with relatively few parameters---the main stylised facts of the implied volatility surface, including the steep short-maturity equity implied volatility skew. 
Calibration of these models has pointed~\cite{gatheral2018volatility,BFG16} towards very small values of the Hurst parameter, around~$0.1$,
implying a H\"older
regularity of the sample paths of~$V$ significantly rougher than in classical stochastic volatility models. 
Along with the absence of Markovianity, this has resulted in several challenges due to the lack of available tools (It\^o calculus, PDEs, large deviations), 
giving rise to streams of research on strong and weak error rates of Monte Carlo and Euler approximations \cite{BHT22,bayer2022weak,gassiat2023weak,friz2022weak,bonesini2023rough}, 
Markovian lifts~\cite{cuchiero2020generalized,cuchiero2019markovian,abi2021linear,bayer2022pricing,hamaguchi2023markovian,hamaguchi2023weak}, asymptotic implied volatility smile approximations~\cite{fukasawa2021rough,takano2025large}, 
moment estimates and functional inequalities for the law of~$S$~\cite{gassiat2019martingale,gasteratos2023transportation}.
One technical issue in rough volatility models lies in the absence of a Stratonovich formulation for the stock price 
(since the quadratic covariation $[V, \log S]$ between the variance and the log price is infinite), which is 
key for Wong-Zakai approximations and for quasi-Monte Carlo and cubature schemes~\cite{ninomiya2008weak,lyons2004cubature}.

A widely adopted toolkit to deal with equations driven by continuous processes with low regularity is Lyons's theory of rough paths~\cite{Lyo98}. 
A rough path \say{above} a multidimensional path (or stochastic process)~$X$ consists of a specification of all its iterated integrals $$\int_{s < u_1 < \ldots < u_n < t} \dif X_{u_1} \otimes \cdots \otimes \dif X_{u_n}$$ up to some order depending on the regularity of~$X$. Such a structure uniquely determines the meaning of an equation driven by~$X$, and does so in a way that makes the map from the enhanced~$X$ to the solution continuous. Despite their broad scope, classical rough path approaches do not apply to rough volatility due to the correlation of the driving noises and the roughness of~$V$. Indeed, even though rough paths are applicable in the regime $H\in(\frac{1}{4}, \half]$,
canonical rough lifts do not exist when $H \leq \frac{1}{4}$~\cite{coutin2002stochastic}. 
Moreover, denoting~$W$ the standard Brownian motion driving~$S$, and~$\rho$ the correlation between~$S$ and~$V$, 
classical Wong-Zakai approximations of $\int V \dif W$ fail since~\cite{BFGJS20}
$\int V^\eps \dif W^\eps \sim\rho \eps^{H-\half}$,
where~$V^\eps$ and~$W^\eps$ are piecewise linear (or mollifier) approximations of~$V$ and~$W$ and~$\sim$ means asymptotic equivalence as $\eps\to 0$.
There, inspired by  Hairer's work on singular stochastic PDEs~\cite{hairer2014theory}, Bayer, Friz, Gassiat, Martin and Stemper used tools from regularity structures to obtain a pathwise formulation that comes with renormalised Wong-Zakai results 
(with convergence recovered by subtracting  diverging quantities).
The generality of this approach comes at the cost of \say{heavy} tools (Schwartz distributions, negative Besov spaces, algebraic renormalisation), 
not easily accessible for practical purposes.
Furthermore, when~$V$ is a stochastic Volterra equation,
the aforementioned Wong-Zakai results become more complicated as the number of diverging terms increases as $H\to 0$,
and only the $H > \frac{1}{4}$ case is fully proved so far~\cite[\S5.2]{BFGJS20}, the case $H \leq \frac{1}{4}$ requiring "a 
Hopf algebraic $[\ldots]$ construction of the structure group".

In this paper, we argue that rough paths can be applied to rough volatility.
Given a $d$-dimensional Brownian motion~$W$
and a one-dimensional adapted process~$X$ (such as a correlated fractional Brownian motions), 
we construct a rough path above~$(X,W)$ (in fact, this is done more in general when~$X$ is a possibly multidimensional adapted rough path).
This definition contains the It\^o integrals in the partial construction of Fukasawa-Takano~\cite{FT22}, 
but crucially also assigns values to the integrals that cannot be made sense of classically, such as $\int W \dif X$, $\int \dif X \dif W \dif X$, and so on.
By imposing integration by parts identities, our approach extends~\cite{DOR15} in the direction of both low regularity and random~$X$. 
On the one hand, 
setting~$W$ multidimensional and~$X$ one-dimensional sidesteps the challenges imposed by $H \leq \frac{1}{4}$.
On the other hand, the divergence issue of the quadratic covariation is not observed thanks to the integrals being It\^o, all the while preserving geometric structure of the rough path. 
We then argue that a natural way to obtain joint dynamics~$(S,V)$
can be obtained by rough differential equations (RDEs) driven by $(X,W)$, enriched with the newly defined rough path structure. 
This is a fundamentally different type of dynamic to Volterra equations, even in the one-dimensional smooth case. 
Indeed, 
given a smooth kernel~$K$ and a smooth path~$Z$, 
compare the ordinary and the Volterra differential equations
\begin{subequations}
\begin{align}
& \text{(Ordinary)} & Y_t^{(1)} & = \displaystyle Y_0^{(1)} +  \int_0^t F(Y_s^{(1)}) \d X_s, \qquad \text{with} \quad X_t = \int_0^t K(t,u) \d Z_u,\label{eq:ODE}\\
& \text{(Volterra)} & Y_t^{(2)} & = \displaystyle Y_0^{(2)} + \int_0^t K(t,s) F(Y_s^{(2)}) \d Z_s.
\label{eq:VDE}
\end{align}
\end{subequations}
Using the Leibniz integral rule
for~\eqref{eq:ODE}
and 
the smoothness of the kernel for~\eqref{eq:VDE}, so that, for $s \leq t$, $K(t,s) = K(s,s) + \int_s^t \partial_1K(v,s) \d v$, we can write
\begin{align*}
    Y_t^{(1)}
    &= Y_0^{(1)} + \int_0^t F(Y_s^{(1)}) K(s,s) \d Z_s + \int_{0 < u < s < t} F(Y_s^{(1)}) \partial_1 K(s,u)\d Z_u \d s,\\
    Y_t^{(2)}
    &= Y_0^{(2)} + \int_0^t F(Y_s^{(2)}) K(s,s)\d Z_s + \int_{0 < u < s < t} F(Y_u^{(2)}) \partial_1K(s,u) \d Z_u \d s.
\end{align*}
While they look similar, these two equations are fundamentally different as~$F(Y)$ is evaluated 
in the double integral
at~$s$ (integrated against $\dif s$) in the first and at~$u$ (integrated against~$\dif Z_u$) in the second.
The problem becomes even trickier in the case of a singular kernel~$K$ as such a comparison cannot even be made.
If one accepts that ODEs are the most common way of understanding dynamics in the smooth setting, it is natural to consider their analogue in the rough setting, which is provided by rough differential equations (RDEs), not by singular Volterra DEs. 
This holds in the technical sense of Wong-Zakai: every RDE is the limit of ODEs driven by smooth paths approximating the rough path in an appropriate topology. In rough volatility, an advantage of RDEs over Volterra DEs is that in the former, the price and vol can be viewed as \emph{jointly} solving the same equation, while in the latter they are solving a \say{mixed It\^o-Volterra equation}~\cite[\S 5.2]{BFGJS20}.

One should nevertheless acknowledge the special role played by Volterra processes and equations in rough volatility:
given by It\^o integrals, their mean and variance are easy to compute.
While some rough volatility models (rough Heston in particular) are underpinned by microstructural justifications~\cite{JR16}, 
this may not, however, be universal, 
and their defining feature really is the irregularity of the volatility sample paths, a feature that can be obtained alternatively with RDEs. 
Recently, in computational finance, 
statistical estimation of parameters has been reconsidered in the light of black-box neural computations, which rely on non-parametric models as well as on fast and reliable numerical schemes;
for example, via the emerging literature on neural SDEs in finance~\cite{CKT20,GSSSZ20,CRW23,CJB24}, amenable for 
ODEs and SDEs, but to a lesser extent for Volterra equations (although see the recent~\cite{neuralVolterra}).
Our goal here is not to argue against established models in rough volatility,
but rather to propose an alternative framework with numerous advantages and which includes the majority of rough volatility models already present in the literature.

In our RDE approach for rough volatility, trajectories of $(S,V)$ are simulated appealing to Wong-Zakai. 
One must identify an approximating sequence $(X^\eps,W^\eps) \to (X,W)$ in rough path topology, so that one can solve ODEs driven by $(X^\eps,W^\eps)$. 
We identify more than one such sequence, all based on the idea of \say{lead-lag approximations} introduced in~\cite{FHL16} to approximate the It\^o rough path, and prove strong rates of convergence. 
We validate our approach with numerical tests and propose a new RDE-volatility model for financial applications.

Our contributions can be summarised as follows:
    \begin{itemize}
        \item In \autoref{sec:ItoLift} we introduce a novel way of performing the joint lift of an adapted (sufficiently integrable) rough path and a Brownian motion, extending~\cite{DOR15}, in \autoref{def:itoLift} and \autoref{thm:itoLift}.
        \item In \autoref{sec:roughVol} we specify the very general construction of the previous section to the case in which the adapted rough path is the canonical lift of a one-dimensional process. Most often, but not always, this will be a low regularity fBm, possibly correlated with the Brownian motion. A new interpretation is given of the rough path terms (\autoref{prop:binomial}). The general RDE used in the rest of paper is proposed in~\eqref{eq:Model}, accompanied by several modelling considerations. We show how this equation embeds many models already considered in the literature (\autoref{subsec: Applications}), as well as a very general way of parametrising new ones.
        \item In \autoref{sec:lead-lag} we discuss the convergence of three different types of lagged approximations. Section~\ref{subsection:LeadLagPiecewise} is devoted to the convergence of piecewise-linear lead-lag approximations with explicit rates. 
        In Section~\ref{subsec:HybridLeadLag}, we extend this convergence result to the case when~$X$ is given by  a hybrid scheme approximation~\cite{bennedsen2017hybrid} of a fractional Brownian motion.  In passing, we also obtain a novel almost-sure convergence result for the hybrid scheme approximation in H\"older topology (Theorem~\ref{thm:hybridalmostsure}).
        Finally, Section~\ref{subsec:LaggedMollifier} deals with lagged mollifier approximations.
        \item In \autoref{sec:SimCal} we numerically validate the aforementioned theoretical results and propose an RDE-based adaptation of the quadratic rough Heston model, which we calibrate to market data. 
    \end{itemize}

While the framework we introduce is utilised as a novel way of generating rough volatility models, we hope and expect it to have much broader applicability to systems governed by non-i.i.d.\ noise.
Before diving into the details,
we fix some frequently used notations and provide an overview of RDE-based models considered in this paper, as well as some of their flexibility and modelling advantages.

\paragraph{General framework, applications overview and frequently used  notations.}

Throughout this paper, $(\Omega,\mathcal{F}_{\bullet},\mathbb{P})$ denotes a filtered probability space satisfying the usual conditions and~$T>0$ a fixed time horizon. 
We write $X\indep Y$ to denote independence of random variables $X, Y$ and often abuse notation by writing "$X\in\mathcal{F}$ " to indicate that~$X$ is $\mathcal{F}$-measurable. 
Given two vector spaces~$\Vv$ and~$\Ww$,
we denote by $\mathcal{L}(\Vv, \Ww)$ the vector space of $\Ww$-valued linear maps defined on $\Vv$.  Furthermore, we use $x \lesssim_a y$ to mean \say{there exists $C>0$, depending on $a$ such that $x \leq C y$}.\\
For $p\geq 1, \gamma \in (0, 1)$ we denote by $(L^p(\Omega, \mathcal{F},\mathbb{P}), |\cdot|_{L^p(
\Omega)} )$ and $(C^\gamma([0,T]; \Vv), |\cdot|_{C^{\gamma}([0,T];\Vv)})$ the Banach spaces of $p$-integrable, real-valued random variables and ${\gamma}$-H\"older continuous $\Vv$-valued paths, defined on the interval $[0,T]$, with their usual norm topologies. When easily understood from the context, we shall omit the domain and co-domain notation and instead write $|\cdot|_{L^p}, |\cdot|_{C^{\gamma}}$. Finally, the set of smooth, compactly supported test functions is denoted by $\mathcal{C}_c^\infty(\mathbb{R})$. The (topological) support of a test function $\phi\in\mathcal{C}_c^\infty(\mathbb{R}) $ is denoted by $\text{supp}(\phi)$.

Our framework (see \autoref{sec:roughVol}) is described by a general class of RDEs given by
\begin{equation}\label{eq:Model}
    \left\{
    \begin{array}{rll}
        \d S_t & = \displaystyle \sigma_\alpha(S_t, V_t, t) \d \bfW_t^\alpha + g(S_t, V_t, t) \d t, & S_0=s_0\in\bbR,\\
        \d V_t & = \displaystyle \tau(S_t, V_t, t) \d \bfX_t + \varsigma_\alpha(S_t, V_t, t) \d \bfW_t^\alpha + h(S_t, V_t, t) \d t, & V_0=v_0\in\bbR^m,
    \end{array}
    \right.
\end{equation}
where here and below we use the Einstein summation convention, and for fixed $m, d,e\in\bbN$, $g: \bbR^{m+1}\times[0,\infty)\rightarrow \bbR$,  $h:\bbR^{m+1}\times[0,\infty)\rightarrow\bbR^m$, $\tau: \bbR^{m+1}\times[0,\infty)\rightarrow\mathcal{L}(\bbR^e;\bbR^m)$ and, for each $\alpha=1,\dots d$, $\varsigma_\alpha: \bbR^{m+1}\times[0,\infty)\rightarrow\bbR^m$ and $\sigma_{\alpha}: \bbR^{m+1}\times[0,\infty)\rightarrow\bbR$ are sufficiently smooth vector fields.  The dynamics of $(S, V)$ are driven by a path $( W, X)$, where $W=(W^\alpha)_{\alpha=1}^d$ is a $d$-dimensional standard Brownian motion on $(\Omega,\mathcal{F}_{\bullet},\mathbb{P})$ ($\mathcal{F}_{\bullet}$ is not necessarily the filtration generated by~$W$),~$X$ an $\bbR^e$-valued path of "low" H\"older regularity and $\bfW, \bfX$ are geometric rough paths over~$W, X$ respectively 
(we refer the reader to to \autoref{sec:ItoLift} for definitions and rough path notations).
For financial purposes, $S$ represents the asset price process and~$V$ the variance process. This general class of asset price models offers significant flexibility in terms of modelling choices. In particular,
\begin{enumerate}
    \item It allows for \emph{different types of correlation between~$S$ and~$V$}: Usually, in (classical and rough) volatility models, the Brownian motion driving~$S$ and the (fractional) one driving~$V$ are correlated to account for the \emph{leverage effect}. 
    The RDE~\eqref{eq:Model} offers another way of achieving this by driving~$V$ both with the same factor used for~$S$ and with another path~$\bfX$.
Therefore the case $X \indep W$ does not preclude~$S$ and~$V$ from being correlated.
Classical correlation can nevertheless be recovered by correlating~$X$ and~$W$ and setting~$\varsigma$ to zero.
The models~\eqref{eq:Model} thus encompass both ways of introducing correlation, as well as any combination of them, for which the construction of \autoref{sec:ItoLift} is necessary.

\item It allows for \emph{path dependency of~$V$ on~$S$}: Allowing the coefficients of~$V$ to depend on~$S$ is not problematic from a mathematical perspective. While atypical from a modelling point of view, such a choice leads to a particular instance of (rough) volatility models in which the spot volatility depends on the past price trajectory. Path-dependent volatility models have been considered in~\cite{GL23} and have been useful in replicating the Zumbach effect~\cite{Zum09,Zum10}, namely the impact of historical prices  on the volatility. 
\item The RDE~\eqref{eq:Model}  \emph{includes many volatility models considered in the literature as special cases}
(and a more detailed discussion on these models and how they embed in our framework is deferred to \autoref{subsec: Applications}):
\begin{enumerate}
\item Classical (local) stochastic volatility models such as Black-Scholes, Bergomi, Heston, Stein-Stein.
\item Rough volatility models such as (multifactor) rBergomi, rHeston, quadratic rHeston models are (at least  partially) recovered from  \eqref{eq:Model} (see also Remark \ref{rem:SVEsvsRDEs} below). 
\item Path-dependent stochastic volatility models such as Guyon-Lekeufack~\cite{GL23}.
\end{enumerate}

\item \emph{Extendability to multi-asset models and volatility with smooth non-Markovian drivers}. Even though we focus exclusively on single-asset models ($S$ is real-valued) it is possible to extend our framework to multi-asset models in which each asset price depends on a single component of~$V$. Moreover,~\eqref{eq:Model} opens the door for a unified study of non-Markovian volatility models that feature both roughness and long-range dependence ($V$ driven by both smooth ($H>\half$) and rough ($H<\half$) fractional Brownian motions). 
More details on such extensions can be found in Remarks~\ref{rem:Multi-asset} and~\ref{rem:XindW} below.
\item Different rough path lifts  have appeared in the literature. The approaches in~\cite{bank2025rough} and~\cite{takano2025large} involve joint rough path lifts of the It\^o-integrated volatility, which is a semimartingale, and the (independent) Brownian noise driving the asset price. In contrast to our work, this is agnostic to the choice of volatility dynamics and hence numerical approximations of asset prices pre-suppose a choice of numerical scheme for simulation of the volatility process. In the present work, we propose modelling the joint dynamics of the asset price and the volatility processes by RDEs. In turn, our proposed Wong-Zakai numerical schemes approximate the joint RDE dynamics by a system of Ordinary Differential Equations with random coefficients.  
\end{enumerate}


\section{The It\^o lift of an adapted rough path}\label{sec:ItoLift}

We begin this section with a gentle introduction to rough paths with inhomogeneous degrees of regularity; we refer to~\cite{Gyu16} for a comprehensive treatment of this topic. Let $A$ be a finite alphabet, the elements of which we refer to as \emph{letters}. $\mathbb R^A$ can be identified with the linear span of $A$, while, denoting $A^\bullet$ the set of words in $A$ (i.e.\ finite sequences of letters), the tensor algebra 
\[
T(\mathbb R^A) = \bigoplus_{n = 0}^\infty (\mathbb R^A)^{\otimes n} = \mathrm{span}(A^\bullet)
\]
can be identified as the linear span of words in the alphabet $A$ (with $(\mathbb R^A)^{\otimes 0} = \mathbb R$ and $(\mathbb R^A)^{\otimes 1} = \mathbb R^A$ the space of letters). 
We recall two important operations on $T(\mathbb R^A)$: the \emph{shuffle} product and the \emph{deconcatenation} coproduct
\begin{align*}
\shuffle \colon T(\bbR^A) \otimes T(\bbR^A) \to T(\bbR^A), \qquad \Delta \colon T(\bbR^A) \to T(\bbR^A) \otimes T(\bbR^A) .
\end{align*}
Each is defined on words and extended linearly. The shuffle of two words is the sum of all the words obtained by mixing their letters without modifying the order of the letters in each word, namely
\[
ab \shuffle cd = abcd + acbd + cabd + acdb + cadb + cdab .
\]
The deconcatenation of a word is the sum over all tensor products of two words obtained by cutting the word into two, including \say{trivial} cuts e.g.\
\[
\Delta (abcd) = 1 \otimes abcd + a \otimes bcd + ab \otimes cd + abc \otimes d + abcd \otimes 1 ,
\]
where we emphasise that $\otimes$ denotes the external tensor product and $1 \in \mathbb R = (\bbR^A)^{\otimes 0}$;
see \cite[\S 1.3 \& 1.4]{Reu93} for precise definitions of both operations.
We call $\mathrm{Sh}(\bbR^A) \coloneqq (T(\bbR^A), \shuffle, \Delta)$ the tensor algebra endowed with shuffle and deconcatenation,
which is a bialgebra \cite[\S~6]{Reu93}.
The algebraic dual of the vector space $T(\bbR^A)$ is $T(\!( \bbR^A )\!) \coloneqq \prod_{n = 0}^\infty (\bbR^A)^{\otimes n}$, the space of formal series of tensors in~$\bbR^A$. 
We denote $\langle \cdot,\cdot \rangle$ the canonical dual pairing
between $T(\bbR^A) \otimes T(\!( \bbR^A )\!)$ and~$ \bbR$, namely the linear extension of the pairing on single words for which $\langle w, z \rangle = 1$ if $w = z$ and zero otherwise. The set 
\[
\mathcal G(\bbR^A) \coloneqq \{x \in T(\!( \bbR^A )\!) \mid \forall u, v \in \mathrm{Sh}(\bbR^A) \ \langle u \shuffle v , x \rangle = \langle u , x \rangle \langle v , x \rangle\}
\]
is called the space of \emph{grouplike elements} and forms a group under tensor multiplication.
Let $| \, \cdot \, | \colon A \to (0,1]$ be a function, which we call \emph{weight} with $|1| \coloneqq 0$, which we extend by $|a_1 \ldots a_n| \coloneqq |a_1| + \ldots + |a_n|$ on words. We denote 
\[
T\lceil \bbR^A \rceil \coloneqq \mathrm{span}\{a_1\ldots a_n \in A^\bullet \mid |a_1 \ldots a_n| \leq 1 \}
\]
the tensor algebra truncated at weight~$1$.
We similarly denote $\mathrm{Sh}\lceil \bbR^A \rceil$ and $\mathcal G\lceil \bbR^A \rceil$, which remains a group under the product which for $v, w \in A^\bullet$ is defined by the ordinary tensor product $v \otimes w$ if $|v| + |w| \leq 1$ and $0$ otherwise \cite[\S~7]{FV10}. We continue to denote $\otimes$ this product and rely on context (i.e.\ whether the factors belong to $T(\bbR^A)$ or $T\lceil \bbR^A \rceil$) to tell it apart from the usual tensor product.

Let now $X \colon [0,T] \to \bbR^A$ be a multidimensional path whose components have regularity governed by the weight function. More precisely, since our paths are samples of a stochastic process, for which regularity is obtained via Kolmogorov's continuity criterion, it is convenient to say that, for $a \in A$, the component $X^a \colon [0,T] \to \bbR$ is $\gamma$-H\"older regular for all $\gamma < |a|$ (note that this is a slight departure form the literature, in which the regularity is exactly $|a|$). In particular, since the weight function is positive, then~$X$ is always continuous. We recall the definition of a rough path, due to~\cite{Lyo98} (\cite{Gyu16} in the inhomogeneous regularity case); $\Delta_T$ denotes the simplex $\{(s,t) \in [0,T]^2 \mid s < t\}$ and we use $1 \in A^\bullet$ to denote the empty word (the generator of $(\bbR^A)^{\otimes 0})$.

\begin{definition}[Rough path]\label{def:rp}
A (geometric) \emph{rough path} above $X \colon [0,T] \to \bbR$, H\"older regular with respect to the weight function~$|\, \cdot \,|$, consists of a function $\bfX \colon \Delta_T \to T\lceil \bbR^A \rceil$, whose coordinates are denoted $\bfX^w_{st} \coloneqq \langle w, \bfX_{st} \rangle \in \mathbb R$ for $w \in \mathrm{Sh}\lceil \bbR^A \rceil$ and $(s,t) \in \Delta_T$, with $\bfX^1 \equiv 1$ and satisfying the following three properties.
    \begin{itemize}
        \item \textbf{Regularity.} For all $w \in A^\bullet$ with $|w| \leq 1$ and $\gamma < |w|$ there exists $C$ such that, for all $0 < s < t < T$,
        $|\bfX^w_{st}| \leq C|t-s|^\gamma$;
        \item \textbf{Integration by parts.} For $u,v \in \mathrm{Sh}\lceil \bbR^A \rceil$ and all $0 < s < t < T$,
        $\bfX^{u \shuffle v}_{st} = \bfX^u_{st} \bfX^v_{st}$;
        \item \textbf{Chen identity.} For any $0 < s < u < t < T$,
        $\bfX_{su} \otimes \bfX_{ut} = \bfX_{st}$.
    \end{itemize}
    A topology on the set of such rough paths can be introduced via the \emph{inhomogeneous rough path distance}
    $$
    d_{\gamma}(\bfX, \bfY):=\sum_{w\in A^\bullet}\sup_{s\neq t\in [0, T]}\frac{|\bfX^w_{st}|}{|t-s|^{\gamma_w}},
    $$ defined for a multi-index $\gamma=(\gamma_1, \dots \gamma_{|A^\bullet|})$ such that for each $w\in A^\bullet, \gamma_w\in (0, |w|)$. Convergence in this topology amounts to convergence in $C^{\gamma_w}-$H\"older norm for each component $\bfX^w$ of the rough path. 
\end{definition}

We comment on the meaning of the above definition. The term $\bfX^{a_1 \ldots a_n}_{st}$ is an abstraction that represents the iterated path integral
\begin{equation}\label{eq:int0}
\int_{s < u_1 < \ldots < u_n < t} \d X^{a_1}_{u_1} \cdots \d X^{a_n}_{u_n} ,
\end{equation}
which is not defined as a Stieltjes integral unless~$X$ is of bounded variation and therefore must be supplied, usually through some theory of stochastic integration (It\^o, Stratonovich, Gaussian processes, etc.). If $a \in A$ is a single letter, $\bfX^a$ is just the ($a$-coordinate of) of path~$X^a$, usually denoted without bold font and called the \emph{trace} of~$\bfX$. These iterated integrals are significant in that they form the building blocks of \emph{rough differential equations} (RDEs)
\begin{equation}\label{eq:defRDE}
\d Y = F_a(Y) \d \bfX^a, \qquad Y_0 = y_0
\end{equation}
where $F_a$ are vector fields in the Euclidean space in which the solution $Y$ is valued, and there is an implicit summation over $a \in A$. This is because a solution to~\eqref{eq:defRDE} (once again, classically meaningless) is defined by the \emph{Davie expansion}~\cite{davie}
\begin{equation}\label{eq:daviedef}
Y_{st} \approx \sum_{\substack{w \in A^\bullet \\ |w| \leq 1}} F^\circ_w(Y_s) \bfX^w_{st},
\end{equation}
where $F^\circ_{a_1 \ldots a_n} \coloneqq F_{a_1} \cdots F_{a_n}$ (iterated composition of vector fields), $Y_{st}$ denotes the increment of the solution $Y_t - Y_s$ (similar notation for increments will be used throughout) and $\approx$ means that the left- and right-hand sides differ by at most $O((t-s)^\theta)$ with $\theta > 1$. This implies that iteratively summing such terms over $[s,t]$ in some partition of $[0,T]$ converges as the step size of the partition tends to zero. The limiting path obtained by this numerical scheme is (one of the equivalent definitions of) the solution to the RDE: note how the bold font for $\bfX$ is kept in~\eqref{eq:defRDE} as a reminder of the fact that different rough paths yield different notions of solutions despite the underlying path being the same (such as  It\^o, Stratonovich, etc.).

Returning to \autoref{def:rp}, the first property is what one would expect from the H\"older regularity of~\eqref{eq:int0} given the H\"older regularity of the components of $X$. The second property states that the product of two iterated integrals can be obtained as a sum of iterated integrals, by interleaving the integrators in each factor. It can be equivalently stated by requiring $\bfX$ to be $\mathcal G \lceil \bbR^A \rceil$-valued. Note that not all integration theories satisfy this rule, most notably It\^o integration does not, because of the quadratic covariation correction in $X^\a_{st} X^\b_{st} = \int_s^t X_{su}^\a \d X_u^\b + \int_s^t X_{su}^\b \d X_u^\a + [X]^{\a \b}_{st}$. Rough paths that satisfy this property are generally called \emph{geometric} since they obey the rules of ordinary calculus; we will only consider geometric rough paths from now on. Although we make extensive use of the It\^o integral throughout, this issue is sidestepped in a way that will be made clear. The Chen identity represents additivity of integration on consecutive time intervals, lifted to the iterated integrals, which turns additivity into multiplicativity. In particular, it implies that the rough path is determined by its values over pairs $(0,t)$, by $\bfX_{st} = \bfX_{0s}^{-1} \otimes \bfX_{0t}$, where the inversion of the first factor is taken in the group $\mathcal G\lceil \bbR^A \rceil$; this is why $\bfX$ may be thought of as a path in $\mathcal G\lceil \bbR^A \rceil$ started at~$1$. 
The Chen identity can be stated dually as
\begin{equation}\label{eq:dualchen}
\bfX_{st}^w = \langle \Delta w , \bfX_{su} \otimes \bfX_{ut} \rangle, \qquad \text{for all } w \in A^\bullet, \ |w| \leq 1,
\end{equation}
where we are viewing the tensor product as \say{external}, $\bfX_{su} \otimes \bfX_{ut} \in T(\R^A) \otimes T(\R^A)$ and $\langle u \otimes v , w \otimes z \rangle \coloneqq \langle u , w \rangle\langle v ,  z \rangle$; this is the form in which it is usually convenient to check it.
Let now~$W$ be a $d$-dimensional Brownian motion on a filtered probability space $(\Omega, \mathcal F_\bullet, \mathbb P)$ and consider~\cite{FH20,FV10} the Stratonovich rough path defined in coordinates by 
$$
\mathbf{W}^{\a\b}_{st} \coloneqq\int_{s}^{t} W_{su}^\a \circ \d W^\b_u.
$$
We now consider an $e$-dimensional stochastic process $X$ with a.s.\ $\gamma$-H\"older continuous sample paths for all $\gamma < H$, with $H \in (0,1)$ fixed, already lifted to a (random) rough path~$\bfX$. 
We wish to consider differential equations driven jointly by $W$ and $X$. For this to have meaning it is not sufficient for $W$ and $X$ to individually be lifted to a rough path: we must have a joint lift. We take $A \coloneqq [d] \sqcup [e]$ (where $[n] \coloneqq \{1,\ldots, n\})$. 
We use Greek letters for coordinates in $[d]$ (for $W$) and latin letters for coordinates in $[e]$ (for $X$). We consider the weight function $|\a| \equiv 1/2$ and $|i|\equiv H$, for $\a \in [d]$ and $i \in [e]$. We now refine our notion of rough path~$\bfX$ so that it respects the filtration and has enough integrability to allow applications of Kolmogorov's continuity theorem. 
That such an~$\bfX$ actually defines a rough path almost surely is deduced as a special case of \autoref{thm:itoLift} below. Note that this definition could easily be extended to accommodate inhomogeneous regularity of the components of $X$ (as opposed to just different regularity of components of $X$ and of $W$), but this is not needed.
\begin{definition}[Adapted $H$-integrable rough path]\label{def:adapted}
An $\mathcal{F}_{\bullet}$-\emph{adapted $H$-integrable (geometric) rough path} (on $[0,T]$, taking values in $\mathbb R^e$) is an $\mathcal{F}_\bullet$-adapted, $\mathcal{G}\lceil \mathbb{R}^e \rceil$-valued stochastic process~$\bfX$, started at~$1$, which satisfies the Chen identity and such that, denoting $\bfX_{st} \coloneqq \bfX_s^{-1} \otimes \bfX_t$,
\[
\sup_{0 \leq s \leq t \leq T}|\bfX^w_{st}|^p_{L^p} \lesssim_{p,T} (t-s)^{p|w|} .
\]
for each $w \in [e]^\bullet$ with $|w| \leq 1$ and $p \in [1,+\infty)$.
\end{definition}

We denote $\overline{\bfX}$ the joint lift of $\bfW$ and~$\bfX$ which we proceed to construct. For this, we leverage the fact that the It\^o integral is defined on very general adapted integrands. For example, this allows us to set $\barX^{i\a}_{st}$ ($\a \in [d]$ and $i \in [e]$) to the It\^o integral $\int_s^t X_{su}^i \d W^\alpha_u$. The term $\barX^{\a i}_{st}$ appears more challenging to define, but note that the shuffle relation
\begin{equation}\label{eq:ibp}
\a i = i \shuffle \a - i \a 
\end{equation}
implies that in order for~$\barX$ to be geometric, we must set $\barX_{st}^{\a i} = X^i_{st} W^\a_{st} - \int_s^t X_{su}^i \d W^\alpha_u$, amounting to an imposition of the classical integration-by-parts identity. This idea is original to~\cite{DOR15} in the case~$\bfX$ deterministic (a special case of \autoref{def:adapted}) and of bounded $p$-variation with $p \in [2,3)$ (corresponding to $H \in (\frac{1}{3},\half]$ here), in which case~\eqref{eq:ibp} is the only relevant shuffle relation. Without~$\bfX$ being deterministic, the integrability assumptions of \autoref{def:adapted} on~$\bfX$ are necessary for our extension. The assumption of adaptedness is also necessary, as dropping it would result in counterexamples such as those in~\cite{Lyo91}. The hypotheses on the coefficients of the equations defined by the joint lift of~\cite{DOR15} can be relaxed thanks to stochastic sewing~\cite{FHL23}. \cite[Definition 1.5]{FZ23} introduces \emph{rough semimartingales}, allowing for general (possibly discontinuous) local martingales instead of Brownian motion. The $p$-variation regularity of the rough path to be jointly lifted is still $p \in [2,3)$, and~$X$, while not necessarily deterministic, is assumed to be controlled by a deterministic reference path. This implies existence of the quadratic covariation corresponding to our $[W,X]$, which rules out the main example considered in \autoref{sec:roughVol} below \cite[Example 1.2]{FZ23} (the first-order integration-by-parts identity of~\eqref{eq:ibp} is briefly mentioned on p.401, but is not used later). We believe it would be interesting to extend the material in this paper to allow~$W$ to be a more general local martingale as done in~\cite{FZ23}; while the main ingredients (like BDG) continue to apply, this would necessitate the use of $p$-variation estimates instead of H\"older regularity ones.

For the lower-regularity case we need the following algebraic lemma. It will be used to argue that, once a rough path is defined over words for which its values are either supplied or can be defined by It\^o integration, imposing integration by parts uniquely results in an extension to the full truncated shuffle algebra over $\mathbb R^{d+e}$.
\begin{lemma}\label{lem:algebra}
    The set
    \begin{equation}\label{eq:B}
        \mathfrak{B} \coloneqq
        \left\{w, z \c, \a \b \mid w, z \in [e]^\bullet \text{ with } |w| \leq 1, |z| \leq \half, 
        \a, \b, \c \in [d] \right\}
    \end{equation}
    shuffle-generates $\mathrm{Sh} \lceil \mathbb{R}^{d+e} \rceil$ and it does so freely modulo the shuffle relations in $\mathbb{R}^d$ and $\mathbb{R}^e$. Namely, calling $\mathcal{B} \coloneqq \mathrm{span}(\mathfrak{B})$ and given any algebra $A$ and linear map $\phi \colon \mathcal{B} \to A$ restricting to algebra maps on $\mathrm{Sh}\lceil \mathbb{R}^{d} \rceil$ and
    ~$\mathrm{Sh}\lceil \mathbb{R}^{e} \rceil$, there exists a unique algebra map $\Phi \colon \mathrm{Sh}\lceil \mathbb{R}^{d+e} \rceil \to A$ such that $\Phi|_{\mathcal{B}} = \phi$.
\end{lemma}

\begin{proof}
    The words that are left to shuffle-generate are those of the form $u \a v$ with $\a \in [d]$ and $u,v \in [e]^\bullet$ with~$v$ non-empty and $|u| + |v| \leq \half$. 
    Setting $v = j_1 \ldots j_n$, we have, splitting the terms in the shuffle product $u\a \shuffle v$ in terms of the number of letters~$j_l$ appearing after~$\a$,
\begin{equation}\label{eq:recursion}
        u \a j_1 \ldots j_n = u \a \shuffle j_1 \ldots j_n - \sum_{k = 1}^n (u \shuffle j_1 \ldots j_k) \a j_{k+1} \ldots j_n,
    \end{equation}
    on which~$\Phi$ is uniquely determined by
    \[
    \Phi(u \a j_1 \ldots j_n) = \phi(u \a)\cdot \phi(j_1 \ldots j_n) - \sum_{k = 1}^n \Phi((u \shuffle j_1 \ldots j_k) \a j_{k+1} \ldots j_n).
    \]
    This defines $\Phi$ inductively as the terms in the sum have $n-1$ or fewer trailing letters in~$[e]$ and proves its uniqueness. To show existence, by truncation and the assumptions on $\phi$, the only shuffle relation yet to be checked is that $\Phi(u\alpha v \shuffle w) = \Phi(u\alpha v)\cdot \phi(w)$ with $u,v,w \in [e]^\bullet$. By \eqref{eq:recursion} and associativity of the shuffle product we have
    \begin{align*}
    u \a j_1 \ldots j_n \shuffle w = u \a \shuffle (j_1 \ldots j_n \shuffle w) - \sum_{k = 1}^n ((u \shuffle j_1 \ldots j_k) \a j_{k+1} \ldots j_n) \shuffle w 
    \end{align*}
    and inductively we reduce to the case of $v$ empty, for which, calling now $w = j_1\ldots j_k$, by definition of $\Phi$,
    \begin{align*}
    \Phi(u\alpha \shuffle w) = \Phi(u\alpha w) + \sum_{k = 1}^n \Phi((u \shuffle j_1 \ldots j_k) \a j_{k+1} \ldots j_n) = \phi(u\alpha)\cdot\phi(w),
    \end{align*}
    concluding the proof.
\end{proof}

\begin{example}\label{ex:iajk}
We give an example of the recursion~\eqref{eq:recursion}:
    \begin{align*}
    i\a jk &= i\a \shuffle  jk - (i \shuffle j)\a k - (i \shuffle  jk)\a \\
    &= i\a \shuffle  jk -  ij \a k -  ji \a k -  ijk\a -  jik\a -  jki\a \\
    &= i\a \shuffle  jk - [ ij \a \shuffle k - [ ij\shuffle k] \a] - [ ji \a \shuffle k - [ ji\shuffle k] \a] -  ijk\a -  jik\a -  jki\a \\
    &= i\a \shuffle  jk - ij \a \shuffle k +  ikj \a+  kij \a -  ji \a \shuffle k +  kji \a.
\end{align*}
\end{example}

\begin{definition}[It\^o lift]\label{def:itoLift}
    Let~$\bfX$ be an adapted $H$-integrable geometric rough path. We define its \emph{It\^o lift} by Stratonovich and It\^o integrals on the remaining elements of $\mathfrak{B}$ as
    \begin{equation}\label{eq:itoL}
    \barX^{\a\b}_{st} \coloneqq \int_s^t W^{\a}_{su}\circ \dif W^{\b}_u, \qquad \barX^{w\c}_{st} \coloneqq \int_s^t \bfX^{w}_{su}\dif W^{\c}_u, \qquad w \in [e]^\bullet, \a, \b, \c \in [d],
    \end{equation}
    and further extending it to $\mathrm{Sh}\lceil \mathbb{R}^{d+e} \rceil$ by \autoref{lem:algebra}, specifically by means of~\eqref{eq:recursion}. Note that the It\^o integral defining~$\barX^{w\c}$ is well defined since $\bfX_{st}$ is square-integrable and $\mathcal F_t$-measurable by \autoref{def:adapted}.
\end{definition}

\begin{theorem}\label{thm:itoLift}
    Let~$\bfX$ be as above. The It\^o lift~$\barX$ of an adapted $H$-integrable geometric rough path~$\bfX$ is the unique (almost surely defined, stochastic) rough path which restricts to~\eqref{eq:itoL} on $\mathfrak{B}$, and is adapted in the sense that $\barX_{st}$ is $\mathcal{F}_t$-measurable. $\overline \bfX$ is H\"older continuous according to the inhomogeneous weighting on $[d]\sqcup [e]$, namely for any $p$ and almost all $\omega \in \Omega$
    \[
    |\overline \bfX{}^w_{st}(\omega)| \lesssim_{p,T,\omega}  (t-s)^l, \qquad l < |w|,
    \]
    with the constant of proportionality a random variable in $L^p(\Omega)$.
\end{theorem}

\begin{proof}
        The uniqueness statement directly follows from the corresponding statement in \autoref{lem:algebra}, since the integration by parts identity states that $\overline{\bfX}_{st}$ is an algebra map $\mathrm{Sh}\lceil \bbR^{d + e}\rceil \to \bbR$ for each $(s,t) \in \Delta_T$.
        The shuffle property can be stated as saying that for each $s,t$ the map $\barX_{st} \colon \mathrm{Sh}\lceil\mathbb{R}^{d+e} \rceil \to \mathbb{R}$ is an algebra morphism: this follows directly from the first part of \autoref{lem:algebra}, since~$\bfW$ and~$\bfX$ are algebra morphisms by assumption and by the integration-by-parts identity for Stratonovich calculus. As for the Chen identity, on words in $\mathfrak{B}$ this follows classically from the fact that It\^o and Stratonovich integration are additive on consecutive time intervals. We must check that it is preserved on the algebraic extension. Letting 
        \[
        ([e] \sqcup [d])^\bullet \ni w = \sum_k \lambda_k b^k_1 \shuffle \cdots \shuffle b^k_{n_k}
        \]
        with $b^i_j \in \mathfrak{B}$, we check the Chen identity in dual form~\eqref{eq:dualchen}:
        \begin{align*}
            \barX^w_{st} &= \sum_k\lambda_k \langle  b^k_1, \barX_{st} \rangle \cdots \langle  b^k_{n_k}, \barX_{st} \rangle \\
            &= \sum_k\lambda_k \langle  \Delta b^k_1, \barX_{su} \otimes \barX_{ut} \rangle \cdots \langle  \Delta b^k_{n_k}, \barX_{su} \otimes \barX_{ut} \rangle \\
            &= \sum_k\lambda_k \langle ( \Delta b^k_1) \shuffle^{\otimes 2} \cdots \shuffle^{\otimes 2} (\Delta b^k_{n_k}), \barX_{su} \otimes \barX_{ut} \rangle \\
            &= \sum_k\lambda_k \langle \Delta( b^k_1 \shuffle \cdots \shuffle  b^k_{n_k}), \barX_{su} \otimes \barX_{ut} \rangle \\
            &= \langle \Delta w, \barX_{su} \otimes \barX_{ut} \rangle,
        \end{align*}
        using the properties of bialgebras and that~$\bfX$ satisfies the Chen identity on words in~$\mathfrak{B}$. Adaptedness follows from the representation $\barX^w_{st} = \sum_k\lambda_k \langle  b^k_1, \barX_{st} \rangle \cdots \langle  b^k_{n_k}, \barX_{st} \rangle$ and adaptedness of~$\bfX$. Regularity follows similarly once it is established on $\barX^{b}, b \in \mathfrak{B}$, and keeping in mind that it holds for the It\^o rough path \cite[Proposition~3.4]{FH20}, it only remains to show it for words of the form~$w$, $w\a$, with $w \in [e]^\bullet$. We only treat the second case as the first is analogous. For $p \geq 2$ fixed, by \autoref{prop:BDG} and the Kolmogorov assumption
        \[
        \mathbb{E}[|\barX^{w\a}_{st}|^p] \lesssim_p (t-s)^{p(|w|+\half)},
        \]
        and thus for any $0 < c < \half+ |w| - \frac{2}{p}$ by \autoref{thm:Kolm}, there exists $J_p \in L^p$ with 
        \[
        \sup_{0 \leq s \leq t \leq T}|\barX^{w\a}_{st}| \leq J_p(t-s)^c.
        \]
        The required regularity now follows from considering an arbitrarily high $p$.
    \end{proof}

\begin{remark}[Non-geometric joint lifts]
    The assumption that~$\bfX$ be geometric is not essential: one could perform the It\^o lift of an adapted branched rough path, with a proof similar to that of \autoref{thm:itoLift}, in which the shuffle Hopf algebra is replaced with the Connes-Kreimer Hopf algebra. Similarly, it would have been possible to define~$\bfW^{\a\b}$ by It\^o integration instead of Stratonovich integration. We chose the latter in order for the resulting rough path to be geometric: this has the advantage of being the limit of smooth approximations, which we consider in \autoref{sec:lead-lag} below, as it is our numerical scheme of choice.
\end{remark}

\begin{example}[It\^o lift of Stratonovich Brownian motion]\label{expl:Bm}
As the only example with~$\barX$ multidimensional (below, it will always be the polynomial lift of a $1$-dimensional path), we consider the case in which~$X$ and~$W$ are both Brownian motions, not necessarily independent, defined on a common filtration. We take $H = \half$ and $\bfX$ to be the Stratonovich lift of $X$. Then we have $\barX^{i\a}_{st} = \int_s^t X^i_{su} \dif W^\a_u$, and using~\eqref{eq:ibp},
    \[
    \barX^{\a i}_{st} = W_{st}^\a X_{st}^i - \int_s^t X^i_{su} \dif W_u^\a = \int_s^t W^\a_{su} \dif X^i_u + [W^\a,X^i]_{st}.
    \]
    In particular, note that if~$X$ and~$W$ are \emph{the same} $d$-dimensional Brownian motion, $\barX^{\a i}_{st} \neq \barX^{i \a}_{st}$ (they differ by the deterministic quantity $t-s$). While this asymmetry may seem odd in the case in which the components~$X$ and~$W$ are identically distributed, the main examples that we consider in the next sections do not have this feature, and thus one should expect $\barX^{\a i}_{st}$ and $\barX^{i \a}_{st}$ to not be equal in law anyway.
    Note also that, since the canonical Stratonovich lift of $(X,W)$ differs at level-$2$ from~$\barX$ by a bounded variation path, equations driven by~$\barX$ can be written as Stratonovich (or It\^o) equations up to changing the drift.
\end{example}

Given $\bfW$, $\bfX$ in \autoref{def:adapted} and the It\^o lift~$\barX$, 
we can now give meaning to equations, such as~\eqref{eq:Model},
jointly driven by~$W$ (in the Stratonovich sense) and the original~$\bfX$. 
Including a drift term is never problematic in rough path theory and just amounts to extending the rough path with an extra letter for the path~$t$, which, thanks to its weight being set to~$1$, never has to be considered inside a word; in terms of the alphabet, this amounts to adding an extra letter $*$ with weight $1$, $\overline{\bfX}{}^*_{st} = t - s$, so that the full alphabet is now $[d] \sqcup [e] \sqcup \{*\}$; from now on this is omitted from the notation. The next proposition focuses on the well-posedness of~\eqref{eq:Model} in a special block form expressing part of the solution as an It\^o integral with drift. Since we always deal with single-asset models, $S$ is taken one-dimensional, but it is helpful to allow~$V$ to be multidimensional to encode additional states besides the volatility (however, whenever left unspecified, $V$ is also taken to be one-dimensional). It is important to note that, even though the RDE is defined in block form, the solution does generally depend on the lift $\overline \bfX$, beyond the individual rough paths $\bfW$ and $\bfX$---see \autoref{rem:ft} below for details.

\begin{proposition}\label{prop:block}
If $(S,V)$ is the solution to the RDE in block form, driven by $\overline \bfX$, with~$S$ real-valued,~$V$ valued in $\mathbb{R}^m$ (as specified in \eqref{eq:Model}), and the coefficients satisfying the assumptions that guarantee (local) existence and uniqueness of solutions \cite[Theorem~10.14]{FV10}, \cite[Theorem~4.3]{Gyu16} (which include differentiability)
\begin{equation}\label{eq:blockRDE}
    \begin{pmatrix}
        \dif S_t \\ \dif V_t
    \end{pmatrix} = \begin{pmatrix} 0 & \sigma & g \\ \tau & \varsigma & h \end{pmatrix}(S_t,V_t,t) \begin{pmatrix}
    \dif \bfX \\ \dif \mathbf W \\ \dif t
\end{pmatrix},
\end{equation}
then $S$ is given by It\^o and Riemann integration as
\[
S_t = S_0 + \int_0^t  \sigma(S_u,V_u,u) \dif W_u + \int_0^t \bigg[\frac 12\sum_\a(\partial_S \sigma_\a \sigma_\a + \partial_V \sigma_\a \varsigma_\a) + g \bigg](S_u,V_u,u) \d u.
\]
In particular, if $g = -\frac 12\sum_\a(\partial_S \sigma_\a \sigma_\a + \partial_V \sigma_\a \varsigma_\a)$, 
then $S$ is a local martingale.
\end{proposition}

\begin{proof}
    We compute the Davie expansion of the solution $S$ (see \cite[\S 10.2]{FV10}, but truncated at inhomogeneous degree) in~\eqref{eq:blockRDE}. For a word $k_1\ldots k_n$ in the alphabet $[e]$ we set recursively $\tau_{k_1\ldots k_n} \sigma_\c \coloneqq \partial_V(\tau_{k_2 \ldots k_n} \sigma_\c)\tau_{k_1}$. We use the Einstein convention over sub/superscript pairs of indices or words (up to the required regularity) and suppress the evaluations of all coefficients at $(S_u,V_u,u)$. We have
    \begin{align*}
    S_{uv} \approx{} &\sigma_\c W^\c_{uv} + (\partial_S \sigma_\b \sigma_\a + \partial_V \sigma_\b \varsigma_\a) \barX^{\a\b}_{uv} +  \tau_w \sigma_\c \barX^{w\c}_{uv} + g \cdot (v-u) \\
    ={} &\sigma_\c W^\c_{uv} + (\partial_S \sigma_\b \sigma_\a + \partial_V \sigma_\b \varsigma_\a)   \int_u^v W_{ur}^\a \dif W_r^\b +  \tau_w \sigma_\c \int_u^v \bfX_{ur}^w \d W^\c_r \\
    {}&+ \frac 12 (\partial_S \sigma_\b \sigma_\a + \partial_V \sigma_\b \varsigma_\a) [W]^{\a\b}_{uv} + g (v-u),
    \end{align*}
    where $\approx$ means that the left- and right-hand sides differ by $\mathcal{O}((v-u)^\theta)$, $\theta > 1$. 
    Since It\^o and rough integrations against the It\^o rough path coincide \cite[Theorem~9.1]{FH20}, this is precisely the Davie expansion of the required It\^o integral with drift.
\end{proof}

\begin{example}[Davie expansion of~$V$ with $\frac{1}{3} < H \leq \half$]\label{ex:davie}
It is instructive to also Davie-expand the second equation; we restrict ourselves to the simplest case of $\frac{1}{3} < H \leq \half$. Using the elementary integration-by-parts identity~\eqref{eq:ibp} (and omitting evaluations),
\begin{align*}
V_{uv} \approx{} &\tau X_{uv} + \varsigma_\c  W^\c_{uv} + h \cdot (v-u) + (\partial_V \varsigma_\b \varsigma_\a + \partial_S \varsigma_\b \sigma_\a)\int_u^v W_{ur}^\a \circ \dif W_r^\b + \partial_V \tau_j \tau_i \bfX_{uv}^{ij} \\
&+ \partial_V \varsigma_\c \tau_k  \int_u^v X^k_{ur} \d W^\c_r + (\partial_S \tau_k \sigma_\c + \partial_V \tau_k \varsigma_\c )\bigg(X_{uv}^k W^\c_{uv} - \int_u^v X^k_{ur} \d W^\c_r \bigg)  \\
={}&\tau X_{uv} + \varsigma_\c  W^\c_{uv} + h \cdot (v-u) + (\partial_V \varsigma_\b \varsigma_\a + \partial_S \varsigma_\b \sigma_\a)\int_u^v W_{ur}^\a \circ \dif W_r^\b + \partial_V \tau_j \tau_i \bfX_{uv}^{ij} \\
&+ [\left(\begin{smallmatrix}0\\ \tau_k\end{smallmatrix}\right),\left(\begin{smallmatrix} \sigma_\c \\ \varsigma_\c \end{smallmatrix}\right)]^V \int_u^v X^k_{ur} \d W^\c_r + (\partial_S \tau_k \sigma_\c + \partial_V \tau_k \varsigma_\c ) X_{uv}^kW^\c_{uv},
\end{align*}
where 
\[
[\left(\begin{smallmatrix}0\\ \tau_k\end{smallmatrix}\right),\left(\begin{smallmatrix} \sigma_\c \\ \varsigma_\c \end{smallmatrix}\right)]^V = \tau_k \partial_V \varsigma_\gamma - \sigma_\gamma \partial_S \tau_k - \varsigma_\gamma \partial_V \tau_k
\]
denotes the $V$-component of the Lie bracket of the two vectors. 
If~$X$ and~$W^\c$ admit finite quadratic co-variation (as in \autoref{expl:Bm}) we can write this as
\begin{align*}
V_{uv} \approx & \tau X_{uv} + \varsigma_\c  W^\c_{uv} + h \cdot (v-u) + (\partial_V \varsigma_\b \varsigma_\a + \partial_S \varsigma_\b \sigma_\a)\int_u^v W_{ur}^\a \circ \dif W_r^\b + \partial_V \tau_j \tau_i \bfX_{uv}^{ij} \\
& + \tau_k \partial_V \varsigma_\c \int_u^v X^k_{ur} \circ \d W_r^\c + (\partial_V \varsigma_\b \varsigma_\a + \partial_S \varsigma_\b \sigma_\a) \int_u^v W_{ur}^\c \circ \d X^k_r\\
 & - \frac 12 [\left(\begin{smallmatrix}0\\ \tau_k\end{smallmatrix}\right),\left(\begin{smallmatrix} \sigma_\c \\ \varsigma_\c \end{smallmatrix}\right)]^V [X^k,W^\c]_{uv}.
\end{align*}
Omitting the Lie bracket term would yield the Davie expansion for the equation interpreted in Stratonovich form. In general, however, all three $\int_u^v X^k_{ur} \circ \d W^\c_r$, $\int_u^v W^\c_{ur} \circ \d X^k_{ur}$ and $[X^k,W^\c]_{uv}$ can be divergent, and the previous expression can be viewed as redistributing this infinite correction in such a way that all resulting terms are finite; the Lie bracket term is necessary for this to happen. The relationship between quadratic variation and antisymmetric $2$-rough path term has already been observed~\cite{FHL16} for semimartingale rough paths, for which this correction is finite. At the lower regularities studied in the next sections, such expansions would greatly increase in complexity.
\end{example}

\begin{remark}[$\mathfrak{B}$ and Lyndon words]
    One may wonder about the relationship between the set $\mathfrak{B}$~\eqref{eq:B} and the set of Lyndon words of weight $\leq 1$, in which we order $[e] < [d]$. A word is Lyndon if it is lexicographically smaller than all of its proper rotations; Lyndon words are of significance in that they are a free set of shuffle generators \cite[\S 6.1]{Reu93}. It is obviously not true that every word in $\mathfrak{B}$ is Lyndon, for example $\a\b$ with $\a \geq \b$. While it is true that words of the form $z\c$ with $z \in [e]^\bullet$ is Lyndon if and only if $z$ is Lyndon, there are Lyndon words that do not belong to $\mathfrak{B}$, such as $i\c j$ with $i < j$, assuming that $e > 1$. If, however, $e = 1$ (the case of interest in the next section), $\mathfrak{B}$ does contain all Lyndon words of weight $\leq 1$: any word with a single letter in $[d]$ and a trailing letter in $[e] = \{1\}$ admits a rotation that is lexicographically smaller than it, and therefore cannot be Lyndon.
\end{remark}

\begin{remark}[Relationship with backward integration]\label{rem:back}
Let us return to the more general setting and recall the backward integration
\begin{equation}\label{eq:back}
    \int Y \cev{\dif}Z \coloneqq L^2\!\lim_{n \to \infty} \sum_{[s,t] \in \pi_n} Y_t Z_{st},
    \end{equation}
    which is equal to $\int Y \dif Z + [Y,Z]$ when the quadratic covariation is well defined. The definition of It\^o lift in \autoref{def:itoLift} is related to backward integration. For example, again by~\eqref{eq:ibp}, we have
    \begin{align*}
    \barX^{\a i}_{st} &= X_t^\a X^i_{st} - L^2\!\lim_{n \to \infty} \sum_{[u,v] \in \pi_n} X_u^i X^\a_{uv}  \\
    &= (X^\a X^i)_{st} - X_{st}^iX_s^\a - L^2\!\lim_{n \to \infty} \sum_{[u,v] \in \pi_n} \big[ (X^\a X^i)_{uv} - X^\a_v X^i_{uv} \big]
    = \int_s^t X^\a_{su} \cev{\dif} X^i_u.
    \end{align*}
    Note that the integral exists as a limit in $L^2$ even if the quadratic covariation $[X^\a,X^i]$ (which only exists if the forward integral $\int_s^t X^\a_{su} \dif X^i_u$ does) may not.
\end{remark}

\section{A rough path for rough volatility}\label{sec:roughVol}

From now on, we assume~$X$ to be a one-dimensional adapted process; 
one of the main (but not the only) examples is that of a fractional Brownian motion with Hurst parameter $H \in(0,\half)$. We reserve $0$ to denote the coordinate of~$X$ and continue to use $1,\ldots,d$ for the coordinates of the Brownian motion. 
Recall that $\overline \bfX$ then represents a rough path above the process $(X, W)$ and we let~$0^n$ denote the string of~$n$ zeros. There is a unique way of lifting~$X$ to a geometric rough path, by taking powers:
\begin{equation}\label{eq:CanocicalLift1dim}
    \bfX^{0^n}_{st} = \frac{X^n_{st}}{n!}.
\end{equation}
\autoref{def:adapted} then becomes a simple condition on the moments of the increments of~$X$. The terms $\barX^w$ with $w \in \mathfrak{B}$ in~\eqref{eq:B} constitute the partial rough path of~\cite{FT22}, and \autoref{thm:itoLift} implies that~$\barX$ is its unique extension to a geometric rough path.
In this case, an alternative description of the remaining terms of~$\barX$ is available. If the paths were deterministic and smooth, we could write
\begin{align*}
\barX^{0^m \a 0^n}_{st} &\text{\say{$=$}} \int_{s < r_1 < \ldots < r_m < u < v_1 < \ldots < v_n < t} \d X_{r_1} \cdots \d X_{r_m} \d W_u^\a \d X_{v_1} \cdots \d X_{v_n} \\
&= \int_s^t \bfX^{0^m}_{su} \bfX^{0^n}_{ut} \d W^\a_u \\
&= \frac{1}{m! n!} \int_s^t X^m_{su} X^n_{ut} \d W^\a_u.
\end{align*}
As written, however, this requires special care since $X^n_{ut}$ is not $\mathcal{F}_u$-measurable, which is needed to consider the It\^o integral. We can however apply the binomial theorem to $X^n_{ut} = (X_t - X_u)^n$ to take~$X_t$, which is not $\mathcal{F}_u$-measurable but constant in $u$, out of the integral (see~\cite[p.801]{BFGJS20} for a similar idea used in a rather different context). The next proposition shows that this yields the same construction as \autoref{def:itoLift}, providing a closed-form expression for the recursion~\eqref{eq:recursion} for this type of~$\bfX$. Note that the original definition of~$\barX$ is still convenient for assessing its regularity, which is less clear using this representation; moreover, it applies to much more general choices of~$\bfX$.
\begin{proposition}\label{prop:binomial} It holds that
\[
\barX^{0^m \a 0^n}_{st} = \sum_{k = 0}^n\frac{X^{n-k}_t}{m! k! (n-k)!} \int_s^t X^m_{su} (-X_u)^k \d W^\a_u.
\]
\end{proposition}

\begin{proof}
We may call $0^m \eqqcolon w$ and forget about its particular form (in fact, this statement applies more generally in the framework of the last section, but with the letters after $\a$ all equal). It is convenient to use the usual notation for It\^o integrals with possibly non-adapted integrands, intended in the sense of limit in~$L^2$ of forward Riemann sums. In the cases of interest here, this does indeed converge thanks to the binomial expansion above. 
Taking~$\barX$ as defined in the statement of the proposition, we then have
\begin{align*} \barX^{w\a}_{st}\barX^{0^n}_{st} &= \frac{X^n_{st}}{n!}\int_s^t \bfX^w_{su} \d W^\a_u \\
 &= \sum_{k = 0}^n \frac{1}{k! (n-k)!} \int_s^t \bfX^w_{su} X^k_{su} X^{n-k}_{ut} \d W^\a_u \\
 &= \sum_{k = 0}^n \barX^{(w \shuffle 0^k)\a 0^{n-k}}_{st},
\end{align*}
where we have written $X_{st}^n = (X_{su} + X_{ut})^n$ and applied the binomial theorem. This shows the terms of~$\barX$ can be computed via the recursion~\eqref{eq:recursion}, and since for the base case $n = 0$ we have $\overline \bfX{}^{0^m \a}_{st} = m!^{-1}\int_s^t X_{st}^m \d W^\a$, we have shown that the two definitions of $\overline \bfX$ agree on all words, the conclusion follows.
\end{proof}

We propose to jointly model the price~$S$ and the volatility~$V$ in a single-asset market by rough differential equations driven by~$\barX$ of the block form~\eqref{eq:blockRDE} (which is equivalent to~\eqref{eq:Model}), with~$X$ one-dimensional. 
For most applications, the choice of the dimension~$d$ of the Brownian motion can be taken to be $1$, but we allow it to be general since this comes at no extra cost. In most volatility models, the dynamics of~$V$, namely the coefficients $\tau, \varsigma$ and $h$, are taken independent of~$S$, 
but we do not impose this as a strict requirement.

The H\"older regularity of~$V$ is the worst of its driving signals, namely $H^-$, unless $\tau = 0$; we are therefore justified in calling this a rough volatility model. It encompasses rough volatility models in which~$V$ is an explicit function of~$X$ (such as rough Bergomi~\cite{BFG16})
by the classical change of variable formula $f(X) = \int\! f'(X) \d \bfX$. We now analyse a couple of special cases.

\begin{remark}[The case $\varsigma = 0$ and $\tau$ not function of $S_t$, and~\cite{FT22}]\label{rem:ft}
    It was pointed out in~\cite{LV07} that not all RDEs depend on (or even require) the full collection of rough path terms. The most extreme case is that of commuting vector fields, in which no rough path term is required at all, beyond the trace. 
    More generally, the rule of thumb is simple: rough path terms that do not appear in the Davie expansion are not needed to define the RDE. In the special case where $\varsigma = 0$ and $\tau$ does not not depend on~$S_t$, 
    a single (fractional) noise term and drift drive the volatility, 
    the proof of \autoref{prop:block} (and the Davie expansion for~$V$) shows that only the terms of the partial rough path in~\cite{FT22} are needed. Note, however, that~\cite{FT22} only considers the case in which~$V$ is one-dimensional and given explicitly as a function of~$X$, or equivalently as an integral $\int\! f(X) \d X$; this is sufficient to cover several models of interest, such as rough Bergomi (as introduced in Section~\ref{subsec: Applications} below). If $\varsigma \neq 0$ the equation for~$V$ is driven by a multidimensional rough path; if $\tau$ depends on $S$ the term $\partial_S \tau \sigma_\a \barX^{\a 0}$ appears in the expansion: in either case, the partial rough path of~\cite{FT22} is no longer sufficient.
\end{remark}

\begin{remark}[The case $X \indep W$ and Gaussian rough paths]\label{rem:XindW}
If~$X$ is a one-dimensional Gaussian process uncorrelated from~$W$, $(X,W)$ can actually be lifted to a classical Gaussian rough path~\cite{coutin2002stochastic,FV10b}. The only difference with the processes considered therein is that the components are not identically distributed, but the analysis carries over. If $Z$ is a multidimensional fractional Brownian motion with independent increments, each component~$k$ with its own Hurst parameter~$H_k$, the condition that guarantees existence of the L\'evy area as a limit in $L^2$ (as well as that of higher terms) is that $H_i + H_j > \half$ for any $i \neq j$. 
This is somewhat similar to the condition for Young integrability, except for the fact that the regularity of integrand and integrator only needs to sum to more than $\half$, not $1$, thanks to stochastic cancellations given by independence of components.
When $H_i=H_j$ for all $i,j$, this reduces to the well-known condition $H_i > \frac{1}{4}$ for all~$i$,
and it is automatically satisfied in our case $Z = (X,W)$ when $H_k \geq \half$ for all but one component. We do not go into the details, which would involve rehashing the theory of Gaussian rough paths, allowing for the components not to be identically distributed. Also, for $(X,W)$ this Gaussian lift is actually a special case of the lift constructed in \autoref{def:itoLift}, since $[X,W^\a] = 0$ by independence. Note, also, that this opens up the interesting possibility of taking some components of the noise to be fBm's with $H > \half$ to model long-term behaviour, for example as was proposed in~\cite{cheridito2001mixed,jacquier2025rough,lacombe2021asymptotics}.
\end{remark}
Equation~\eqref{eq:blockRDE} is very flexible in modelling many features of the volatility and price process, some of which are alternative to one another. For example, correlation between~$S$ and~$V$ can either be introduced as done usually, by considering~$X$ and~$W$ to be correlated, or by violating the conditions of \autoref{rem:ft} and obtaining the dependence via the equation (in which case one can take $X \indep W$ as in \autoref{rem:XindW}). The latter choice, parametrised by functions, is much more flexible; of course one may also choose a combination of the two.

\subsection{Applications to volatility modelling}\label{subsec: Applications}

Let us now illustrate how our framework~\eqref{eq:Model} (equivalently~\eqref{eq:blockRDE}) encompasses different classes of well-known volatility models as special cases. Before doing so, we make an important remark on the smoothness assumptions of the coefficients:

\begin{remark}[Unbounded and non-smooth vector fields]\label{rem:square-roots} 
Several well-known volatility models in this section (such as the rough Heston model) feature square-root coefficients that are not smooth enough to guarantee global well-posedness.
Nevertheless, our local well-posedness result in Proposition~\ref{prop:block} can still be applied in this setting. 
In this case, unique (pathwise) RDE solutions are only guaranteed to exist locally, up to explosion time. For example, if $\sigma(S, V,t)=\sqrt{V}$ then $S_t(\omega)$  exists for all $t\leq\tau(\omega)=\inf\{t>0: V_t(\omega)=0\}$.
A related discussion on the reconciliation of square-root models and rough path considerations can be found in \cite[\S5.1]{BFGJS20}. Of course, the above comments are not specific to square-root coefficients and they apply (mutatis mutandis) to models with smooth vector fields with unbounded derivatives.
\end{remark}

Throughout the rest of this section we shall, unless otherwise stated, set $e=1$,
and consider all dynamics under the risk-neutral measure
assuming no interest rates. 

\begin{itemize}
    \item[a)] The first class that we consider is that of \emph{classical stochastic volatility models}: 
    \begin{equation*}
        \left\{
        \begin{array}{rll}
        \displaystyle \frac{\d S_t}{S_t} & =  \xi(t,S_t)\zeta(\Vv_t)\d W_t, & S_0=s_0\in\bbR,\\
         \Vv_t & = \displaystyle \Vv_0 + \int_{0}^{t}f_1(u,\Vv_u)\d u + \int_{0}^{t} f_2(\Vv_u)
         \Big(\rho \d W_u + \overline\rho\d B_u\Big), & \Vv_0 = v_0\in\bbR,
        \end{array}
        \right.
    \end{equation*}

     with~$B$ and~$W$ two independent Brownian motions, $\rho \in [-1,1]$, $\overline{\rho}:=\sqrt{1-\rho^2}$, and~$\xi$ accounting for a local volatility component.
     This class is obtained by setting $d=m=1$, 
     $\sigma_1(s, v, t) = s\xi(t, s)\zeta(v)$,
     $X=B$, $H=\half$, 
     $\tau(s, v, t)=\overline{\rho} f_2(v)$, $\varsigma_1(s, v, t)=\rho f_2(v)$ for functions $\xi:[0,\infty)\times \bbR\rightarrow \bbR, f_2:\bbR\rightarrow\bbR$. 
     Since $X$ is one-dimensional, $\bfX$ is taken to be the unique geometric rough-path lift as defined in~\eqref{eq:CanocicalLift1dim} and for $n\leq 2$. Clearly, since $B$ has finite moments of all orders, the moment conditions of Definition~\ref{def:adapted} are satisfied by standard moment bounds and~$\bfX$ is an adapted $H$-integrable rough path.
     In view of Proposition~\ref{prop:block} and Example~\ref{ex:davie}, and since~$X, W$ are independent Brownian motions, stochastic integrals in~\eqref{eq:Model} are interpreted in Stratonovich sense. Accounting for the Stratonovich-It\^o conversion in $\mathcal{V}, S$ then amounts to choosing respectively $$h(s, v, t)=f_1(t,v)-\half\big(  \tau\partial_v\tau+ \varsigma_1\partial_v\varsigma_1\big)(s,v,t)= f_1(t, v)-\half f_2'(v)f_2(v),$$
      for a function $f_1:[0,\infty)\times\bbR\rightarrow\bbR$, and 
\begin{align*}
g(s,v,t)
 & = -\half(\sigma_1\partial_s\sigma_1+\varsigma_1\partial_v\sigma_1)(s,v,t)\\
 & =-\half[s\xi^2\zeta^2+s^2\zeta^2\xi\partial_s\xi+\rho f_2s\xi\zeta'](s,v,t).
\end{align*}
In particular, the following classical models can be recovered:
    \begin{itemize}
        \item The Black-Scholes model~\cite{black1973pricing} is obtained by setting $f_1=f_2=0$, $\xi\equiv 1$, $\zeta(v)\equiv\sigma$ for some $\sigma\in(0, \infty)$.
        The drifts $h,g$ are given by $h\equiv 0$ and $g(s,v,t)=-\half \sigma^2 s$. All the involved vector fields are infinitely-many times continuously differentiable, hence the assumptions of Proposition~\ref{prop:block} are satisfied (up to explosion times).
        
        \item The Heston model~\cite{heston1993closed} with mean-reversion speed $\lambda>0$, vol-of-vol $\nu>0$ and mean-reversion level $\overline{v}>0$ is recovered with $\xi\equiv 1$, 
        $f_1(t, v)=-\lambda (v-\overline{v})$, 
        $f_2( v)=\nu\sqrt{v}$, $\zeta(v) = \sqrt{v}$. The drifts $h,g$ are given by $h(s,v,t)=f_1(t,v)-\tfrac{1}{4}\nu^2$ and $g(s,v,t)=-\half s(  v+\half \rho\nu)$. Apart from $f_1, h, g$ the rest of the vector fields feature square-roots whose derivatives are singular at the origin. Hence, Proposition~\ref{prop:block} can only guarantee existence of unique local solutions (up to explosion time); see Remark~\ref{rem:square-roots}.
        
        \item The Bergomi model~\cite{bergomi2005smile} is recovered with $\xi\equiv 1$, $f_1= 0$, $f_2(v)\equiv v$, $\zeta(v) = \mathrm{e}^{v}$. 
        The drifts $h,g$ are given by $h(s,v,t)=-\half v$ and $g(s,v,t)=-\half s(\mathrm{e}^{2v}+\rho v\mathrm{e}^{v})$. 
        Again, the involved vector fields are smooth and satisfy the assumptions of Proposition~\ref{prop:block} (at least up to explosion times).
        
        \item The Stein-Stein model~\cite{stein1991stock} is given by $\xi\equiv 1$, $f_1(t, v)=-\lambda (v-\overline{v})$, $f_2(v)=1$, $\zeta(v) = \sqrt{v}$.  The drifts $h,g$ are given by $h(s,v,t)=f_1(t,v)$ and $g(s,v,t)=-\half s( v+\tfrac{1}{2\sqrt{v}})$. Again, due to the presence of square-roots, Proposition~\ref{prop:block} only yields local solutions (defined up to explosion time); see Remark~\ref{rem:square-roots}.
    \end{itemize}

    \item[b)] \emph{Classical local stochastic volatility models} trivially fit in the above framework by setting
    $\sigma_{1}(s,v,t) = s\xi(t,s,v)$ for some (smooth) function $\xi$, letting $\zeta \equiv 1$ and choosing any stochastic process for~$\Vv$ as in~a).

\item[c)] Popular rough volatility models can be (partially) encompassed within our framework (see Remark \ref{rem:SVEsvsRDEs} for a clarification of the word\textit{ partially}): 
\begin{itemize}
    \item[1)] The rough Heston model \cite{el2019characteristic,guennoun2018asymptotic}  is given by 
    \begin{equation*}
\left\{
\begin{array}{rl}
\displaystyle \frac{\d S_t}{S_t} & = \sqrt{\Vv_t}\d W_t,\quad S_0=s_0\in\bbR,\\
 \Vv_t & = \displaystyle  \Vv_0
 + \int_{0}^{t}K(t-u)\Big(
 -\lambda(\Vv_u-\overline{v})\d u + \sqrt\Vv_u(\rho \d W_u + \overline{\rho}\d B_u)\Big),
\end{array}
\right.
\end{equation*}
where $\nu, \overline{v}, \lambda>0$, $\rho \in [-1,1]$,
$\overline{\rho}:=\sqrt{1-\rho^2}$ and $K(u)=u^{H-\half}/\Gamma(H+\half)$ on $(0,\infty)$, 
with $H\in (0, \half)$, where~$\Gamma$ denotes the Gamma function. To recover this model from~\eqref{eq:Model} we set  $m=1$, $d=1$, $\sigma_1(s, v, t)=s\sqrt{v}, \tau\equiv 1$, $\varsigma_1\equiv 0$, $h\equiv 0$ and take~$X$ to be a path that is equal in law to the unique weak solution~$\Vv$ of the variance equation. The associated adapted $H$-integrable rough path $\bfX$ is thus given by~\eqref{eq:CanocicalLift1dim} and the moment assumptions of Definition~\ref{def:adapted} are satisfied since $\mathcal{V}$ has finite moments of all orders~\cite[Lemma 3.1]{abi2019affine}.
The Stratonovich-It\^o conversion in the equation for $S$ amounts to choosing a constant drift $g(s,t,v)=-\half\sigma_1\partial_s\sigma_1=-\half sv$. Square-root coefficients are not restrictive for the existence of local solutions per Remark~\ref{rem:square-roots}.
    
\item[2)] The rough Bergomi model, introduced in~\cite{BFG16}, is given by 
     \begin{equation*}
\left\{
\begin{array}{rl}
\displaystyle \frac{\d S_t}{S_t} & = \displaystyle
\sqrt{v_0}\exp\left\{\nu \Vv_t-\frac{\nu^2 t^{2H}}{2\Gamma(H+\half)^2}\right\}\d W_t,\quad S_0 = s_0\in\bbR,\\
 \Vv_t & = \displaystyle
 \frac{1}{\Gamma(H+\half)}\int_0^t(t-s)^{H-\half}\Big(\rho \d W_s + \sqrt{1-\rho^2}\ \d B_s\Big),
\end{array}
\right.
\end{equation*}
where $\nu, v_0>0$, $H\in (0, \half)$, $\rho \in [-1,1]$.
To recover this model from~\eqref{eq:Model}, set  $m=1$, $d=1$, $\sigma_1(s, v, t) = s\sqrt{v_0}\exp\left\{\nu v-\frac{\nu^2 t^{2H}}{2\Gamma(H+\half)^2}\right\}$, $\tau\equiv 1$, $\varsigma_1\equiv0$, 
$h\equiv 0$ and $X=\Vv$.
The $H$-integrable adapted rough path~$\bfX$ is given by~\eqref{eq:CanocicalLift1dim} and the moment conditions of Definition~\ref{def:adapted} are satisfied since $X$ is a Gaussian process with almost surely $\gamma$-H\"older continuous paths for any $\gamma<H$.
The Stratonovich-It\^o conversion in the equation for $S$ amounts to choosing a drift $g(s,t,v)=-\half\sigma_1\partial_s\sigma_1=-\half\nu s v_0\exp\left\{2\nu v-\frac{\nu^2 t^{2H}}{\Gamma(H+\half)^2}\right\}$.
Finally, all the associated vector fields are (locally) smooth and satisfy the assumptions of Proposition~\ref{prop:block} per Remark~\ref{rem:square-roots}.  

    \item[3)] Multi-factor rough Bergomi models with one fractional and multiple Markovian volatility factors:
    let $N\in\bbN$, $H\in(0, \half)$, 
    $\rho=(\rho_1,\dots, \rho_N)\in [-1, 1]^N$, such that $\sum_i\rho_i^2=1$ and consider
        \begin{equation*}
\left\{
\begin{array}{rll}
\displaystyle \frac{\d S_t}{S_t} & = \displaystyle 
f(t, \Vv_t) \sum_{i=1}^{N}\rho_i          \d W^i_t,& S_0=s_0\in\bbR,\\
 \Vv^i_t & =
 \displaystyle \frac{1}{\Gamma(H_i+\half)}\int_0^t(t-s)^{H_i-\half} \d W^i_s,
 & i=1,\dots, N,
\end{array}
\right.
\end{equation*}
where $H_1=H$ and $H_i=\half$ for $i\neq 1$.
The volatility function $f:\bbR^N\rightarrow\bbR$ reads
$$
f^2(t,v) := \sum_{j=1}^{N}\chi_j \exp\left\{\frac{1}{2}\left(\nu_jv^j-\frac{\nu_j^2 t^{2H_j}}{2\Gamma^2(H_j+\half)} \right)\right\}, 
\quad (t, v)\in[0,\infty)\times\bbR^N,
$$
and
$ \{\chi_j\}_{j=1}^{N}, \{\nu_j\}_{j=1}^{N}\subset\bbR^+$ are suitable parameters. 
This family fits into the general framework after setting 
$m=d=N, h\equiv 0$, $\tau = (1,0,\dots, 0)$, $\sigma_\alpha:\bbR^{N+1}\times [0,\infty)\rightarrow \bbR$ be given by $\sigma_\alpha(s,v,t)=\rho_\alpha sf(t,v), \alpha=1,\dots, N$, $\varsigma\in\mathcal{L}(\bbR^{N};\bbR^{N})$ a constant, square matrix with zeroes on the first row, ones on the diagonal element of the $i$-th row for $i\ne 1$ and zero on every other entry, $X=\Vv^1$
 and $W^{i}=\Vv^i$ when $i\ne 1$. 
 The associated $H$-integrable adapted rough path $\bfX$ is defined exactly as in 2). The Stratonovich-It\^o conversion in the equation for $S$ amounts to choosing a drift $g(s,t,v)=-\half\sum_{\alpha}(\sigma_\alpha\partial_s\sigma_\alpha-\varsigma_\alpha\partial_{v^\alpha}\sigma_\alpha)$. Finally all associated vector fields are (locally) smooth and satisfy the assumptions of Proposition~\ref{prop:block} per Remark~\ref{rem:square-roots}.

 \item[4)] The quadratic rough Heston model, proposed in~\cite{GJR20} and further investigated in~\cite{RZ21},
 is an example of a continuous-path model that achieves joint SPX-VIX smile calibration,
 and reads
       \begin{equation*}
    \begin{array}{rl}
    \displaystyle\frac{\dif S_t}{S_t} &=  \displaystyle- \half \left(a(\Vv_t - b)^2 + c\right) \dif t+\displaystyle \sqrt{a(\Vv_t - b)^2 + c} \dif W_t,
    \quad S_0 = s_0>0, \\ 
    \Vv_t  &= \displaystyle v_0+ \frac{\lambda}{\Gamma(H+\half)} \int_0^t K(t-s)\left[(\theta_{s}- \Vv_s)\dif s+ \eta\left[a(\Vv_s - b)^2 + c\right]^{\half} \dif  W_s\right],
    \end{array}
\end{equation*}
with $v_0,a, b, c, \lambda, \eta>0$, $H\in(0,\half)$,$\theta:\bbR^+\rightarrow\bbR$ a suitably chosen deterministic function and $K(u) = u^{H-\half}$. 
This model is recovered from~\eqref{eq:Model} by setting $m=1$, $d=1$, $\sigma_1(s, v, t) = s\sqrt{a(v - b)^2 + c}$, $\tau=1$, $\varsigma_1=0$, $h\equiv0$ and taking~$X$ to be a (one-dimensional) path that is equal in law to the unique weak solution~$\Vv$ of the Volterra SDE given above. 

 As above, since $X$ is one-dimensional, $\bfX$ is given by~\eqref{eq:CanocicalLift1dim} with $n\leq \lceil1/H\rceil$. Furthermore, $\mathcal{V}$ solves a stochastic Volterra equation with globally Lipschitz-continuous coefficients (since $c>0$) that have at most linear growth. Hence, in view of~\cite[Theorems~3.1 and~3.3]{zhang2010stochastic}, $$\mathbb{E}\bigg[\sup_{s\neq t}\frac{|X_{s,t}|^p}{|t-s|^{p\gamma}}\bigg]=\mathbb{E}\bigg[\sup_{s\neq t}\frac{|\mathcal{V}_{s,t}|^p}{|t-s|^{p\gamma}}\bigg]<\infty,
 $$
for any $p\geq 1, \gamma<H$
and $\bfX$ is an adapted $H$-integrable rough path per Definition~\ref{def:adapted}.
The Stratonovich-It\^o conversion in the equation for~$S$ amounts to choosing a drift $g(s,t,v)=-\half (a(v - b)^2 + c)-\half\sigma_1\partial_s\sigma_1=-(a(v - b)^2 + c)$. Finally, the vector fields involved in the dynamics of $S$ are smooth and thus the assumptions of Proposition~\ref{prop:block} are once again satisfied.
\end{itemize}

\begin{remark}\label{rem:SVEsvsRDEs}
   In contrast to  classical stochastic volatility models, it is worth noting that rough volatility models with $\mathcal{V}$ solving non-trivial stochastic Volterra equations (such as rough Heston and quadratic rough Heston above) are only partially encompassed by our framework. Indeed, in such cases, the volatility dynamics are not described by the RDE \eqref{eq:Model}; instead, we take $X$ to be the unique (local) solution of the corresponding Volterra equation and then lift the latter to a rough path. This is not surprising since, as argued in Section \ref{sec:Intro}, Volterra equations are genuinely different from RDEs.
\end{remark}

\item[c)] 
Path-dependent volatility models, suggested by Hobson and Rogers~\cite{hobson1998complete}, have received recent impetus, notably in~\cite{guyon2014path,parent2023ewma}, and take the following form:
$$
\frac{\d S_t}{S_t} = \sigma\left(S_{u \in [0,t]}\right) \d W_t,
$$
in which the asset price at a given time~$t$ depends on its historical path (and its quadratic variation) up to time~$t$. 
As mentioned there, models in which the Brownian motion driving the price~$S$ and the noise driving the volatility~$V$ have correlation $\rho=\pm 1$ feature full path-dependence.
Apart from these fully correlated (or anti-correlated) examples,~\eqref{eq:Model} allows for different types of path-dependent volatility models, even in the case when~$X$ and~$W$ are independent. 
Path-dependence can be introduced in the form of price-dependent volatility dynamics. Indeed, one can directly check that, in~\eqref{eq:Model}, $V$ can be written as a map of the path $\{S_t; t\in[0,T]\}$.
In turn, the price dynamics inherit this property via feedback from the driving vector fields. Even though this is a particular type of path-dependence, it is sufficient to capture Zumbach-type effects~\cite{Zum09,Zum10} 
as pointed out in~\cite{GL23}.
A particular path-dependent volatility model is the one by Guyon and Lekeufack~\cite{GL23}, 
which (in the particular case $\Delta=0$) reads
\begin{equation*}
    \left\{
    \begin{aligned}
        \displaystyle \frac{\d S_t}{S_t} & = \sqrt{\beta_{0} + \beta_{1}R_{1,t} + \beta_{2}\sqrt{R_{2,t}}}\d W_t, & S_0=s_0\in\bbR,\\
        \displaystyle \begin{pmatrix}R_{1,t}\\R_{2,t}\end{pmatrix} 
        & = \displaystyle \int_0^t\begin{pmatrix}K_1(u,t)\\0\end{pmatrix}\left(\beta_{0} + \beta_{1}R_{1,u} + \beta_{2}\sqrt{R_{2,u}}\right) \d W_u,\\
        &+ \int_0^t \begin{pmatrix}0\\
        K_2(u,t)\end{pmatrix} \left(\beta_{0} + \beta_{1}R_{1,u} + \beta_{2}\sqrt{R_{2,u}}\right)^2\d u,
    \end{aligned}
    \right.
\end{equation*}
where $K_1, K_2:[0,T]^2\rightarrow\bbR^+$ are convolution-type kernels (such as shifted power-laws or sums of exponentials) 

one of which (say $K_2$) is non-singular.
This model is recovered from~\eqref{eq:Model} by setting  $e=m=2$, $d=1$, $H\in (0, \half)$ be the H\"older regularity of the "rough" component $R_{1}$, $\sigma_1(s, v^1, v^2, t)=s\sqrt{\beta_0+\beta_1v^1+\beta_2\sqrt{v^2}}, \tau\equiv 1$, $\varsigma_1\equiv 0$, $h\equiv 0$. We embed the Volterra-type volatility dynamics by letting $\Vv:=(\Vv^1, \Vv^2)\equiv (R_1, R_2)$ be a path that is equal in law to the unique (probabilistically weak) solutions of the above Volterra SDEs and setting $X\equiv \mathcal{V}$. 
Since the volatility process $\mathcal{V}$ is an affine Volterra process, $X^1=\mathcal{V}^1, X^2=\mathcal{V}^2$ have finite moments of all orders~\cite[Lemma 3.1]{abi2019affine}.
As a consequence, 
\begin{equation}\label{eq:int}
\bfX^w_{st}=\int_{s < u_1 < \ldots < u_n < t} \d X^{a_1}_{u_1} \cdots \d X^{a_n}_{u_1}
\end{equation}
are well-defined Young integrals for any word $w=a_1\dots a_n$ from the alphabet $\{1, 2\}$ with $|w|\leq1$ (where the weights of each letter are given by $|1|=H, |2|=1$).
From these observations it follows that $\bfX$ is an adapted $H$-integrable rough path per Definition~\ref{def:adapted}.

The Stratonovich-It\^o conversion in the equation for $S$ amounts to choosing a drift $g(s,t,v)=-\half\sigma_1\partial_s\sigma_1$.
We remark here that this model presents the only example in which we consider a two-dimensional path~$X$ driving the volatility. This particular extension is well accommodated within our framework since the dynamics of $R_2$ feature only Riemann integrals and hence its path regularity is higher  than that of Brownian motion. Finally, since~$\sigma_1$ features square-roots, Proposition~\ref{prop:block} yields existence of local solutions per Remark~\ref{rem:square-roots}.
\end{itemize}

We conclude this section with a remark on potential extensions of our framework to multi-asset models.

\begin{remark}[Multi-asset model]\label{rem:Multi-asset}
Note that~\eqref{eq:Model} can also be extended to an $n$-asset model, namely an RDE for $(S^1,V^1;\ldots; S^n,V^n)$. Here each~$V^p$ would be driven by its own~$X^p$, and the correlation structure of $(X,W)$ could still be arbitrary, as long as the coefficients for~$V^p$, specifically $\tau^p$, only depends on~$V^p$ and not the other components of~$V$. This guarantees that undefined terms $\bfX^{pq}$ never appear in the expansion. For each~$p$, the dynamics of~$S^p$ should not depend on~$V^q$ with $q \neq p$, but may still depend on~$S^q$ for arbitrary~$q$. This could be important to allow the model to reflect the causal influence that the different asset prices exert on one another. We leave it to the reader to check that the resulting Davie expansion only has terms that are well defined, in the same way that they are in the single-asset case.
\end{remark}

\section{Lead-lag approximations}\label{sec:lead-lag}

In principle, as any rough path, \autoref{def:itoLift} automatically comes with a numerical approximation scheme, the Davie scheme. The terms of the rough path can be calculated even more explicitly thanks to \autoref{prop:binomial}. However, the Davie expansion, especially that for~$V$, contains many terms. Not only is this cumbersome to write out, but it also contains derivatives of up to order $\lfloor H^{-1}\rfloor$. This leads to numerical schemes that are highly prone to being ill-conditioned, especially in cases where the volatility function~$\sigma$ is not differentiable near zero (such as the square root).
For this reason we choose to approximate our RDEs via Wong-Zakai-type approximations, that is by solving ODEs driven by smoothed noise which converges to~$\barX$ in rough path metric. The challenge is to find the sequence $(X^\eps, W^\eps)$ which achieves this convergence: it is necessary to smoothly approximate the It\^o integrals $\int_s^t X^m_{su} \d W_u^\a$. For this we borrow the idea of \emph{lead-lag approximations} from~\cite{FHL16}, first defined in the case in which the integrand is also a Brownian motion. We do not consider Hoff processes as done therein, rather we focus on the two best known ways of approximating an irregular path: piecewise-linear interpolation and convolution with a rescaled mollifier, as in~\cite{BCF18} where it is shown that the L\'evy area of the piecewise lead-lag approximation of (i.i.d.) $(\halff,\half] \ni H$-fBm captures the divergent quadratic variation. 

This section is organised as follows. Section~\ref{subsection:LeadLagPiecewise} is devoted to the convergence of piecewise-linear lead-lag approximations with explicit rates (Theorem~\ref{thm_leadlag}). In Section~\ref{subsec:HybridLeadLag}, and in particular in Theorem~\ref{thm:HybridLLconvergence}, we show that such approximations also converge to the correct iterated It\^o integrals when~$X$ is given by  a hybrid scheme approximation of a fractional Brownian motion. This provides a rigorous justification for the numerical simulations of Section~\ref{sec:SimCal}.  In passing, we also obtain a novel almost-sure convergence result of the hybrid scheme approximation in H\"older topology
(Theorem~\ref{thm:hybridalmostsure}).
Finally, in Section~\ref{subsec:LaggedMollifier} we study lagged mollifier approximations. The main convergence result of this section is Theorem~\ref{thm:LaggedMollifierApproximations}.
Before we proceed to the main body of our analysis, we collect here a few useful observations regarding the main results.

\begin{remark} \ 
\begin{enumerate}
    \item Our lead-lag and lagged mollifier approximation results (Theorems~\ref{thm_leadlag} and~\ref{thm:LaggedMollifierApproximations}) are true for any one-dimensional, H\"older continuous Gaussian process~$X$ that corresponds to the first level of the geometric rough path $\bfX$ in Definition~\ref{def:adapted}. More importantly, the Gaussian assumption is not necessary for these results to hold and is used to simplify moment estimates via the hypercontractivity property (Lemma~\ref{lem:Hypercontractivity}). In fact, since $\bfX$ is always well defined for one-dimensional paths, one can take~$X$ to be the unique (probabilistically) weak solution of a one-dimensional, fully-nonlinear Volterra process with finite moments of all orders, singular kernels and non-zero correlation with~$W$.
    As explained in Section~\ref{subsec: Applications}, such choices arise naturally in several rough volatility models.

    \item In Section~\ref{subsec:HybridLeadLag}, we take $X$ to be a type II-fractional Brownian motion~\eqref{eq:fBm} as the most important example of a Brownian semi-stationary process that satisfies Definition~\ref{def:adapted} and can also approximated by the hybrid scheme~\eqref{eq:Xhybrid}. More examples of such processes can be found in~\cite{bennedsen2017hybrid} where the hybrid scheme was originally introduced. 

    \item Throughout the rest of this section and for the sake of simplicity we take~$W$ to be a one-dimensional standard Brownian motion correlated with~$X$.
    All of the aforementioned results continue to hold if~$W$ is replaced by a $d$-dimensional standard Brownian motion.

      \item Finally, we emphasise here that one can replace~$W$ by the sum $\rho W+\overline{\rho}\overline{W}$ where $\overline{W}$ is a Brownian motion independent of~$W$, $\rho\in(-1, 1)$ and  $\rho^2+\overline{\rho}^2=1$. All the aforementioned approximation results hold true in this case. We choose here to take $\rho=1$ since the fully correlated case is of greater interest when it comes to pathwise approximation results. Indeed, this is the case which requires suitable renormalisation as explained in~\cite{BFGJS20} via the language of regularity structures. 
      It can be easily deduced from Lemma~\ref{lem:zeroRenormalisation} that, when~$W$ and~$X$ are independent, simple (namely non-lagged) mollifiers and piecewise linear approximations converge without further adjustments. Thus, for the sake of lighter notation, we take advantage of this simplification.  
    \end{enumerate}
    
\end{remark}

We conclude with a remark on terminology:

\begin{remark}[Uniform partitions]\label{rem:uniformterminology} 
We shall call a partition $\pi=\{0=t_0<t_1<\dots<t_n=T\}$ of $[0, T]$ uniform (or has uniform mesh) if the points $\{t_i\}_{i=0}^{a(n)}$ are equidistant, that is $t_{i+1}-t_i=t_{j+1}-t_j$ for all $i, j=0,\dots, {a(n)}$. In particular, the phrase "uniform mesh" allows for dyadic partitions 
(Remark~\ref{rem:dyadic}) and is not to be confused with the particular choice $t_{i+1}-t_i=\frac{T}{n}$ for all $i$.
\end{remark}


\subsection{Delayed piecewise linear approximation}\label{subsection:LeadLagPiecewise}
Let $T>0$, $n\in\N$ and $\pi=\{0=t_0<t_1<\dots<t_{a(n)}=T\}$ 
a partition of $[0,T]$ with uniform mesh $|t_{k+1}-t_k|=\Delta=\Delta(n){=\frac{1}{a(n)}}>0$ (not to be confused with the deconcatenation coproduct from Section~\ref{sec:ItoLift}), for  $k=0,\dots, a(n) - 1$.
Let~$W$ be a one-dimensional Brownian motion and~$X$ a one-dimensional, adapted $H$-integrable rough path in the sense of Definition~\ref{def:adapted} (in particular~$X$ corresponds to the first level of the geometric rough path~$\bfX$).
Define the (lead) piecewise linear approximation of~$W$ as
\begin{align}\label{eq:Wlead}
    & W^\Delta_t :=\sum_{k=0}^{a(n)-1}\bigg(W_{t_{k}}+\frac{(t-t_{k})}{\Delta}W_{t_k, t_{k+1}}\bigg)\mathbf{1}_{[t_k, t_{k+1})}(t)\;,\;\;t\in [0,T),\\
    & W^\Delta_T=W^\Delta_{t_n}
    :=W_T=W_{t_n},\nonumber
\end{align}
 and the lagged approximation of $X$
\begin{equation}\label{eq:Xlagged}
    \begin{aligned}
        & \widetilde{X}^\Delta_t
        :=0\cdot\mathbf{1}_{[t_0, t_{1})}(t)+ \sum_{k=0}^{a(n)-2}\bigg(X_{t_{k}}+\frac{(t-t_{k+1})}{\Delta}X_{t_k, t_{k+1}}\bigg)\mathbf{1}_{[t_{k+1}, t_{k+2})}(t)\\
        & \qquad = 0\cdot\mathbf{1}_{[t_0, t_{1})}(t)+ \sum_{k=1}^{a(n)-1}\bigg(X_{t_{k-1}}+\frac{(t-t_{k})}{\Delta}X_{t_{k-1}, t_k}\bigg)\mathbf{1}_{[t_{k}, t_{k+1})}(t)\;,\;\; t\in [0,T),\\      &\widetilde{X}^\Delta_T=\widetilde{X}^\Delta_{t_n}
        :=X_{t_{n-1}},
    \end{aligned}
\end{equation} 
and note that

\begin{equation}\label{eq:WdotLead}
    \dot{W}^\Delta_t=\frac{1}{\Delta}\sum_{k=0}^{a(n)-1}W_{t_k, t_{k+1}}\mathbf{1}_{[t_k, t_{k+1})}(t)\;,\;\;t\in [0,T).
\end{equation}
Next consider, for $[T_1, T_2]\subset[0,T], m\in\bbN$, the two-parameter process
\begin{equation}\label{eq:ImIntegrals}
	\begin{aligned}
		I_{T_1, T_2}^m
        :=\int_{T_1}^{T_2}(X_{T_1, t})^m\dif W_t,
	\end{aligned}
\end{equation}
and its corresponding lead-lag approximation
\begin{align}\label{eq:ImLeadLag}
\widetilde{I}^{\Delta,m}_{T_1, T_2}:= \int_{T_1}^{T_2} (\widetilde{X}^\Delta_{T_1, t})^m  \dif W^{\Delta}_t.
\end{align}

\begin{remark} As mentioned in the beginning of this section, lead-lag approximations were first introduced in~\cite{FHL16} for a Brownian motion~$X$. 
We note here that lead-lag approximations are built "on top" of other approximations of the underlying paths. In particular, the lead-lag approximations~\eqref{eq:ImLeadLag} that we use here are based on regular piecewise linear approximations of~$X$ and~$W$.
In contrast, \cite{FHL16} uses lead-lag approximations based on axis-directed piecewise linear approximations. Hence, our approximations are different but similar in spirit to those from the aforementioned work.
\end{remark}

The first main result of this section is given below and it shows convergence of lead-lag approximations in rough path topology (as given in Definition~\ref{def:rp}) with explicit rates of convergence.
\begin{theorem}[Convergence of lead-lag approximations]\label{thm_leadlag}
    For any $0<\eta\leq H$, $\gamma=mH+\half-\eta$, we have
\begin{equation}\label{eq_rate_conv}
        \mathbb{E}\left[\sup_{T_1,T_2\in[0,T]}\frac{|\widetilde{I}_{T_1,T_2}^{\Delta,m}-I^m_{T_1,T_2}|}{|T_2-T_1|^\gamma}\right]\leq C\Delta^{\eta}.
\end{equation}
 Moreover, if the mesh satisfies  
\begin{equation}\label{eq:meshSummabilityLeadLag}
        \sum_{n\in\N}\Delta(n)^{\eta} = \sum_{n\in\N}\frac{1}{a(n)^\eta}<\infty ,
    \end{equation}
    then 
    $|\widetilde{I}^{\Delta,m}-I^m|_{\mathcal{C}^\gamma}$ tends to zero almost surely
    as $n$ tends to infinity.
\end{theorem}

\begin{remark}\label{rem:tradeoff}
Theorem~\ref{thm_leadlag} shows a trade-off between the H\"older exponent $\gamma$ and the rate of convergence $\eta$. A similar trade-off is already present in the classical case $m=1$ and $H=\half$~\cite[Proposition~13.21]{FV10}. 
\end{remark}

\begin{remark}\label{rem:dyadic} Condition~\eqref{eq:meshSummabilityLeadLag} is satisfied, for example, if we take dyadic partitions $\Delta(n)=T2^{-n}$ (i.e. $a(n)=\frac{2^n}{T}$).
\end{remark}

\begin{proof}[Proof of Theorem~\ref{thm_leadlag}]

We limit ourselves to show an $L^2$-bound of the following form:
\begin{align}\label{eq_L2_bound}
    |\widetilde{I}_{T_1,T_2}^{\Delta,m}-I^m_{T_1,T_2}|^2_{L^2(\Omega)} 
    \lesssim |T_2-T_1|^{2H(m-1)+1}\Delta^{2H}.
\end{align}
Indeed, by Lemma~\ref{lem:Hypercontractivity}, this automatically implies that, for each $p\geq 1$,
\begin{align*}
     |\widetilde{I}_{T_1,T_2}^{\Delta,m}-I^m_{T_1,T_2}|^p_{L^p(\Omega)}  
    \lesssim |T_2-T_1|^{pH(m-1)+\frac{p}{2}}\Delta^{pH},
\end{align*}
from which the rate in~\eqref{eq_rate_conv} follows.

For ease of presentation and following the strategy from~\cite[Proposition~13.21]{FV10}, the proof is divided in three cases depending on the location of $T_1$ and $T_2$ within the partition. Kolmogorov's continuity criterion \autoref{thm:Kolm} then furnishes the desired bounds with the supremum inside the expectation.

\paragraph{Case a): $T_1, T_2$ are partition points}

Start with the case when $T_1$ and $T_2$ are two arbitrary points on the partition, namely $0 \leq T_1= t_{j} \leq t \leq T_2 = t_{j+\ell} \leq T$, for some $j \in\{0, \dots, a(n) - \ell\}$, $\ell \in \{1, \dots, a(n) \}$. 
Let us notice that, in particular, this implies that $T_2-T_1 = \ell \Delta$.
Using the expression for $\widetilde{X}^\Delta_{t}$ 
in~\eqref{eq:Xlagged} and the fact that $t_{j}$ and $t_{j+\ell}$ belong to the time grid, we write
(with $t_{-1}:=0$)
\begin{equation*}\label{eq:PowerDecomposition}
\begin{aligned}
\left(\widetilde{X}^\Delta_{T_1,t}\right)^m
        & = 
        \sum_{k={0}}^{\ell}\bigg(X_{t_{j-1}, t_{j+k-1}}+\frac{(t-t_{j+k})}{\Delta}X_{t_{j+k-1}, t_{j+k}}\bigg)^m \mathbf{1}_{[t_{j+k}, t_{j+k+1})}(t)\\
        & = 
    \sum_{k=0}^{\ell}\bigg[ \sum_{i=0}^m \binom{m}{i}\frac{(t-t_{j+k})^{m-i}}{\Delta^{m-i}}(X_{t_{j-1}, t_{j+k-1}})^i (X_{t_{j+k-1}, t_{j+k}})^{m-i}\bigg] \mathbf{1}_{[t_{j+k},t_{j+k+1})}(t).
    \end{aligned}
\end{equation*}
This expression can be justified as follows. From~\eqref{eq:Xlagged}, we have
\begin{align*}
\widetilde{X}^\Delta_{T_1,t}
& := \widetilde{X}^\Delta_t - \widetilde{X}^\Delta_{t_j}\\
    & = \sum_{k=0}^{a(n)-2}\bigg(X_{t_{k}}+\frac{(t-t_{k+1})}{\Delta}X_{t_k, t_{k+1}}\bigg)\mathbf{1}_{[t_{k+1}, t_{k+2})}(t) - X_{t_{j-1}}\\
    & = \left(\sum_{k=0}^{a(n)-2}X_{t_{k}}\mathbf{1}_{[t_{k+1}, t_{k+2})}(t)- X_{t_{j-1}}\right) + \sum_{k=0}^{a(n)-2}\frac{(t-t_{k+1})}{\Delta}X_{t_k, t_{k+1}}\mathbf{1}_{[t_{k+1}, t_{k+2})}(t).
\end{align*}
Since
$t_j\leq t\leq t_{j+\ell}$ then
$
1= \sum_{u=0}^{\ell}\mathbf{1}_{[t_{u+j}, t_{u+j+1})}(t)
$ and thus the first bracket reads
\begin{align*}
    \sum_{k=j-1}^{j+\ell-1}X_{t_{k}}\mathbf{1}_{[t_{k+1}, t_{k+2})}(t)
     - X_{t_{j-1}}
    & = \sum_{u=0}^{\ell}X_{t_{u+j-1}}\mathbf{1}_{[t_{u+j}, t_{u+j+1})}(t)
     - X_{t_{j-1}},\\
    & = \sum_{u=0}^{\ell}X_{t_{u+j-1}}\mathbf{1}_{[t_{u+j}, t_{u+j+1})}(t) - \sum_{u=0}^{\ell}X_{t_{j-1}}
     \mathbf{1}_{[t_{u+j}, t_{u+j+1})}(t)\\
    & = \sum_{u=0}^{\ell}\Big( X_{t_{u+j-1}} - X_{t_{j-1}}\Big)
    \mathbf{1}_{[t_{u+j}, t_{u+j+1})}(t)\\
    &= \sum_{u=0}^{\ell} X_{t_{j-1}, t_{j+u-1}}
    \mathbf{1}_{[t_{j+u}, t_{j+u+1})}(t),
\end{align*}
with the change of variable $u=k-(j-1)$ in the first line.
Note that the term $u=\ell$ in the first line is only required to account for the case $t=t_{j+\ell}$.
Similarly, with the change of variable $u=k-(j-1)$, we can write
\begin{align*}
    \sum_{k=0}^{a(n)-2}\frac{(t-t_{k+1})}{\Delta}X_{t_k, t_{k+1}}\mathbf{1}_{[t_{k+1}, t_{k+2})}(t)
    & = \sum_{k=j-1}^{j+\ell-1}\frac{(t-t_{k+1})}{\Delta}X_{t_k, t_{k+1}}\mathbf{1}_{[t_{k+1}, t_{k+2})}(t)\\
    & = \sum_{u=0}^{\ell}\frac{(t-t_{j+u})}{\Delta}X_{t_{j+u-1}, t_{j+u}}\mathbf{1}_{[t_{j+u}, t_{j+u+1})}(t).
\end{align*}
Then, the corresponding lead-lag approximation reads
\begin{align*}
    \widetilde{I}^{\Delta,m}_{T_1, T_2}
    &= \int_{T_1}^{T_2} \left(\widetilde{X}^\Delta_{T_1, t}\right)^m  \dif W^{\Delta}_t \\
    & = \sum_{k=0}^{\ell-1}\left[ \sum_{i=0}^m \binom{m}{i}\frac{1}{m-i+1}(X_{t_{j-1}, t_{j+k-1}})^i (X_{t_{j+k-1}, t_{j+k}})^{m-i}\right]
    W_{t_{j+k}, t_{j+k+1}}.
\end{align*}
We proceed with the proof of~\eqref{eq_L2_bound}. The key idea is to show that the lead-lag approximation behaves asymptotically as a left-point Riemann sum approximation of the It\^o integral $I^m$. To this end, we add and subtract
\begin{align*}
    L(I)^{\Delta,m}_{T_1, T_2}
    : = \int_{T_1}^{T_2} (L(X)^\Delta_{T_1, t})^m  \dif W^{\Delta}_t
    & = \sum_{k=1}^{\ell-1}(X_{t_j,t_{j+k}})^m W_{t_{j+k}, t_{j+k+1}},
\end{align*}
with $L(X)^\Delta_{t_j, t} := \sum_{k=1}^{\ell-1} X_{t_j,t_{j+k}}\mathbf{1}_{[t_{j+k}, t_{j+k+1})}(t)$. Thus, we reduce the problem of estimating the $L^2$-difference between the original process and its approximations to the following two terms:
\begin{align*}
\left|\widetilde{I}^{\Delta,m}_{T_1,T_2}-I^m_{T_1,T_2}\right|^{2}_{L^2(\Omega)} \nonumber
    & \leq 2\Big|\underbrace{L(I)^{\Delta,m}_{T_1, T_2}-\widetilde{I}^{\Delta,m}_{T_1,T_2}}_{=:R^{\Delta,m}_{T_1,T_2}}\Big|^2_{L^2(\Omega)}
    + 2\left|I^m_{T_1,T_2}-L(I)^{\Delta,m}_{T_1, T_2}\right|^2_{L^2(\Omega)}.
\end{align*}
Regarding the second term, it is straightforward to see that it converges to $0$ at the correct rate (the one in~\eqref{eq_L2_bound}). 
Indeed, exploiting that, for any $m \in \mathbb{N}$ and $a,b \in \mathbb{R}$, 
\begin{align}\label{eq:pol_m_dec}
    a^m-b^m = (a-b) p_{m-1}(a,b),
\end{align}
with $p_{m-1}$ a homogeneous polynomial of degree $m-1$, we obtain
\begin{equation*}
    \begin{aligned}
        \left|I^m_{T_1,T_2}-L(I)^{\Delta,m}_{T_1, T_2}\right|^2_{L^2(\Omega)}
        & = \int_{t_j}^{t_{j+\ell}} \mathbb{E}\left[|X_{t_j,t}^m-(L(X_{t_j,t})^{\Delta})^m|^2\right]\dif t        \\&=\sum_{k=0}^{\ell-1}\int_{t_{j+k}}^{t_{j+k+1}}\mathbb{E}\left[|(X_{t_j,t})^m-(X_{t_j,t_{j+k}})^m|^2\right]\dif t\\
        & =\sum_{k=0}^{\ell-1}\int_{t_{j+k}}^{t_{j+k+1}}\mathbb{E}\left[|(X_{t_j,t}-X_{t_j,t_{j+k}})^2 p_{m-1}(X_{t_j,t},X_{t_j,t_{j+k}} )|^2\right]\dif t\\
        &\leq \sum_{k=0}^{\ell-1}\int_{t_{j+k}}^{t_{j+k+1}}\mathbb{E}\left[|(X_{t}-X_{t_{j+k}})^4\right]^{\frac{1}{2}} \mathbb{E}\left[p_{m-1}(X_{t_j,t},X_{t_j,t_{j+k}} )^4\right]^{\frac{1}{2}}\dif t\\
        & \leq C_H \sum_{k=0}^{\ell-1}\int_{t_{j+k}}^{t_{j+k+1}}(t-{t_{j+k}})^{2H} (T_2-T_1)^{2H(m-1)}\dif t\\
        & =  C_H (T_2-T_1)^{2H(m-1)} \sum_{k=0}^{\ell-1}\int_{t_{j+k}}^{t_{j+k+1}}(t-{t_{j+k}})^{2H} \dif t
        \\&=  \frac{C_H}{2H+1}(T_2-T_1)^{2H(m-1)+1} \Delta^{2H}
        \lesssim (T_2-T_1)^{2H(m-1)+1} \Delta^{2H},
    \end{aligned}
\end{equation*}
where in the last identity we exploited the fact that by construction $T_2-T_1 = \ell \Delta$.
To conclude we have to show that $R^{\Delta, m}_{T_1,T_2}$ vanishes as $\Delta \to 0$ with the rate in~\eqref{eq_L2_bound}.
Exploiting its explicit representation and writing $"\in \mathcal{G}"$ for $\mathcal{G}$-measurability of a random variable and $\indep \mathcal{G}$ for independence, then
{
\begin{align}\label{eq:est_R_1}
        |&R^{\Delta, m}_{T_1,T_2}|^2_{L^2(\Omega)}
        = \left|L(I)^{\Delta,m}_{T_1, T_2}-\widetilde{I}^{\Delta,m}_{T_1,T_2}\right|^2_{L^2(\Omega)} \\
        & = \mathbb{E}\Bigg[\Bigg( \sum_{k=1}^{\ell-1}\left\{\sum_{i=0}^m \frac{\binom{m}{i}}{m-i+1}X_{t_{j-1},t_{j+k-1}}^i X_{t_{j+k-1}, t_{j+k}}^{m-i} -X_{t_j,t_{j+k}}^m\right\}W_{t_{j+k}, t_{j+k+1}}\Bigg)^2\Bigg] \nonumber \\
        &=2\sum_{1\leq h <k\leq \ell-1}\mathbb{E}\Bigg[\underbrace{\left\{\sum_{i=0}^m \frac{\binom{m}{i}}{m-i+1}X_{t_{j-1},t_{j+h-1}}^i
        X_{t_{j+h-1}, t_{j+h}}^{m-i} -X_{t_{j},t_{j+h}}^m\right\}}_{\in \mathcal{F}_{t_{j+k}}}\cdot   \nonumber\\
        &\cdot \overbrace{\left\{\sum_{i=0}^m \frac{1}{m-i+1}\binom{m}{i}
        X_{t_{j-1},t_{j+k-1}}^i
        X_{t_{j+k-1}, t_{j+k}}^{m-i} -X_{t_j,t_{j+k}}^m\right\}W_{t_{j+h}, t_{j+h+1}}}^{\in\mathcal{F}_{t_{j+k}}}\overbrace{W_{t_{j+k}, t_{j+k+1}}}^{\indep\mathcal{F}_{t_{j+k}}}\Bigg]  \nonumber \\  
        & + \sum_{k=1}^{\ell-1}\mathbb{E}\Bigg[\overbrace{\left\{\sum_{i=0}^m \frac{\binom{m}{i}}{m-i+1}
        X_{t_{j-1},t_{j+k-1}}^i X_{t_{j+k-1}, t_{j+k}}^{m-i}
        -X_{t_{j},t_{j+k}}^m\right\}^2}^{\in\mathcal{F}_{t_{j+k}}}\overbrace{W_{t_{j+k}, t_{j+k+1}}^2}^{\indep\mathcal{F}_{t_{j+k}}}\Bigg]. \nonumber
    \end{align}}
Thus, an application of the tower property together with the identity in~\eqref{eq:pol_m_dec}, yields
{\footnotesize
\begin{align}\label{eq:est_R_2}
        &|R^{\Delta, m}_{T_1,T_2}|^2_{L^2(\Omega)}
        =2\sum_{1\leq h <k\leq \ell-1}\mathbb{E}\Bigg[\left\{\sum_{i=0}^m \frac{\binom{m}{i}}{m-i+1}
        X_{t_{j-1},t_{j+h-1}}^i X_{t_{j+h-1}, t_{j+h}}^{m-i} -X_{t_{j},t_{j+h}}^m\right\} \\
        &\qquad\cdot \left\{\sum_{i=0}^m \frac{\binom{m}{i}}{m-i+1}X_{t_{j-1},t_{j+k-1}}^i
        X_{t_{j+k-1}, t_{j+k}}^{m-i} - X_{t_j,t_{j+k}}^m\right\} W_{t_{j+h}, t_{j+h+1}}\overbrace{\mathbb{E} [ W_{t_{j+k}, t_{j+k+1}}| \mathcal{F}_{t_{j+k}}] }^{=0}\Bigg]\nonumber \\ 
        & \qquad + \sum_{k=1}^{\ell-1}\mathbb{E}\Bigg[\left\{\sum_{i=0}^m \frac{\binom{m}{i}}{m-i+1}
        X_{t_{j-1},t_{j+k-1}}^i
        X_{t_{j+k-1}, t_{j+k}}^{m-i} - X_{t_{j},t_{j+k}}^m\right\}^2 \overbrace{\mathbb{E}[W_{t_{j+k}, t_{j+k+1}}^2| \mathcal{F}_{t_{j+k}}]}^{=\Delta}\Bigg]\nonumber \\
        & = \Delta \sum_{k=1}^{\ell-1}\mathbb{E}\Bigg[\left\{\sum_{i=0}^m \frac{\binom{m}{i}}{m-i+1}
        X_{t_{j-1},t_{j+k-1}}^i X_{t_{j+k-1}, t_{j+k}}^{m-i} -X_{t_{j},t_{j+k}}^m\right\}^2\Bigg]\nonumber \\
        & = \Delta \sum_{k=1}^{\ell-1}\mathbb{E}\Bigg[\left\{\sum_{i=0}^{m-1} \frac{\binom{m}{i}}{m-i+1}
        X_{t_{j-1},t_{j+k-1}}^i X_{t_{j+k-1}, t_{j+k}}^{m-i} + X_{t_{j-1},t_{j+k-1}}^m - X_{t_{j},t_{j+k}}^m\right\}^2\Bigg]\nonumber \\
        & = \Delta \sum_{k=1}^{\ell-1}\mathbb{E}\Bigg[\Bigg\{\sum_{i=0}^{m-1} \frac{\binom{m}{i}}{m-i+1}
        X_{t_{j-1},t_{j+k-1}}^i X_{t_{j+k-1}, t_{j+k}}^{m-i}
         + p_{m-1}(X_{t_{j-1},t_{j+k-1}},X_{t_{j},t_{j+k}})(X_{t_{j-1},t_{j}}-X_{t_{j+k-1},t_{j+k}})\Bigg\}^2\Bigg] \nonumber.
    \end{align}}
We estimate this remainder with Cauchy-Schwarz inequality in the following way:
\begin{align*}
        &\left|R^{\Delta, m}_{T_1,T_2}\right|^2_{L^2(\Omega)}\\
        & \leq 2^m \Delta \sum_{k=1}^{\ell-1}\mathbb{E}\Bigg[\Bigg\{\sum_{i=0}^{m-1} \left(\frac{\binom{m}{i}}{m-i+1}\right)^2 X_{t_{j-1},t_{j+k-1}}^{2i}
        X_{t_{j+k-1}, t_{j+k}}^{2(m-i)}  \\
        & \qquad + p_{m-1}(X_{t_{j-1},t_{j+k-1}},X_{t_{j},t_{j+k}})^2(X_{t_{j-1},t_{j}}-X_{t_{j+k-1},t_{j+k}})^2\Bigg\}^2\Bigg]\\
        & \leq 2^m \Delta \sum_{k=1}^{\ell-1}\Bigg\{\sum_{i=0}^{m-1} \left(\frac{\binom{m}{i}}{m-i+1}\right)^2\mathbb{E}\big[X_{t_{j-1},t_{j+k-1}}^{4i}\big]^{\frac{1}{2}}\mathbb{E}\big[X_{t_{j+k-1}, t_{j+k}}^{4(m-i)}\big]^{\frac{1}{2}} \\
        & \qquad + \mathbb{E}\big[p_{m-1}(X_{t_{j-1},t_{j+k-1}},X_{t_{j},t_{j+k}})^4\big]^{\frac{1}{2}}\mathbb{E}\big[(X_{t_{j-1},t_{j}}-X_{t_{j+k-1},t_{j+k}})^4\big]^{\frac{1}{2}}\Bigg\}\\
        & \lesssim \Delta \sum_{k=1}^{\ell-1}\Bigg\{\sum_{i=0}^{m-1} \left(\frac{\binom{m}{i}}{m-i+1}\right)^2 (T_2-T_1)^{2Hi}\mathbb{E}\big[X_{t_{j+k-1}, t_{j+k}}^{4(m-i)}\big]^{\frac{1}{2}} \\
        & \qquad + (T_2-T_1)^{2H(m-1)}\bigg(\mathbb{E}\big[X_{t_{j-1},t_{j}}^4\big] + \mathbb{E}\big[X_{t_{j+k-1},t_{j+k}}^4\big]\bigg)^{\frac{1}{2}}\Bigg\}\\
        & \lesssim \Delta \sum_{k=1}^{\ell-1}\Bigg\{\sum_{i=0}^{m-1} \left(\frac{\binom{m}{i}}{m-i+1}\right)^2 (T_2-T_1)^{2Hi}\Delta^{2(m-i)} + (T_2-T_1)^{2H(m-1)}\Delta^{2H}\Bigg\}\\
        &\lesssim (T_2-T_1) \Bigg\{\sum_{i=0}^{m-1} \left(\frac{\binom{m}{i}}{m-i+1}\right)^2 (T_2-T_1)^{2Hi}\Delta^{2(m-i)} + (T_2-T_1)^{2H(m-1)}\Delta^{2H}\Bigg\}\\
        &\lesssim  (T_2-T_1)^{1+2H(m-1)} \Delta^{2H},
    \end{align*}
where we also used the fact that, for all $t\geq s\geq 0, p\geq 1$, the Gaussian random variable $X_{s,t}$ satisfies $|X_{s,t}|_{L^p(\Omega)}\lesssim |X_{s,t}|_{L^2(\Omega)}$ and again the fact that $T_2-T_1 = \ell \Delta$.

\paragraph{Case b): $T_1, T_2$ are on the same partition interval}

Consider the case when $T_1$ and $T_2$ are two arbitrary points belonging to the same interval in the partition, but do not necessarily belong to the partition, namely $t_{j}\leq T_1 \leq t \leq T_2 \leq t_{j+1}$, for some $j \in\{0, \dots, n-1\}$ (recall that $t_{-1}:=0$).
Recalling the explicit expression for~$\widetilde{X}^\Delta_{t}$ in~\eqref{eq:Xlagged} and exploiting that $t_{j}\leq T_1 \leq t \leq T_2 \leq t_{j+1}$ on our interval of interest, we have
\begin{equation*}
\widetilde{X}^\Delta_{T_1,t} 
    = \widetilde{X}^\Delta_{t} - \widetilde{X}^\Delta_{T_1}
    = \left[X_{t_{j-1}} + \frac{t-t_j}{\Delta}X_{t_{j-1},t_j}\right] - \left[X_{t_{j-1}} + \frac{T_1-t_j}{\Delta}X_{t_{j-1},t_j}\right] 
    = \frac{t-T_1}{\Delta}X_{t_{j-1},t_j},
\end{equation*}
and so, in particular, we obtain
\begin{align*}
\widetilde{I}^{\Delta,m}_{T_1, T_2}
 = \int_{T_1}^{T_2} (\widetilde{X}^\Delta_{T_1,t})^m  \dif W^{\Delta}_t
 & = \int_{T_1}^{T_2}\left( \frac{t-T_1}{\Delta}X_{t_{j-1},t_j}\right)^m \frac{W_{t_{j},t_{j+1}}}{\Delta} \dif t\\
 & = \frac{(T_2-T_1)^{m+1}}{(m+1)\Delta^{m+1}} X_{t_{j-1},t_j}^m W_{t_{j},t_{j+1}}.
\end{align*}
Thus, we derive the following upper bound on the second moment of the difference:
\begin{align*}
\big|\widetilde{I}^{\Delta,m}_{T_1,T_2}-I^m_{T_1,T_2}&\big|^{2} _{L^2(\Omega)}
    \leq 2\big|\widetilde{I}^\Delta_{T_1,T_2}\big|^2_{L^2(\Omega)}+  2\big|I^m_{T_1,T_2}\big|^2_{L^2(\Omega)}
    \\\nonumber 
    & = 2\mathbb{E} \left[\left(\frac{(T_2-T_1)^{m+1}}{(m+1)\Delta^{m+1}} 
    X_{t_{j-1},t_j}^m W_{t_{j},t_{j+1}} \right)^2\right] + 2\mathbb{E} \left[\left(\int_{T_1}^{T_2} X_{T_1, t}^m  \dif W_t\right)^2\right]\\ \nonumber
    & = 2\frac{(T_2-T_1)^{2(m+1)}}{(m+1)^2\Delta^{2(m+1)}} \mathbb{E}\left[X_{t_{j-1},t_j}^{2m}\right]\mathbb{E} \left[W_{t_{j},t_{j+1}}^{2}\right] + 2\int_{T_1}^{T_2} \mathbb{E} \left[X_{T_1, t}^{2m} \right]  \dif t\\ \nonumber
    & \lesssim \frac{(T_2-T_1)^{2(m+1)}}{\Delta^{2(m+1)}}  (t_{j}-t_{j-1})^{2Hm} (t_{j+1}-t_{j}) + \int_{T_1}^{T_2} (t-T_1)^{2Hm}  \dif t\\ \nonumber
    & \lesssim (T_2-T_1)^{2(m+1)}\Delta^{2Hm+1-2(m+1)} + (T_2-T_1)^{2Hm+1}\\ \nonumber
    & =\left(\frac{T_2-T_1}{\Delta}\right)^{2m+1}\Delta^{2mH} (T_2-T_1)+ (T_2-T_1)^{2Hm+1}\\ \nonumber
    & \lesssim \Delta^{2mH} (T_2-T_1),
\end{align*}
where we exploited the conditional independence of the increments and It\^o isometry. 

\paragraph{Case c): $T_1, T_2$ are on different partition intervals}

Finally, we consider the case of two arbitrary extremal points, 
$0 < t_j < T_1 < t_{j+1} < \dots < t_{j+\ell -1} < T_2 < t_{j+\ell} <T$, with $\ell \in \{1, \dots,a(n)-1\}$ and $j\in\{0,\dots,a(n)-\ell\}$ (recall that $t_{-1}:=0$). Let us notice that, in particular, this implies that $T_2-T_1 > \ell \Delta$. We exploit this property several times in the following computations.
Similarly to what we have done for case a), we exploit the expression for $\widetilde{X}^\Delta_{t}$ in~\eqref{eq:Xlagged} to write
\begin{align*}
(\widetilde{X}^\Delta_{T_1,t})^m
        & = \left(\frac{(t-T_1)}{\Delta}X_{t_{j-1}, t_{j}}\right)^m \mathbf{1}_{[t_{j}, t_{j+1})}(t)\\
        &  + \sum_{k=1}^{\ell-1}\left[X_{t_{j-1}, t_{j+k-1}}+\frac{(t-t_{j+k})}{\Delta}X_{t_{j+k-1}, t_{j+k}} -\frac{(T_1-t_{j})}{\Delta}X_{t_{j-1}, t_{j}} \right]^m \mathbf{1}_{[t_{j+k}, t_{j+k+1})}(t),
\end{align*}

so that the lead-lag approximation in this case reads
\begin{align*}	\widetilde{I}^{\Delta,m}_{T_1, T_2}
    & = \int_{T_1}^{T_2} (\widetilde{X}^\Delta_{T_1, t})^m  \dif W^{\Delta}_t\\ \nonumber
    & = \frac{1}{m+1}\frac{(t_{j+1}-T_1)^{m+1}}{\Delta^{m+1}}X_{t_{j-1},t_j}^m W_{t_{j},t_{j+1}}\\ \nonumber
    & + \sum_{k=1}^{\ell-2}\left\{ \sum_{i=0}^m \frac{\binom{m}{i}}{m-i+1}
    \left(X_{t_{j-1}, t_{j+k-1}}-\frac{(T_1-t_{j})}{\Delta}X_{t_{j-1}, t_{j}}\right)^i X_{t_{j+k-1}, t_{j+k}}^{m-i}\right\} W_{t_{j+k}, t_{j+k+1}}\\ \nonumber
    & +  \sum_{i=0}^m \frac{\binom{m}{i}}{m-i+1}\frac{(T_2-t_{j+\ell-1})^{m-i+1}}{\Delta^{m-i+1}}\bigg(X_{t_{j-1}, t_{j+\ell-2}}-\frac{(T_1-t_{j})}{\Delta}X_{t_{j-1}, t_{j}}\bigg)^i\cdot\\&\cdot X_{t_{j+\ell-2}, t_{j+\ell-1}}^{m-i} W_{t_{j+\ell-1}, t_{j+\ell}}.
\end{align*}

We proceed similarly to case a) by adding and subtracting the term 
\begin{align*}
    L(I)^{\Delta,m}_{t_{j}, t_{j+\ell-1}}
    : = \int_{t_{j}}^{t_{j+\ell-1}} (L(X)^\Delta_{t_j, t})^m  \dif W^{\Delta}_t
    & = \sum_{k=1}^{\ell-2}X_{t_j,t_{j+k}}^m W_{t_{j+k}, t_{j+k+1}},
\end{align*}
with ${L(X)}^\Delta_{t_j, t} := \sum_{k=1}^{\ell-1} X_{t_j,t_{j+k}}\mathbf{1}_{[t_{j+k}, t_{j+k+1})}(t)$,
so that, exploiting the triangle inequality, we reduce the problem of estimating the $L^2$-difference of the original process and its approximations to the following:
\begin{align*}
\left|\widetilde{I}^{\Delta,m}_{T_1,T_2}-I^m_{T_1,T_2}\right|^{2}_{L^2(\Omega)} \nonumber
    & \leq 2\Big|\underbrace{L(I)^{\Delta,m}_{t_{j}, t_{j+\ell-1}}-\widetilde{I}^{\Delta,m}_{T_1,T_2}}_{=:\widetilde{R}^{\Delta,m}_{T_1,T_2}}\Big|^2_{L^2(\Omega)}
    +  2\left|I^m_{T_1,T_2}-L(I)^{\Delta,m}_{t_{j}, t_{j+\ell-1}}\right|^2_{L^2(\Omega)}.
\end{align*}
We start by proving that the second term converges to $0$ with the right regularity and speed as $\Delta\to 0$. Indeed, exploiting once more the identity in~\eqref{eq:pol_m_dec} we have
{\footnotesize
\begin{align*}
& \Big|I^m_{T_1,T_2}-L(I)^{\Delta,m}_{t_{j}, t_{j+\ell-1}}\Big|^2_{L^2(\Omega)}\\
& = \int_{T_1}^{t_j}\overbrace{\mathbb{E}[X_{T_1,t}^{2m}]}^{\lesssim(t-T_1)^{2mH}}\dif t + \sum_{k=0}^{\ell-2}\int_{t_{j+k}}^{t_{j+k+1}}\underbrace{\mathbb{E}\left[|X_{T_1,t}^m - X_{t_j,t_{j+k}}^m|^2\right]}_{=\mathbb{E}[|(X_{T_1,t}-X_{t_j,t_{j+k}})^2 p_{m-1}(X_{t_j,t},X_{t_j,t_{j+k}} )|^2] }\dif t  + \int_{t_{j+\ell-1}}^{T_2}\overbrace{\mathbb{E} [X_{T_1,t}^{2m}]}^{\lesssim(t-T_1)^{2mH}}\dif t\\
& \lesssim \underbrace{(t_{j}-T_1)^{2mH+1}}_{\lesssim \Delta^{2mH+1}} \\
        & \quad + \sum_{k=0}^{\ell-2}\int_{t_{j+k}}^{t_{j+k+1}}\overbrace{\left\{ \overbrace{\mathbb{E}\left[X_{T_1,t_j}^4\right]}^{\lesssim(t_j-T_1)^{4H}}+  \overbrace{\mathbb{E}\left[X_{t_{j+k},t}^4\right]}^{\lesssim(t-t_{j+k})^{4H}}\right\}^{\frac{1}{2}} }^{\lesssim \Delta^{2H}}\underbrace{\mathbb{E}\left[p_{m-1}(X_{t_j,t},X_{t_j,t_{j+k}} )^4\right]^{\frac{1}{2}}}_{\lesssim(T_2-T_1)^{2(m-1)H}}\dif t \\
        & \quad + \underbrace{(T_2-T_1)^{2mH+1}-(t_{j+\ell-1}-T_1)^{2mH+1}}_{\lesssim(T_2-t_{j+\ell-1})^{2mH+1}\lesssim\Delta^{2mH+1}} \\
        & \lesssim \Delta^{2mH+1} + (T_2-T_1)^{2(m-1)H} \ell \Delta^{2H+1} 
        \lesssim (T_2-T_1)^{2(m-1)H+1}\Delta^{2H}.
\end{align*}
}
Then, we decompose $\widetilde{R}^{\Delta,m}_{T_1,T_2}$ in three contributions:
{\footnotesize
\begin{align*}
    &\widetilde{R}^{\Delta,m}_{T_1,T_2}
     = \int_{T_1}^{T_2} (\widetilde{X}^\Delta_{T_1, t})^m  \dif W^{\Delta}_t\\ \nonumber
    & = \frac{1}{m+1}\frac{(t_{j+1}-T_1)^{m+1}}{\Delta^{m+1}}X_{t_{j-1},t_j}^m W_{t_{j},t_{j+1}}\\ \nonumber
    & + \overbrace{\sum_{k=1}^{\ell-2}\left\{ \sum_{i=0}^m \frac{\binom{m}{i}}{m-i+1}\left(X_{t_{j-1}, t_{j+k-1}}-\frac{(T_1-t_{j})}{\Delta}X_{t_{j-1}, t_{j}}\right)^i X_{t_{j+k-1}, t_{j+k}}^{m-i} - X_{t_j,t_{j+k}}^m\right\} W_{t_{j+k}, t_{j+k+1}}}^{=:{R}^{\Delta,m}_{T_1,T_2}}\\ \nonumber
    & +  \sum_{i=0}^m \frac{\binom{m}{i}}{m-i+1}\frac{(T_2-t_{j+\ell-1})^{m-i+1}}{\Delta^{m-i+1}}\left(X_{t_{j-1}, t_{j+\ell-2}}-\frac{(T_1-t_{j})}{\Delta}X_{t_{j-1}, t_{j}}\right)^i X_{t_{j+\ell-2}, t_{j+\ell-1}}^{m-i} W_{t_{j+\ell-1}, t_{j+\ell}}.
\end{align*}}
Thus, in order to bound the $L^2$-norm of this remainder we consider these three terms separately.
We start with the first term. 
Exploiting the conditional independence of the Brownian increments we obtain 
\begin{align*}
    \bigg|\frac{1}{m+1}& \frac{(t_{j+1}-T_1)^{m+1}}{\Delta^{m+1}} X_{t_{j-1},t_j}^m W_{t_{j},t_{j+1}}\bigg|^{2}_{L^2(\Omega)}
    \\& = \left(\frac{1}{m+1}\right)^{2}\frac{(t_{j+1}-T_1)^{2(m+1)}}{\Delta^{2(m+1)}}\mathbb{E}[X_{t_{j-1},t_j}^{2m}]\mathbb{E}[ W_{t_{j},t_{j+1}}^2]
    \lesssim \Delta^{2Hm+1}.
\end{align*}
Then, we move on to consider the third term.
Exploiting the conditional independence of increments as above together with Cauchy-Schwarz inequality, we obtain
{\footnotesize
\begin{align*}
    & \left|
    \sum_{i=0}^m \frac{\binom{m}{i}}{m-i+1}\frac{(T_2-t_{j+\ell-1})^{m-i+1}}{\Delta^{m-i+1}}
    \left(X_{t_{j-1}, t_{j+\ell-2}}-\frac{(T_1-t_{j})}{\Delta}X_{t_{j-1}, t_{j}}\right)^i X_{t_{j+\ell-2}, t_{j+\ell-1}}^{m-i} W_{t_{j+\ell-1}, t_{j+\ell}}\right|^{2}_{L^2(\Omega)} \\
    & = \mathbb{E}\left[\left|\sum_{i=0}^m \frac{\binom{m}{i}}{m-i+1}\frac{(T_2-t_{j+\ell-1})^{m-i+1}}{\Delta^{m-i+1}}
    \left(X_{t_{j-1}, t_{j+\ell-2}}-\frac{(T_1-t_{j})}{\Delta}X_{t_{j-1}, t_{j}}\right)^i 
    X_{t_{j+\ell-2}, t_{j+\ell-1}}^{m-i} \right|^2\right]\mathbb{E}[W_{t_{j+\ell-1}, t_{j+\ell}}^2]\\
    & \lesssim \Delta \left\{\sum_{i=0}^m \left(\frac{\binom{m}{i}}{m-i+1}\right)^2\frac{(T_2-t_{j+\ell-1})^{2(m-i+1)}}{\Delta^{2(m-i+1)}}\mathbb{E}\left[\left(X_{t_{j-1}, t_{j+\ell-2}}-\frac{(T_1-t_{j})}{\Delta}X_{t_{j-1}, t_{j}}\right)^{2i} X_{t_{j+\ell-2}, t_{j+\ell-1}}^{2(m-i)}\right] \right\}\\
    & \lesssim \Delta \left\{\sum_{i=0}^m \left(\frac{\binom{m}{i}}{m-i+1}\right)^2\mathbb{E}\left[\left(X_{t_{j-1}, t_{j+\ell-2}}-\frac{(T_1-t_{j})}{\Delta}X_{t_{j-1}, t_{j}}\right)^{4i}\right]^{\frac{1}{2}} \mathbb{E}\left[X_{t_{j+\ell-2}, t_{j+\ell-1}}^{4(m-i)}\right]^{\frac{1}{2}} \right\}\\
    & \lesssim \Delta \left\{\sum_{i=0}^m \left(\frac{\binom{m}{i}}{m-i+1}\right)^2(T_2 -T_1)^{2Hi}\Delta^{2H(m-i)} \right\}
    \lesssim \Delta(T_2 -T_1)^{2Hm}.
\end{align*}}
Finally, to conclude we have to prove that $R^{\Delta, m}_{T_1,T_2}$ vanishes, as $n\to\infty$, in H\"older norm with the correct rate.
This term is completely analogous to the term in case a). Indeed, with computations analogous to the ones performed in~\eqref{eq:est_R_1}-\eqref{eq:est_R_2} together with the Cauchy-Schwarz inequality, we write
{\footnotesize   \begin{align*}\label{eq:est_R_3}
        \left|R^{\Delta, m}_{T_1,T_2}\right|^2_{L^2(\Omega)}
        & = \Delta \sum_{k=1}^{\ell-2}\mathbb{E}\Bigg[\left\{\sum_{i=0}^m \frac{\binom{m}{i}}{m-i+1}
        \left(X_{t_{j-1},t_{j+k-1}}+\frac{T_1-t_j}{\Delta}X_{t_{j-1},t_j}\right)^i X_{t_{j+k-1}, t_{j+k}}^{m-i} - X_{t_{j},t_{j+k}}^m\right\}^2\Bigg]\\
        & = \Delta \sum_{k=1}^{\ell-2}\mathbb{E}\Bigg[\Bigg\{\sum_{i=0}^{m-1} \frac{\binom{m}{i}}{m-i+1}
        \left(X_{t_{j-1},t_{j+k-1}}+\frac{T_1-t_j}{\Delta}X_{t_{j-1},t_j}\right)^i X_{t_{j+k-1}, t_{j+k}}^{m-i} \\
        & \qquad + \underbrace{p_{m-1}\left(X_{t_{j-1},t_{j+k-1}}+\frac{T_1-t_j}{\Delta}X_{t_{j-1},t_j},X_{t_{j},t_{j+k}}\right)}_{=:p_{m-1}}\left(\frac{t_{j+1}-T_1}{\Delta}X_{t_{j-1},t_j}-X_{t_{j+k-1},t_{j+k}}\right)\Bigg\}^2\Bigg]\\
        & \leq 2^m \Delta \sum_{k=1}^{\ell-2}\mathbb{E}\Bigg[\sum_{i=0}^{m-1} \left(\frac{\binom{m}{i}}{m-i+1}\right)^2
        \left(X_{t_{j-1},t_{j+k-1}}+\frac{T_1-t_j}{\Delta}X_{t_{j-1},t_j}\right)^{2i} X_{t_{j+k-1}, t_{j+k}}^{2(m-i)} \\
        & \qquad + p_{m-1}^2\left(\frac{t_{j+1}-T_1}{\Delta}X_{t_{j-1},t_j}-X_{t_{j+k-1},t_{j+k}}\right)^2\Bigg]\\
        & \leq 2^m \Delta \sum_{k=1}^{\ell-2}\Bigg\{\sum_{i=0}^{m-1} \left(\frac{\binom{m}{i}}{m-i+1}\right)^2\mathbb{E}\big[\left(X_{t_{j-1},t_{j+k-1}}+\frac{T_1-t_j}{\Delta}X_{t_{j-1},t_j}\right)^{4i}\big]^{\frac{1}{2}}\mathbb{E}\big[X_{t_{j+k-1}, t_{j+k}}^{4(m-i)}\big]^{\frac{1}{2}} \\
        & \qquad + \mathbb{E}\big[p_{m-1}^4\big]^{\frac{1}{2}}\mathbb{E}\Bigg[\left(\frac{t_{j+1}-T_1}{\Delta}X_{t_{j-1},t_{j}}-X_{t_{j+k-1},t_{j+k}}\right)^4\Bigg]^{\frac{1}{2}}\Bigg\}\\
        & \lesssim \Delta \sum_{k=1}^{\ell-2}\Bigg\{\sum_{i=0}^{m-1} \left(\frac{\binom{m}{i}}{m-i+1}\right)^2 (T_2-T_1)^{2Hi}\Delta^{2(m-i)} + (T_2-T_1)^{2H(m-1)}\Delta^{2H}\Bigg\}\\
        & = (T_2-T_1) \Bigg\{\sum_{i=0}^{m-1} \left(\frac{\binom{m}{i}}{m-i+1}\right)^2 (T_2-T_1)^{2Hi}\Delta^{2(m-i)} + (T_2-T_1)^{2H(m-1)}\Delta^{2H}\Bigg\}\\
        &\lesssim  (T_2-T_1)^{1+2H(m-1)} \Delta^{2H}.
    \end{align*}}
Thus, the proof of estimate~\eqref{eq_L2_bound} for case~c) is complete.
It remains to verify that the desired almost sure convergence is true. Since we have an explicit rate of convergence and provided that~\eqref{eq:meshSummabilityLeadLag} is true, this assertion follows from Chebyshev's inequality and the Borel-Cantelli lemma. The reader is referred to the proof of Lemma~\ref{lem:HybridPointwisePartitionPoints} where a similar argument is employed in the context of hybrid scheme approximations.
\end{proof}

\begin{remark}[Comparison of lead-lag and left-point Riemann sum approximations] The reader might notice that in the proofs of Theorem~\ref{thm_leadlag} we use the fact that such approximations converge (in probability) to the associated It\^o integrals. We emphasise here that lead-lag approximations provide smooth approximations of the rough path $\barX$ which extends $(\bfX, W)$. Consequently, RDEs driven by $\barX$ can be approximated and simulated by (random) ODEs driven by the corresponding lead-lag approximations. Of course, the latter cannot be achieved by simple left-point Riemann sum approximations of It\^o integrals.    
\end{remark}

We conclude this section with an observation on the rate of convergence.
\begin{remark}\label{rem:alpha_Hremark} 
A careful look at the previous proof shows that the constraint $\eta \leq H$ in the rate of convergence only arises in case~c). 
For a) and b) it suffices to take $\eta \in (0, mH +\half]$. Similar considerations appear in the classical case~\cite[Proposition~13.21]{FV10} with $m=1$, $H=\half$.
\end{remark}

\subsection{Hybrid lead-lag approximations}\label{subsec:HybridLeadLag}

Throughout this section the path~$X$ is taken to be a type-II (or Riemann-Liouville) fractional Brownian motion
\begin{equation}\label{eq:fBm}
    X_t=\sqrt{2\zeta+1}\int_0^t(t-s)^{\zeta}\dif W_s,\; 
    \qquad  
    t\geq 0,
\end{equation}
with $\zeta:=H-\half$ and Hurst parameter $H\in (0, 1)$.
For our numerical simulations in Section~\ref{sec:SimCal} we shall mainly consider Rough Differential Equations driven by the pair $(X,W)$. Since the covariance of the Gaussian vector $(X,W)$ is explicit, one can simulate the pair by the classical Cholesky method. We choose a more efficient method to simulate $X$ by using the so-called hybrid scheme, initially introduced in~\cite{bennedsen2017hybrid}. The latter requires significantly smaller computational costs~\cite[Remark 3.1]{gassiat2023weak} and is in fact the current state-of-the-art for simulating non-Markovian, convolution-type Volterra processes such as $X$.

The hybrid scheme relies on approximating the It\^o integral in~\eqref{eq:fBm} by a standard, left-point Riemann sum away from the diagonal at $s=t$,
which is where the integrand is singular when $H<\half$.
To be precise, we let $T>0$,  $n\in\N$ and $\pi=\{0=t_0<t_1<\dots<t_{a(n)}=T\}$ a partition of $[0,T]$ with uniform mesh $t_{1}-t_0=\Delta>0$. The hybrid scheme (with truncation level $\kappa=1$) approximation of $X$ reads 
\begin{equation}\label{eq:Xhybrid}
\left\{
\begin{array}{rl}
    \G X_{0} & :=0,\\
    \G X_{t_k} & := \displaystyle \sqrt{2\zeta+1}\left(
    \int_{t_{k-1}}^{t_k}(t_k-s)^\zeta \dif W_s + \sum_{i=2}^{k}b^*_i W_{t_{k-i}, t_{k-i+1} }\right),
    \quad k=1,\dots, {a(n)},
    \end{array}
    \right.
\end{equation} 
where the weights, chosen to minimise the mean-square-error of the scheme, are given by $ b^*_i=\Delta^{-1}\int_{t_{i-1}}^{t_i} s^\zeta \dif s$,  $i=2,\dots, k
$ and $\G X$ is then extended to a continuous process on $[0,T]$ by piecewise linear interpolation between partition points.

Our goal in this section is to establish that lead-lag (piecewise linear) approximations of the integrals $I^m$ in~\eqref{eq:ImIntegrals}  with $X$ replaced by $\G X$ (or hybrid lead-lag approximations) converge in the rough path topology from Definition~\ref{def:rp}. In doing so we obtain: i) rigorous justifications for our numerical simulations (Theorem~\ref{thm:HybridLLconvergence}),  ii) a novel almost-sure convergence result for the hybrid scheme in H\"older topology (Theorem~\ref{thm:hybridalmostsure}).

Before we proceed to the main body of this section, we shall introduce some  additional notation. 
The hybrid lead-lag  approximations of $I^m$ are given by
\begin{equation}\label{eq:hybridleadlag}
		\begin{aligned}
			\G \widetilde{I}_{T_1, T_2}^{\Delta, m}=\int_{T_1}^{T_2} (\G \widetilde{X}_{T_1,t}^\Delta)^m \dif W^\Delta_t,
		\end{aligned}
	\end{equation} 
 where 
\begin{equation}\label{eq:hybridlaggedpiecewiselinear}
\left\{
\begin{array}{rl}
\G \widetilde{X}_t^\Delta
& = \displaystyle 0\mathbf{1}_{[t_0, t_{1}]}(t)+ \sum_{k=1}^{n-1}\bigg(\G X_{t_{k-1}}+\frac{(t-t_{k})}{\Delta}\G X_{t_{k-1}, t_k}\bigg)\mathbf{1}_{[t_{k}, t_{k+1})}(t)\;,\;\;t\in [0,T),\\
 & \G \widetilde{X}_T^\Delta= \G X_{t_{n-1}},
 \end{array}
 \right.
\end{equation} 
is the lagged hybrid approximation of $X$. Using notation from Section~\ref{subsection:LeadLagPiecewise}, $W^\Delta$, $\dot{W}^\Delta$, $\widetilde{X}^\Delta$, $\widetilde{I}^{\Delta, m}$ are the lead approximation of~$W$, its time-derivative, the lagged approximation of $X$ and the lead-lag piecewise linear approximation of the integrals $I^m$ in~\eqref{eq:ImIntegrals}, as in~\eqref{eq:Wlead},~\eqref{eq:WdotLead},~\eqref{eq:Xlagged},~\eqref{eq:ImLeadLag}, respectively.

We have already proved in Theorem~\ref{thm_leadlag} that for any $\gamma<mH+\half$,
$$
\E\left[|\widetilde{I}^{\Delta, m}- \widetilde{I}^{ m}|_{C^\gamma}\right]\longrightarrow 0,
$$
with an explicit $\gamma$-dependent rate. Our main strategy for showing that hybrid lead-lag approximations also converge to the correct limit lies in establishing that  

$$
\E\left[|\widetilde{I}^{\Delta, m}- \G\widetilde{I}^{\Delta, m}   |_{C^\gamma}\right]\longrightarrow 0.
$$
This is the subject of Theorem~\ref{thm:HybridLLconvergence}:
to obtain such a strong mode of convergence, we first prove almost sure convergence for the hybrid scheme at partition points.

\begin{lemma}\label{lem:HybridPointwisePartitionPoints}
    Let $H\in(0,1)$ and $\pi^n$ be a sequence of partitions of $[0,T]$ with uniform mesh $\Delta(n)\rightarrow 0$ as $n\to\infty$. Then, for each $t_k\in\pi^n$,
    \begin{align}\label{Eq:L^2boundspartpoints}
       \E\left[|\G X^n_{t_k}-X_{t_k}|^2\right] \leq C \Delta(n)^{2H}, 
    \end{align}
    where the constant $C$ is independent of $k$. 
    Moreover, if the mesh satisfies  
    \begin{equation}\label{eq:meshSummability}
        \sum_{n\in\N}\Delta(n)^{2H}{=\sum_{n\in \N}a(n)^{-2H}}<\infty ,
    \end{equation}
    then, for all $k\in\N$, we have 
    $|\G X^n_{t_k}-X_{t_k}|\longrightarrow 0$ almost surely as~$n$ tends to infinity.
\end{lemma}
\begin{proof}
Ignoring, for the moment,  the constant $\sqrt{2\zeta+1}$ we  write  
\begin{equation*}
    \begin{aligned}
       \G X^n_{t_k}- X_{t_k}
       &=\sum_{i=2}^{k}\Delta^{-1}\bigg(\int_{t_{i-1}}^{t_i} s^\zeta \dif s\bigg) W_{t_{k-i}, t_{k-i+1} }-\int_{0}^{t_{k-1}} (t_k-s)^\zeta \dif W_s  \\
       &= \left(\sum_{i=2}^{k}\Delta^{-1}\bigg(\int_{t_{i-1}}^{t_i} s^\zeta \dif s\bigg) W_{t_{k-i}, t_{k-i+1} }-\sum_{i=2}^{k} t^\zeta_{i-1} W_{t_{k-i}, t_{k-i+1} } \right)\\
       &\qquad + \left(\sum_{i=2}^{k} t^\zeta_{i-1} W_{t_{k-i}, t_{k-i+1} }-\sum_{i=2}^{k} t^\zeta_{i} W_{t_{k-i}, t_{k-i+1} } \right)\\
       &\qquad + \left(\sum_{i=2}^{k} t^\zeta_{i} W_{t_{k-i}, t_{k-i+1} }-\int_0^{t_{k-1}}(t_k-s)^\zeta \dif W_s \right) 
       = :\jo + \jj + \jjj.
    \end{aligned}
\end{equation*}
For $\jo$, we set $F(t)=\int_0^{t}s^\zeta \dif s$ and use the Taylor estimate 
    \begin{equation*}   \begin{aligned}
        \bigg|\Delta^{-1}\bigg(\int_{t_{i-1}}^{t_i} s^\zeta \dif s\bigg)-t^\zeta_{i-1}\bigg|
        &=\bigg|\frac{F(t_{i-1}+\Delta)-F(t_{i-1})}{\Delta}-F'(t_{i-1})\bigg|\\
        &\leq \frac{\Delta}{2}\sup_{s\in[t_{i-1}, t_{i-1}+\Delta   ]}|F''(s)|
        = \frac{|\zeta|\Delta}{2}t_{i-1}^{\zeta-1}.
    \end{aligned}
    \end{equation*}
From this, a conditioning argument (which takes care of cross-terms) and It\^o's isometry, we have 
\begin{equation*}
\E \left[I^2\right]
\leq \sum_{i=2}^{k} \frac{\zeta^2\Delta^{2}}{4}(t_{i-1})^{2\zeta-2}\Delta
=\frac{\zeta^2\Delta^3}{4}\Delta^{2H-3}\sum_{i=2}^{k}\big(i-1\big)^{2H-3}
\lesssim \Delta^{2H}\sum_{i=1}^{\infty}
i^{2H-3},
\end{equation*}
which converges since $3-2H>1$. 

As for $\jj$, the mean-value inequality implies
\begin{equation*}
\E \left[ \jj^2\right]
\leq \sum_{i=2}^{k} \bigg(t_{i-1}^\zeta- t_{i}^\zeta  \bigg)^2\E\left[W_{t_{k-i}, t_{k-i+1}}^2\right]
\lesssim \Delta\sum_{i=2}^{k}t_{i-1}^{2(\zeta-1)}(t_{i-1}-t_{i})^2
\lesssim \Delta^{2H},
\end{equation*}
where we used the same argument as above in the last inequality. 

Finally we re-index the sums that appear in $\jjj$ so that
\begin{equation*}
    \begin{aligned}
        \jjj=\sum_{i=0}^{k-2} (t_{k-i})^{ \zeta} W_{t_{i},t_{i+1} }-\int_0^{t_{k-1}}(t_k-s)^\zeta \dif W_s.
    \end{aligned}
\end{equation*}
Since we are using the uniform partition, $t_{k-i}=t_k-t_i$, and $\jjj$ is the error from the left-point Riemann sum approximation of the It\^o integral. 
Thus, It\^o isometry yields
\begin{align*}
 \E \left[\jjj^2\right]
 & =\sum_{i=0}^{k-2}\E\bigg[\int_{t_i}^{t_{i+1}} \bigg((t_{k-i})^{\zeta}-(t_k-s)^\zeta\bigg) \dif W_s\bigg]^2\\
 & \leq \sum_{i=0}^{k-2}\int_{t_i}^{t_{i+1}}\bigg((t_{k}-t_i)^{\zeta}-(t_k-s)^\zeta\bigg)^2\dif s\\
 & \leq (\zeta-1)^2\sum_{i=0}^{k-2}\int_{t_i}^{t_{i+1}}(t_k-s)^{2(\zeta-1)}(s-t_i)^2\dif s\\
& \lesssim \Delta^3 \sum_{i=0}^{k-2}(t_k-t_{i+1})^{2(\zeta-1)}
\lesssim \Delta^{2H}\sum_{i=1}^{\infty}i^{2H-3},
\end{align*}
where the third line follows from the mean-value inequality.
The proof of the first statement of Lemma~\ref{lem:HybridPointwisePartitionPoints}, namely~\eqref{Eq:L^2boundspartpoints}, is complete upon combining the estimates for $\jo$, $\jj$ and $\jjj$, together with the well-known upper bound $\left(\sum_{i=1}^N a_i\right)^2 \leq 2^N \sum_{i=1}^n a_i^2$ which holds for all $N\in\N$.

To obtain almost sure convergence, we fix $\eps>0$ and observe that~\eqref{Eq:L^2boundspartpoints} and Chebyshev's inequality yield
\begin{equation}\label{eq:cc}
    \begin{aligned}
       \sum_{n\in\N}\bbP\bigg(  |\G X^n_{t_k}- X_{t_k}|>\eps    \bigg)\leq C\eps^{-2}\sum_{n\in\N}\Delta(n)^{2H}<\infty,
    \end{aligned}
\end{equation}
for a constant $C>0$ independent of $\Delta$ and $k$, where in the last line we exploited the summability assumption on the partition as in~\eqref{eq:meshSummability}.

An application of the Borel-Cantelli lemma then yields
$$
\Pr\bigg(\limsup_{n\to\infty} |\G X^n_{t_k}-X_{t_k}|>\epsilon\bigg)=0,
$$ 
for all $\eps>0$, and therefore $\lim_{n\to\infty}(\G X^n_{t_k}-X_{t_k})=0$ $\bbP$-almost surely. The reader is referred to~\cite[Chapter 5, Theorem~3.1]{gut2006probability} for a proof of this assertion (in the language of this reference,~\eqref{eq:cc} is called \textit{complete convergence}). \end{proof}

\begin{remark}
Condition~\eqref{eq:meshSummability} is satisfied for the dyadic partitions $\Delta(n)=T/2^{n}$.
\end{remark}

We now obtain uniform bounds for the hybrid scheme in H\"older topology. 

\begin{lemma}\label{lem:HybridHolderBounds} 
    Let $H\in(0,1)$.  
    There exists $C>0$ such that, for all $n, k,m\in\N$, $p\geq 1$, 
    \begin{equation}\label{eq:HybridTightnessEstimate}
        \E\left[|\G X^n_{t_k,t_{m}}|^p\right] \leq C(t_{m}-t_{k})^{pH}.
    \end{equation}
    Moreover, for all $\gamma<H$, there exists a random constant $K>0$ with finite moments of all orders such that, for each $n\in\N$, and $t,s\in[0,T]$,
    \begin{align*}
        | \G X^n_{t}-\G X^n_{s}  |\leq K( t-s)^{\gamma},
        \quad\text{almost surely}.
    \end{align*}
\end{lemma}

\begin{proof} From Lemma~\ref{lem:HybridPointwisePartitionPoints} and H\"older regularity of $X$ we have
\begin{equation*}
    \begin{aligned}
        \E\left[|\G X^n_{t_k,t_{m}}|^2\right]
        &\leq 3\E\left[|\G X^n_{t_{m}}-X_{t_{m}}|^2\right]
        +3\E\left[|\G X^n_{t_{k}}-X_{t_{k}}|^2\right]
        +3\E\left[|X_{t_{k},t_{m} }|^2\right]\\
        & \lesssim  \Delta^{2H}+  (t_{m}-t_{k})^{2H}
        \lesssim (t_{m}-t_{k})^{2H},
    \end{aligned}
\end{equation*}
up to constants independent of $n, k, m$. 
From Lemma~\ref{lem:Hypercontractivity}, $\E\left[|\G X^n_{t_k,t_{m}}|^p\right]\lesssim (t_{m}-t_{k})^{pH}$
for any $p\geq 2$,
up to constants that depend on $p$ as well (but still not on $n,k,m$).
Finally, since the hybrid approximation is defined via piecewise linear interpolation between partition points, the same estimate holds by replacing $t_m,t_k$ with arbitrary $s,t\in [0,T]$.
A similar argument can be found in~\cite[Theorem~3.4]{horvath2024functional}. Kolmogorov's continuity criterion \autoref{thm:Kolm} yields the conclusion.
\end{proof}

In Lemma~\ref{lem:HybridPointwisePartitionPoints}, we have shown that the hybrid scheme approximation  $\{\G X^n\}_{n\in\N}$ converges to $X$ pointwise almost surely at partition points (provided that the mesh of the partition is fine enough). Moreover, as shown in Lemma~\ref{lem:HybridHolderBounds}, for almost every $\omega\in\Omega, \gamma<H$, the sequence  $\{\G X^n(\omega)\}_{n\in\N}\subset C^\gamma([0,T])$ is uniformly bounded. Below we show that these two statements are sufficient to pass to almost sure convergence of the sequence to $X$ in H\"older topology.

\begin{theorem}[Almost sure convergence of the hybrid scheme]\label{thm:hybridalmostsure}
    Let $X, \G X$ as in~\eqref{eq:fBm},~\eqref{eq:Xhybrid}, $T>0, H\in(0,1)$. For each $\gamma<H$ then 
    $$
        \lim_{n\to\infty}|\G X^n-X|_{C^\gamma[0,T]}=0,        
    $$
    almost surely and in $L^p(\Omega)$, for any $p\geq 1$. 
\end{theorem}

\begin{proof}  

By interpolation~\cite[Exercise 2.9]{FH20}, uniform H\"older bounds and pointwise (almost sure) convergence for all $t\in[0, T]$ imply convergence in H\"older topology. In fact, pointwise convergence at partition points is sufficient for convergence of the first level of a rough path.  
From Lemma~\ref{lem:HybridPointwisePartitionPoints} and the uniform H\"older bounds in Lemma~\ref{lem:HybridHolderBounds} we conclude that, for all $t\in[0,T]$,
$  \G X^n_{t}\rightarrow X_{t}$,  almost surely, hence we obtain convergence in H\"older topology. 
As for the $L^p$ convergence, notice from Lemma~\ref{lem:HybridHolderBounds} that, for all $p\geq 1$,
\begin{align*}
    \sup_{n\in\N}\E\big[|\G X^n-X|^p_{C^\gamma([0,T])}\big]<\infty,
\end{align*}
and the theorem follows by the dominated convergence theorem.
\end{proof}

We turn our attention to hybrid lead-lag approximations of the integrals $I^m_{T_1, T_2}$.

\begin{theorem}[Convergence of hybrid lead-lag approximations]\label{thm:HybridLLconvergence}
    Let 
    $T>0$, $H\in(0,1)$, $ m\in\N$ be such that $mH+\half<1$ and $\gamma\in (0, mH+\half)$.
    For $n\in\N$, let $\pi^n=\{t^n_k\}_{k=0,\dots, {a(n)}}$ be a partition of $[0,T]$ with uniform mesh $\Delta(n):=|t^n_1-t^n_{0}|{=a(n)^{-1}}$, such that $\lim_{n\to\infty}\Delta(n)=0$.
    Moreover, let   $\G\widetilde{I}^{\Delta(n), m}$  as in~\eqref{eq:hybridleadlag}
    denote the hybrid lead-lag approximation of the iterated integral $I^{m}$ in~\eqref{eq:ImIntegrals} along the partition $\pi^n$. 
    For all $p\geq 1$, we have
    \begin{align*}
        \lim_{n\to\infty}\sup_{T_1\neq T_2\in[0,T]}\frac{\big|\G\widetilde{I}^{\Delta(n), m}_{T_1,T_2}-I^m_{T_1,T_2}\big|}{|T_2-T_1|^\gamma}=0, \qquad \text{ in } L^p(\Omega).
    \end{align*}
\end{theorem}

\begin{proof} We prove the convergence by comparing the lead-lag approximation~\eqref{eq:ImLeadLag}
to the hybrid lead-lag approximation~\eqref{eq:hybridleadlag}.
To this end, fix $T_1<T_2\in[0,T]$ and write

\begin{equation}\label{eq:LeadLagHybridLeadLagDifference}
    \frac{\big|\G\widetilde{I}^{\Delta, m}_{T_1,T_2}-I^m_{T_1,T_2}\big|}{|T_2-T_1|^\gamma}\leq \frac{\big|\G\widetilde{I}^{\Delta, m}_{T_1,T_2}-\widetilde{I}^{\Delta, m}_{T_1,T_2}\big|}{|T_2-T_1|^\gamma}+\frac{\big|\widetilde{I}^{\Delta, m}_{T_1,T_2}-I^{m}_{T_1,T_2}\big|}{|T_2-T_1|^\gamma}.
\end{equation}
In view of Theorem~\ref{thm_leadlag}, we have
\begin{equation}   \label{eq:LeadLagLpconvergence}
    \E \bigg[\bigg(\sup_{T_1\neq T_2\in[0,T]}\frac{\big|\widetilde{I}^{\Delta, m}_{T_1,T_2}-I^m_{T_1,T_2}\big|}{|T_2-T_1|^\gamma}\bigg)^p\bigg]\lesssim \Delta(n)^{p\eta}\longrightarrow 0,
\end{equation}
as $n\to\infty$, for any $\eta<H$ (see Remark~\ref{rem:alpha_Hremark})
It remains to show that the same is true for the first term on the right-hand side of~\eqref{eq:LeadLagHybridLeadLagDifference}. To this end, consider three cases depending respectively  on whether $T_1, T_2$ are partition points, lie on the same partition interval or belong to different partition intervals. Throughout the proof and for sake of lighter notation we drop the superscript~$n$ from partition points and write  $t_k^n=:t_k$.\\

\noindent \textbf{Case a): $T_1, T_2\in\pi^n$ are partition points} Let $0 \leq T_1= t_{j} \leq t \leq T_2 = t_{j+\ell} \leq T$, for some $j \in\{0, \dots, {a(n)} - \ell\}$, $\ell \in \{1, \dots, {a(n)}-1\}$. 
The corresponding lead-lag and hybrid lead-lag approximations read
\begin{align*}
\widetilde{I}^{\Delta,m}_{T_1, T_2}
   & = \int_{T_1}^{T_2} (\widetilde{X}^\Delta_{T_1, t})^m  \dif W^{\Delta}_t
     = \sum_{k=0}^{\ell-1}\bigg( \sum_{i=0}^m \frac{\binom{m}{i}}{m-i+1}
     X_{t_{j-1}, t_{j+k-1}}^i X_{t_{j+k-1}, t_{j+k}}^{m-i}\bigg) W_{t_{j+k}, t_{j+k+1}},\\
\G\widetilde{I}^{\Delta,m}_{T_1, T_2}
    &= \int_{T_1}^{T_2} (\G\widetilde{X}^\Delta_{T_1, t})^m  \dif W^{\Delta}_t
     \\&= \sum_{k=0}^{\ell-1}\bigg( \sum_{i=0}^m \frac{\binom{m}{i}}{m-i+1}(\G X_{t_{j-1}, t_{j+k-1}})^i (\G X_{t_{j+k-1}, t_{j+k}})^{m-i}\bigg) W_{t_{j+k}, t_{j+k+1}},
\end{align*}
respectively. Hence, rearranging terms we obtain
{\footnotesize 
\begin{equation}\label{eq:Casea)estimate}
    \begin{aligned}
    & \G\widetilde{I}^{\Delta,m}_{T_1, T_2}-\widetilde{I}^{\Delta,m}_{T_1, T_2}\\
    &  =\sum_{k=0}^{\ell-1} \sum_{i=0}^m \frac{\binom{m}{i}}{m-i+1}\bigg[(\G X_{t_{j-1}, t_{j+k-1}})^i (\G X_{t_{j+k-1}, t_{j+k}})^{m-i}
    - X_{t_{j-1}, t_{j+k-1}}^i X_{t_{j+k-1}, t_{j+k}}^{m-i}\bigg] W_{t_{j+k}, t_{j+k+1}}\\
    &
   =\sum_{k=0}^{\ell-1} \sum_{i=0}^m \frac{\binom{m}{i}}{m-i+1}\bigg[ (\G X_{t_{j-1}, t_{j+k-1}})^i\bigg((\G X_{t_{j+k-1}, t_{j+k}})^{m-i} - X_{t_{j+k-1}, t_{j+k}}^{m-i}    \bigg)   \bigg]W_{t_{j+k}, t_{j+k+1}}\\&
   +\sum_{k=0}^{\ell-1} \sum_{i=0}^m \frac{\binom{m}{i}}{m-i+1}\bigg[   X_{t_{j+k-1}, t_{j+k}}^{m-i}\bigg((\G X_{t_{j-1}, t_{j+k-1}})^i - X_{t_{j-1}, t_{j+k-1}}^i   \bigg)   \bigg]W_{t_{j+k}, t_{j+k+1}}.
   \end{aligned}
\end{equation}
}
The differences that appear on the last display can be estimated by writing
\begin{align}\label{eq:HybridPowerDifference}
    &( \G X_{t_{j+k-1}, t_{j+k}})^{m-i}-  X_{t_{j+k-1}, t_{j+k}}^{m-i}\\ \nonumber
    &=( \G X_{t_{j+k-1}, t_{j+k}}-X_{t_{j+k-1}, t_{j+k}})p_{m-i-1}\big( \G X_{t_{j+k-1}, t_{j+k}},  X_{t_{j+k-1}, t_{j+k}}   \big),
\end{align}
where $p_{m-i-1}(x,y)$ is a homogeneous polynomial of degree $m-i-1$ in $x,y$. Then, from the H\"older continuity of $X$ and the uniform (in $n$) H\"older estimates for $\G X$ from Lemma~\ref{lem:HybridHolderBounds} we have, for any $\gamma'<H$,
\begin{align}\label{eq:estGXpower}
    & \bigg|( \G X_{t_{j+k-1}, t_{j+k}})^{(m-i)} - X_{t_{j+k-1}, t_{j+k}}^{m-i}\bigg|\\
    &\lesssim \big|\G X-X\big|_{C^{\gamma'}}\max\left\{ \big|\G X\big|^{m-i-1}_{C^{\gamma'}},\big| X\big|^{m-i-1}_{C^{\gamma'}}\right\}
    (t_{j+k}-t_{j+k-1})^{(m-i)\gamma'}.
\end{align}
and similarly
\begin{align*}
    \bigg|\bigg((\G X_{t_{j-1}, t_{j+k-1}})^i&- X_{t_{j-1}, t_{j+k-1}}^i   \bigg)\bigg|\\&\lesssim \big|\G X-X\big|_{C^{\gamma'}}\max\left\{ \big|\G X\big|^{i-1}_{C^{\gamma'}},\big| X\big|^{i-1}_{C^{\gamma'}}\right\}
    (t_{j+k}-t_{j+k-1})^{i\gamma'}.
 \end{align*}
 
By working on $L^2(\Omega)$ and remembering that cross-terms over~$k$ vanish due to independence of the disjoint increments of~$W$,
that the increments of $X$ and~$W$ are independent and the tower property, we obtain
{\footnotesize 
\begin{align}\label{eq:star}
\nonumber
    &\E \Bigg[ \left\{\sum_{k=0}^{\ell-1} \sum_{i=0}^m \frac{\binom{m}{i}}{m-i+1}\bigg[ (\G X_{t_{j-1}, t_{j+k-1}})^i\bigg\{(\G X_{t_{j+k-1}, t_{j+k}})^{m-i}- X_{t_{j+k-1}, t_{j+k}}^{m-i}    \bigg\}   \bigg]W_{t_{j+k}, t_{j+k+1}}\right\}^2\Bigg]\\
    & \nonumber=\sum_{k=0}^{\ell-1} \sum_{i=0}^m \frac{(m+1)\binom{m}{i}^2}{(m-i+1)^2}\E\Bigg[\Bigg\{ (\G X_{t_{j-1}, t_{j+k-1}})^i\bigg\{(\G X_{t_{j+k-1}, t_{j+k}})^{m-i}- X_{t_{j+k-1}, t_{j+k}}^{m-i}    \bigg\} W_{t_{j+k}, t_{j+k+1}}\Bigg\}^2\Bigg]\\
    &\lesssim \Delta \sum_{k=0}^{\ell-1} \sum_{i=0}^m \bigg(\frac{\binom{m}{i}}{m-i+1}\bigg)^2\E\Bigg[(\G X_{t_{j-1}, t_{j+k-1}})^{2i} \bigg((\G X_{t_{j+k-1}, t_{j+k}})^{m-i} - X_{t_{j+k-1}, t_{j+k}}^{m-i}    \bigg)^2 \Bigg].
\end{align}
}
Now, in order to handle the terms involving $X$, we exploit~\eqref{eq:estGXpower} and the Cauchy-Schwarz inequality along with the first estimate in Lemma~\ref{lem:HybridHolderBounds}. Thus, we can further bound the last expression in~\eqref{eq:star} by
\begin{align*}
    &\Delta \sum_{k=0}^{\ell-1} \sum_{i=0}^m \Delta^{2(m-i)\gamma'}  \E\Bigg[(\G X_{t_{j-1}, t_{j+k-1}})^{2i} \big|\G X-X\big|_{C^{\gamma'}}^2\max\left\{ \big|\G X\big|^{m-i-1}_{C^{\gamma'}},\big| X\big|^{m-i-1}_{C^{\gamma'}}\right\}^2\Bigg]\\
    & \lesssim \Delta^{1+2m\gamma'}  \big|\big|\G X-X\big|_{C^{\gamma'}}\big|^{2}_{L^4(\Omega)} 
    \E\Bigg[ \max\left\{ \big|\G X\big|^{m-1}_{C^{\gamma'}},\big| X\big|^{m-1}_{C^{\gamma'}}\right\}^8\Bigg]^{\frac{1}{4}}\cdot\\&\cdot\sum_{k=0}^{\ell-1} \sum_{i=0}^m \Delta^{-2i\gamma'}  \E\Bigg[(\G X_{t_{j-1}, t_{j+k-1}})^{8i}\Bigg]^{\frac{1}{4}}  \\
    &\lesssim \Delta^{1+2m\gamma'}  \big|\big|\G X-X\big|_{C^{\gamma'}}\big|^{2}_{L^4(\Omega)} \sum_{k=0}^{\ell-1} \sum_{i=0}^m \Delta^{-2i\gamma'}  t_k^{2iH}  \\
    &\lesssim \Delta^{1+2m\gamma'}  \big|\big|\G X-X\big|_{C^{\gamma'}}\big|^{2}_{L^4(\Omega)} (T_2-T_1)^{2mH}\Delta^{-2m\gamma'}\ell  \\
    & =\big|\big|\G X-X\big|_{C^{\gamma'}}\big|^{2}_{L^4(\Omega)}(T_2-T_1)^{2mH+1},
    \end{align*}
where the final upper bound on the double sum holds since 
$t_k\leq T_2-T_1$.
With these arguments and similar ones to handle the second sum on the right-hand side of~\eqref{eq:Casea)estimate}, along with hypercontractivity (Lemma~\ref{lem:Hypercontractivity}), then
\begin{equation*}
    \left|\G\widetilde{I}^{\Delta(n), m}_{T_1,T_2}-\widetilde{I}^{\Delta, m}_{T_1,T_2}\right|_{L^p(\Omega)}\lesssim \left|\left|\G X-X\right|_{C^{\gamma'}}\right|_{L^4(\Omega)}(T_2-T_1)^{mH+\half} . 
\end{equation*}

\noindent \textbf{Case b): $T_1, T_2$ are on the same partition interval.}
Let $t_{j}\leq T_1 \leq t \leq T_2 \leq t_{j+1}$, for some $j \in\{0, \dots, {a(n)}-1\}$.
Recalling the explicit expression in~\eqref{eq:hybridlaggedpiecewiselinear} we see that in our interval of interest
\begin{align*}
    \G\widetilde{X}^\Delta_{T_1,t} 
    &= \G\widetilde{X}^\Delta_{t} - \G\widetilde{X}^\Delta_{T_1} = \left(\G X_{t_{j-1}} + \frac{t-t_j}{\Delta}\G X_{t_{j-1},t_j}\right) - \left(\G X_{t_{j-1}} + \frac{T_1-t_j}{\Delta}\G X_{t_{j-1},t_j}\right) \\
    &= \frac{t-T_1}{\Delta}\G X_{t_{j-1},t_j},
\end{align*}
and so, in particular, we obtain
\begin{align*}
    \G\widetilde{I}^{\Delta,m}_{T_1, T_2}
    &= \int_{T_1}^{T_2} (\G \widetilde{X}^\Delta_{T_1,t})^m  \dif W^{\Delta}_t
    = \int_{T_1}^{T_2}\left( \frac{t-T_1}{\Delta}\G X_{t_{j-1},t_j}\right)^m \frac{W_{t_{j},t_{j+1}}}{\Delta} \dif t \\
    & = \frac{(T_2-T_1)^{m+1}}{(m+1)\Delta^{m+1}} \left(\G X_{t_{j-1},t_j}\right)^m W_{t_{j},t_{j+1}}.
\end{align*}  
Thus, we have
\begin{equation*}
\begin{aligned}
\G\widetilde{I}^{\Delta,m}_{T_1, T_2}-\widetilde{I}^{\Delta,m}_{T_1, T_2}=\frac{(T_2-T_1)^{m+1}}{(m+1)\Delta^{m+1}} \bigg(\left(\G X_{t_{j-1},t_j}\right)^m - X_{t_{j-1},t_j}^m\bigg)W_{t_{j},t_{j+1}}.
\end{aligned}
\end{equation*}
Notice that here we do not need estimates as in~\eqref{eq:HybridPowerDifference}. Instead we use the independence of $X, 
\G X$ and~$W$ on the above intervals, along with the uniform bounds from Theorem~\ref{thm:hybridalmostsure} to
write
\begin{align*}
& \left|\G\widetilde{I}^{\Delta,m}_{T_1, T_2}-\widetilde{I}^{\Delta,m}_{T_1, T_2}\right|_{L^2(\Omega)}\\
&\leq\frac{(T_2-T_1)^{m+1}}{(m+1)\Delta^{m+1}}|W_{t_{j},t_{j+1}}|_{L^2(\Omega)}
    \bigg( \left|(\G X_{t_{j-1},t_j})^m\right|_{L^2(\Omega)}+\left|X_{t_{j-1},t_j}^m\right|_{L^2(\Omega)}  \bigg) \\
    &\lesssim \frac{(T_2-T_1)^{m+1}}{(m+1)\Delta^{m+1}}\Delta^{mH+\half}
    \lesssim (T_2-T_1)^{mH+\half-\eta}\Delta^\eta,
\end{align*}
where we also used that $T_2-T_1\leq \Delta$.
Here $\eta$ is an arbitrary non-negative constant that satisfies $\eta<m(H+\half)$. 
From Lemma~\ref{lem:Hypercontractivity} it follows that for any $p\geq 1$,
$$ \E\left[\left|\G\widetilde{I}^{\Delta,m}_{T_1, T_2}-\widetilde{I}^{\Delta,m}_{T_1, T_2}\right|^p\right]\lesssim  (T_2-T_1)^{p(mH+\half-\eta)}\Delta^{p\eta}.   $$

\noindent \textbf{Case c): $T_1, T_2$ are on different  partition intervals.} 
For $\ell \in \{1, \dots,{a(n)}-1\}$, $j\in\{0,\dots,{a(n)}-\ell\}$,  let $0 < t_j < T_1 < t_{j+1} < \dots < t_{j+\ell -1} < T_2 < t_{j+\ell} <T$.
Similarly to what we have done for case a), we exploit the expression for $\widetilde{X}^\Delta_{t}$ in~\eqref{eq:Xlagged} to write
{\footnotesize 
\begin{equation*}
\left(\G \widetilde{X}^\Delta_{T_1,t}\right)^m
 = 
\sum_{k=1}^{\ell}\bigg(\G X_{t_{j-1}, t_{j+k-1}}+\frac{(t-t_{j+k})}{\Delta}\G X_{t_{j+k-1}, t_{j+k}} +\frac{(T_1-t_{j})}{\Delta}\G X_{t_{j-1}, t_{j}} \bigg)^m \mathbf{1}_{[t_{j+k}, t_{j+k+1})}(t).
\end{equation*} 
}
As in the proof of Theorem~\ref{thm_leadlag}, we decompose the hybrid lead-lag approximation as
{\footnotesize
\begin{equation*}
\begin{aligned}
	&\G\widetilde{I}^{\Delta,m}_{T_1, T_2}
     = \int_{T_1}^{T_2} (\G \widetilde{X}^\Delta_{T_1, t})^m  \dif W^{\Delta}_t \\&
     = \frac{1}{m+1}\frac{(t_{j+1}-T_1)^{m+1}}{\Delta^{m+1}}\G X_{t_{j-1},t_j}^m W_{t_{j},t_{j+1}}\\&
   + \sum_{k=1}^{\ell-2}\sum_{i=0}^m \frac{\binom{m}{i}}{m-i+1}\bigg(\G X_{t_{j-1}, t_{j+k-1}}-\frac{(T_1-t_{j})}{\Delta}\G X_{t_{j-1}, t_{j}}\bigg)^i ( \G X_{t_{j+k-1}, t_{j+k}})^{m-i} W_{t_{j+k}, t_{j+k+1}}  \\&
 +  \sum_{i=0}^m \frac{\binom{m}{i}}{m-i+1}\frac{(T_2-t_{j+\ell-1})^{m-i+1}}{\Delta^{m-i+1}}\bigg(\G X_{t_{j-1}, t_{j+\ell-2}}-\frac{(T_1-t_{j})}{\Delta}\G X_{t_{j-1}, t_{j}}\bigg)^i (\G X_{t_{j+\ell-2}, t_{j+\ell-1}})^{m-i} W_{t_{j+\ell-1}, t_{j+\ell}}\\&
 =:\G A+\G B+\G C,
\end{aligned}
\end{equation*}
 }
where $A, B, C$ correspond to the terms in the lead-lag approximation (with~$\G X$ substituted by~$X$).
The first term corresponds to Case~b) in the sense that the limits of integration are points of the same partition interval $[t_j, t_{j+1}]$.
The second and third terms are similar to what we dealt with in Case~a), where the limits of integration are respectively given by the partition points $\{t_{j+1}, t_{j+\ell-1}\}$ and $\{t_{j+\ell-1}, t_{j+\ell}\}$.
The difference lies in the presence of the base point~$T_1$ which does not coincide with the lower bound of integration since we are not using Chen's relation but rather decomposing the domain of integration. 
Nevertheless, the estimates we obtain for these three terms are in complete analogy to the previous two cases.

Indeed, for the first term we can use the independence of increments between the~$X$ and~$W$ terms along with the fact that $t_{j+1}-T_1<\Delta$ to obtain 
\begin{equation}\label{eq:caseC)term1}
    \begin{aligned}
    |\G B-B|_{L^2(\Omega)}&=
        \bigg|\frac{(t_{j+1}-T_1)^{m+1}}{\Delta^{m+1}}\bigg(\G X_{t_{j-1},t_j}^m- X_{t_{j-1},t_j}^m\bigg) W_{t_{j},t_{j+1}}\bigg|_{L^2(\Omega)}\\&\lesssim \Delta^{mH+\half}\leq (T_2-T_1)^{mH+\half-\eta}\Delta^{\eta},
    \end{aligned}
\end{equation}
where, as in Case b), $\eta$ is an arbitrary non-negative constant such that $\eta<m(H+\half)$.
By Lemma~\ref{lem:Hypercontractivity} we can replace the $L^2$-norm on the left-hand side by the $L^p$-norm for any $p\geq 1$ to obtain the desired estimate.

For the second and third term note the following. First, the ratios $\frac{(T_1-t_{j})}{\Delta}$ and $\frac{T_2-t_{j+\ell-1}}{\Delta}$ do not depend on the summation index and are bounded above by~$1$.
Then, the cross terms that appear when raising to power~$2$ vanish on expectation. 
This can be seen by conditioning and taking advantage of the independence of $W_{s,t}$ on~$\mathcal{F}_{s}$.  Finally, the expectation of the diagonal terms that appear when raising in power~$2$ splits in the product of expectations of $\G X$ and~$W$ terms due to independence. 
Thus, arguing as in Case a), we obtain the estimates
\begin{equation*}
\begin{aligned}
    |\G B-B|_{L^2(\Omega)}+|\G C-C|_{L^2(\Omega)}\lesssim \big|\big|\G X-X\big|_{C^{\gamma'}}\big|_{L^4(\Omega)}(T_2-T_1)^{m\gamma'+\half}.
\end{aligned}
\end{equation*}
Combining the latter with ~\eqref{eq:Casea)estimate} and using hypercontractivity (Lemma~\ref{lem:Hypercontractivity}) we deduce that, for all $p\geq 1$,
\begin{equation}\label{eq:HybridLeadLagLpRate}
    \begin{aligned}
    \left|\G\widetilde{I}^{\Delta,m}_{T_1, T_2}-\widetilde{I}^{\Delta,m}_{T_1, T_2}\right|_{L^p(\Omega)}
    &\lesssim  \big|\big|\G X-X\big|_{C^{\gamma'}}\big|_{L^4(\Omega)}(T_2-T_1)^{m\gamma'+\half}+(T_2-T_1)^{mH+\half-\eta}\Delta^{\eta}\\
    &\leq (T_2-T_1)^{m\gamma'+\half}\max\bigg\{ \big|\big|\G X-X\big|_{C^{\gamma'}}\big|_{L^4(\Omega)}, \Delta^{m(H-\gamma')}   \bigg\}, 
    \end{aligned}
\end{equation}
where we chose $\eta=m(H-\gamma')$ and the constants do not depend on $n$.\\

Note that this last estimate holds in all three cases (with this choice of $\eta$). By Kolmogorov's continuity criterion \autoref{thm:Kolm} for the two parameter process 
$(T_1,T_2)\mapsto \G\widetilde{I}^{\Delta,m}_{T_1, T_2}-\widetilde{I}^{\Delta,m}_{T_1, T_2}$,
we deduce that
\begin{equation*}
\E\left[\left(\sup_{T_1\neq T_2\in[0,T]}\frac{\big|\G\widetilde{I}^{\Delta(n), m}_{T_1,T_2}-I^m_{T_1,T_2}\big|}{|T_2-T_1|^\gamma}\right)^p\right]
        \lesssim \max\bigg\{ \big|\big|\G X-X\big|_{C^{\gamma'}}\big|_{L^4(\Omega)}, \Delta^{m(H-\gamma')}   \bigg\}^p,
\end{equation*}
for all
$p\geq 1$ and 
$\gamma<m\gamma'+\half$,
and the right-hand side vanishes as $n\to\infty$ by virtue of Theorem~\ref{thm:hybridalmostsure}. Since $\gamma'$ can be taken arbitrarily close to~$H$, the result follows by combining the latter with~\eqref{eq:LeadLagLpconvergence}.
\end{proof}

\begin{remark}[On the almost sure convegence of hybrid lead-lag approximations] In Theorems~\ref{thm:hybridalmostsure} and~\ref{thm_leadlag} we showed that the hybrid scheme approximation of $X$ and the lead-lag approximations of the iterated integrals~$I^m, m\in\N$ converge almost surely in appropriate H\"older topologies. 
For the former, we proved almost sure convergence at partition points and uniform H\"older bounds (Lemma~\ref{lem:HybridHolderBounds}). The interpolation result from~\cite[Exercise~2.9]{FH20}
then allows us to conclude almost sure convergence in H\"older topology. Such a strategy relies on the use of Chen's relation for the second level and is not directly applicable to prove almost sure convergence of hybrid lead-lag approximations for $I^m$.

For the latter, we obtained explicit rates of convergence for H\"older norms in $L^p, p\geq 1$. 
These allow us to obtain almost sure convergence provided that the rate of convergence (and hence the mesh of the partition) vanishes sufficiently fast as $n\to\infty$. 
Such arguments are equally insufficient to deduce almost sure convergence of hybrid lead-lag approximations. 
Indeed, from a glance at~\eqref{eq:HybridLeadLagLpRate},  the $L^p$-rate of convergence in H\"older norm depends on the $L^4$-rate of convergence of $\big|\G X-X\big|_{C^{\gamma'}}$ to~$0$.
To the best of our knowledge,  such a strong rate is an open problem and beyond the scope of the present work. Even though this prevents us from proving almost sure convergence, we were still able to prove convergence in probability. This mode of convergence is both typical for piecewise linear approximations of geometric rough paths and sufficient for our purposes.
\end{remark}

\subsection{Lagged mollifier approximations}\label{subsec:LaggedMollifier}

Throughout this section we fix a (one-dimensional) Wiener process~$W$ and the first level $X$ of the one-dimensional geometric rough path from Definition~\ref{def:adapted}.
Moreover, we let $\varphi \in \mathcal{C}_c^\infty(\mathbb{R})$ be a smooth test function such that $\text{supp}(\varphi)\subset(-1,1)$ and $\int_{\mathbb{R}}\varphi(x)\dif x =1$.
For $\eps\in (0,1), x\in\R, T>0$  we set $\varphi_\eps(x):= \frac{1}{\eps}\varphi(x/\eps)$ and consider the mollifier approximations of the paths $X$ and~$W$ on the interval $[0,T]$: 
\begin{align}\label{eq:mollifiedpath}
    & X^\eps_t := (\varphi_\eps * X)(t)
    = \int_{\mathbb{R}}\varphi_\eps(t-s)X_s \dif s
    = \int_{t-\eps}^{t+\eps}\varphi_\eps(t-s)X_s \dif s,\\ \nonumber
    & W^\eps := (\varphi_\eps * W).
\end{align}
In this section, we consider lagged mollifier approximations of the integrals
$I_{T_1, T_2}^m:=\int_{T_1}^{T_2} X_{s,r}^m \dif W_r$, with  $0\leq T_1\leq T_2\leq T$, $m\in\N$,
given by
\begin{equation*}
		\begin{aligned}			\widetilde{I}_{T_1, T_2}^{\eps, m}
        :=\int_{T_1}^{T_2} (\widetilde{X}_{T_1,r}^\eps)^m \dif W^\eps_r,
		\end{aligned}
	\end{equation*}
where $W^\eps$ is the (lead) standard mollifier approximation from~\eqref{eq:mollifiedpath}, 
\begin{equation}\label{eq:mollifierlag}
    \widetilde{X}^\eps_t:=X^\eps_{t-2\eps}=\int_{t-3\eps}^{t-\eps}\varphi_\eps(t-2\eps-s)X_s \dif s\;, t \in [0,T], 
\end{equation} 
is a lagged mollifier approximation of $X$ and $\widetilde{X}_{T_1,r}^\eps:= \widetilde{X}_{r}^\eps-\widetilde{X}_{T_1}^\eps$.
Our main convergence result below--proved at the end of the section--comes with explicit rates of convergence, and is preceded by an important auxiliary lemma.

\begin{theorem}[Convergence of lagged mollifier approximations]\label{thm:LaggedMollifierApproximations} Let $T>0$, $  m\in\N$ with $mH+\half\in(0,1)$ and $\eta\in (0, H)$. For each $\gamma<mH+\half$, $p\geq 1$, there exists $C>0$ such that, for all $\eps\in (0,1)$,
\begin{equation}\label{eq:MollifierApproxHolderNorm}
\E\left[\left(\sup_{T_1\neq T_2\in[0,T]}\frac{|\widetilde{I}_{T_1, T_2}^{\eps, m}-I_{T_1,T_2}^m|}{|T_2-T_1|^\gamma}\right)^p\right]\leq C \eps^{pH}.
\end{equation}
In particular, $|\widetilde{I}^{\eps, m}-I^m|_{\mathcal{C}^\gamma}\rightarrow 0$ in probability as $\eps\to 0$.
\end{theorem}

In~\cite{BFGJS20}, it was shown that, if $X, W$ are fully correlated, the non-lagged mollifier approximations 
\begin{align*}
    I_{T_1, T_2}^{\eps, m}=\int_{T_1}^{T_2} (X_{T_1,r}^\eps)^m \dif W^\eps_r
\end{align*}
do not converge to the integrals $I^{m}_{T_1, T_2}$. 
In particular, a key consequence of~\cite[Theorem~1.4]{BFGJS20} states that
\begin{align*}
    I_{0,t}^{\eps, m}-I_{0,t}^{m}\sim\eps^{H-\half},
\end{align*}
as $\eps\to 0$. By exploiting the theory of regularity structures, it is possible to show that convergence is restored after subtracting a diverging renormalisation term of the same order.
In contrast, Theorem~\ref{thm:LaggedMollifierApproximations} shows that no renormalisation is required if one chooses to approximate $I^{m}$ by lagged mollifier approximations. This is a consequence of the fact that the increments of the lagged and lead approximations $\widetilde{X}^\eps, W^{\eps}$ are uncorrelated. Due to this probabilistic cancellation, renormalisation constants that appear in standard mollifier approximations are no longer present in our framework. This heuristic explanation is rigorously justified in the next lemma.

\begin{lemma}\label{lem:zeroRenormalisation}
Fix $T>0$. 
For $\eps\in(0,1)$, $m\in\N$ and $0\leq T_1\leq T_2\leq T$, then  
\begin{equation*}
		\begin{aligned}			\widetilde{I}_{T_1, T_2}^{\eps, m}&
		=\int_{T_1}^{T_2}(\widetilde{X}_{T_1,r}^\eps)^m \deltaup  W^\eps_r
        =\sum_{k=0}^{m}\binom{m}{k}(\widetilde{X}^\eps_{T_1})^{m-k}\int_{T_1}^{t}(\widetilde{X}_{r}^\eps)^k \deltaup  W^\eps_r,
		\end{aligned}
\end{equation*}
holds almost surely, where $\deltaup$ denotes Skorokhod integration.
\end{lemma}

\begin{proof}  We only prove the first equality since the second is a simple consequence of the binomial identity.
In view of the calculations in~\cite[Lemma 3.12]{BFGJS20} we have
	\begin{equation*}
		\begin{aligned}			\widetilde{I}_{T_1, T_2}^{\eps, m}&:=\int_{T_1}^{T_2} (\widetilde{X}_{T_1,r}^\eps)^m \dot{W}^\eps_r \dif r\\&
			=\int_{T_1}^{T_2}(\widetilde{X}_{T_1,r}^\eps)^m \diamond\dot{W}^\eps_r \dif r+ m\int_{T_1}^{T_2}\mathbb{E}[ \widetilde{X}_{T_1,r}^\eps \dot{W}^\eps_r
			](\widetilde{X}_{T_1,r}^\eps)^{m-1}\dot{W}^\eps_r\dif r\\&
			=\int_{T_1}^{T_2}(\widetilde{X}_{T_1,r}^\eps)^m \deltaup  W^\eps_r+m\int_{T_1}^{T_2}\mathbb{E}[ \widetilde{X}_{T_1,r}^\eps \dot{W}^\eps_r
			](\widetilde{X}_{T_1,r}^\eps)^{m-1}\dot{W}^\eps_r\dif r,
		\end{aligned}
	\end{equation*}
	where $\diamond, \deltaup$ denote Wick product and Skorokhod integration respectively, and refer the reader to~\cite[Chapters 3,15]{janson1997gaussian} for relevant definitions and details about the Wick product and Skorokhod integration.

    We claim now that for each $0\leq s\leq t \leq T$ and $\eps\in(0,1)$, 
    \begin{equation}
       \label{eq:zerorenormalisation}
       \mathbb{E}\left[ \widetilde{X}_{s,t}^\eps \dot{W}^\eps_t \right]
        = \mathbb{E}\left[  \widetilde{X}_{s,t}^\eps\mathbb{E}[  \dot{W}^\eps_t |\mathcal{F}_{t-\eps} ] \right]
        = 0.
    \end{equation}
    Indeed, in view of~\eqref{eq:mollifierlag}, $\widetilde{X}_{s,t}$ is $\mathcal{F}_{t-\eps}$-measurable. Moreover, from~\eqref{eq:mollifiedpath}  we have
\begin{align*}
    \dot{W}^\eps_t 
    = \frac{1}{\eps^2}\int_{\mathbb{R}}\varphi'\left(\frac{t-s}{\eps}\right)W_s \dif s
    =\frac{1}{\eps^2}\int_{t-\eps}^{t+\eps}\varphi'\left(\frac{t-s}{\eps}\right)W_s \dif s, \qquad t \in [0,T].
\end{align*}
It remains to show that $H^\eps_t:=\mathbb{E}[\dot{W}^\eps_t|\mathcal{F}_{t-\eps}]$ is identically null. To this end, note that
\begin{align*}
    H^\eps_t
    &=\mathbb{E}[\dot{W}^\eps_t|\mathcal{F}_{t-\eps}]
    \\&= \frac{1}{\eps^2}\mathbb{E}\bigg[\int_{t-\eps}^{t+\eps}\varphi'\left(\frac{t-s}{\eps}\right)W_s \dif s|\mathcal{F}_{t-\eps}\bigg]\\
    &=  \frac{1}{\eps^2}\mathbb{E}\bigg[\frac{1}{\eps^2}\int_{t-\eps}^{t+\eps}\varphi'\left(\frac{t-s}{\eps}\right)(W_s-W_{t-\eps}) \dif s|\mathcal{F}_{t-\eps}\bigg]
    +\frac{1}{\eps^2}\int_{t-\eps}^{t+\eps}\varphi'\left(\frac{t-s}{\eps}\right) W_{t-\eps} \dif s\\
    &= \frac{1}{\eps^2}\int_{t-\eps}^{t+\eps}\varphi'\left(\frac{t-s}{\eps}\right)\mathbb{E}[W_s-W_{t-\eps}|\mathcal{F}_{t-\eps}] \dif s
    +\frac{1}{\eps^2}W_{t-\eps}\int_{t-\eps}^{t+\eps}\varphi'\left(\frac{t-s}{\eps}\right)  \dif s \\
    & = 0 + \frac{1}{\eps}W_{t-\eps}(\varphi(1)-\varphi(-1))
    =0.
\end{align*}
\end{proof}

The previous lemma is the main ingredient for the proof of Theorem~\eqref{thm:LaggedMollifierApproximations}, 
which we now develop.
	
	\begin{proof}[Proof of Theorem~\ref{thm:LaggedMollifierApproximations}]

In view of Lemma~\ref{lem:zeroRenormalisation} and from the arguments of~\cite[Proof of~(42)]{BFGJS20}, then
	\begin{equation*}
		\begin{aligned}
        & \left|\widetilde{I}_{T_1, T_2}^{\eps, m}-I_{{T_1},{T_2}}^m\right|^2_{L^2(\Omega)}\\
        & = \mathbb{E}\bigg[ \int_s^{T_2} \deltaup W_r\int_{\R}\dif u\; \varphi_\eps(u-r)\bigg\{ (\widetilde{X}_{{T_1},u}^\eps)^m - X_{{T_1},r}^m    \bigg\}  \bigg]^2\\
        &
		\lesssim \int_s^{T_2} \dif r\int_{\R}\dif u\; \varphi_\eps(u-r)\mathbb{E}\left[\big| (\widetilde{X}_{{T_1},u}^\eps)^m-  X_{{T_1},r}^m  \big|^2\right]\\
        &
		= \int_s^{T_2} \dif r\int_{\R}\dif u\; \varphi_\eps(u-r)\mathbb{E}\bigg[\big|\widetilde{X}_{{T_1},u}^\eps-X_{{T_1},r} \big|^2 p_{m-1}\big(\widetilde{X}^\eps_{{T_1},u}, X_{{T_1},r}\big)^2\bigg]\\
        &
		= \int_s^{T_2} \dif r\int_{r-\eps}^{r+\eps}\dif u\; \varphi_\eps(u-r)\mathbb{E}\bigg[\big|\widetilde{X}_{{T_1},u}^\eps-X_{{T_1},r} \big|^2 
        p_{m-1}\big(\widetilde{X}^\eps_{{T_1},u}, X_{{T_1},r}\big)^2\bigg],
		\end{aligned}
	\end{equation*}
	where the second line follows from "It\^o's isometry for Skorokhod integrals"~\cite[Proposition~1.3.1]{Nua06} (the integrand of the Skorokhod integral is adapted hence the term involving Malliavin derivatives vanishes) and Jensen's inequality and $p_{m-1}(x,y) = \sum_{j=0}^{m-1} x^jy^{m-1-j}$.
	Since $\varphi$ is non-negative, symmetric and of unit mass we have 
	\begin{equation*}
		\begin{aligned}
			\widetilde{X}_{{T_1},u}^\eps-X_{{T_1},r}&=\int_\R\bigg[\varphi_\eps(u-2\eps-z)-\varphi_\eps({T_1}-2\eps-z)\bigg]X_z\dif z- X_{{T_1},r}\\&
			=\int_\R\varphi_\eps(z)(X_{u-2\eps-z}-X_{{T_1}-2\eps-z})\dif z- \int_\R\varphi^\eps({T_1})X_{{T_1},r}\dif z\\&
			=\int_\R\varphi_\eps(z)\left( X_{u-2\eps-z}-X_{r}  \right)\dif z-\int_\R\varphi_\eps(z)
            \left( X_{{T_1}-2\eps-z}-X_{{T_1}} \right)\dif z.
		\end{aligned}
	\end{equation*}
	From the latter we obtain for any $\eps<H$ the pathwise estimate 
    \begin{align*}
        |\widetilde{X}_{{T_1},u}^\eps-X_{{T_1},r}|
        &\leq \int_\R\varphi_\eps(z) |X_{u-2\eps-z}-X_r|\dif z +\int_\R\varphi_\eps(z)| X_{{T_1}-2\eps-z}-X_{{T_1}}|\dif z\\
        &\leq |X|_{C^\eta[0,T]} \int_\R\varphi_\eps(z)\left(|u-r-2\eps-z|^\eta +|2\eps+z|^\eta \right)\dif z\\
        &=|X|_{C^\eta[0,T]}\eps^{-1}\int_{-\eps}^\eps\left( |u-r-2\eps-z|^\eta +|z+2\eps|^\eta\right)\varphi(z/\eps)\dif z\\
        &=|X|_{C^\eta[0,T]}\int_{-1}^1\left( |u-r-2\eps-\eps z'|^\eta +|\eps z'+2\eps|^\eta\right)\varphi(z')\dif z'\\
        &\lesssim |X|_{C^\eta[0,T]}|\varphi|_{\infty}\eps^{-1}\left( |u-r|^{\eta+1}+ \eps^{\eta+1} \right).
    \end{align*}

Moreover, from the H\"older continuity of $X$ we have 
\begin{align*}
    |p_{m-1}\big(\widetilde{X}^\eps_{{T_1},u}, X_{{T_1},r}\big)|\lesssim   |X|^{m-1}_{C^\eta[0,T]}\sum_{j=0}^{m-1}|u-{T_1}|^{j \eta}|r-{T_1}|^{(m-1-j)\eta}.
\end{align*}
In view of these estimates we obtain the "worst-case" bound	
\begin{equation*}
\begin{aligned}
  & \left|\widetilde{I}_{T_1, T_2}^{\eps, m} - I_{{T_1},{T_2}}^m\right|^2_{L^2(\Omega)}\\
  & \lesssim \left(1+ \mathbb{E}\left[|X|^{2m}_{C^\eta[0,T]}\right] \right)\cdot\\&\cdot
  \sum_{j=0}^{m-1}\int_s^{T_2} \dif r\int_{r-\eps}^{r+\eps}\dif u\; \varphi_\eps(u-r)
  \frac{|u-r|^{2\eta+2}+ \eps^{2\eta+2}}{\eps^{2}}|u-{T_1}|^{2j \eta}|r-{T_1}|^{2(m-1-j)\eta}
\\&	\lesssim  \eps^{2\eta}|\varphi|_{\infty}\big(1+ \mathbb{E} |X|^{2m}_{C^\eta[0,T]} \big)({T_2}-{T_1})^{2m\eta+1}.
\end{aligned}
\end{equation*}
Furthermore, from Lemma~\ref{lem:Hypercontractivity} we deduce that
\begin{align*}
    \left|\widetilde{I}_{T_1, T_2}^{\eps, m}-I_{{T_1},{T_2}}^m\right|^p_{L^p(\Omega)}
    \lesssim \eps^{p\eta} ({T_2}-{T_1})^{p(m\eta+\half)}.
\end{align*}
Using probabilistic estimates instead of the pathwise arguments above we can set $\eta=H$. 
Finally, we invoke Kolmogorov's continuity criterion for rough paths (\autoref{thm:Kolm}, see also~\cite[Theorem~3.1]{FH20}, or the standard application of the Garcia-Rodemich-Rumsey lemma~\cite[Theorem~A1]{FV10b}) to obtain the desired bound.
\end{proof}

\begin{remark} Even though lagged mollifier approximations are not implementable in practice (as they require mollification with a smooth kernel), the motivation here is to showcase that mollifier approximations of the rough path~$\barX$ are valid without the need for renormalization. This provides a direct comparison to~\cite[\S3.2]{BFGJS20}.  
\end{remark}

\section{Simulation and calibration}\label{sec:SimCal}

The results of the previous section naturally combine with the continuity of the solution map $\overline \bfX \mapsto (S,V)$ in rough path theory~\cite{Lyo98} to yield a numerical scheme to solve the RDE~\eqref{eq:Model}. Namely, we consider piecewise-linear/mollifier lead-lag approximations of $(X,W)$, call them $(X^\eps,W^\eps)$ (the fact that $X$ is lagged with respect to~$W$ is dropped from the notation, since it is the only case of interest) and solve the ODE
\begin{equation}\label{eq:roughVolRDE}
    \begin{cases}
        \dot S^\eps_t = \sigma_\a(S^\eps_t, V^\eps_t, t) \dot W_t^{\a,\eps} + g(S^\eps_t, V^\eps_t, t) , \\
        \dot V^\eps_t = \tau(S^\eps_t, V^\eps_t, t) \dot X^\eps_t + \varsigma_\a(S^\eps_t, V^\eps_t, t) \dot W_t^{\a,\eps} + h(S^\eps_t, V^\eps_t, t) .
    \end{cases}
\end{equation}
using a conventional ODE solver. We used the \texttt{Python} package \texttt{Diffrax}\cite{diffrax,kidger2021on} which leverages the \texttt{JAX} framework for vectorisation.
This method is a simpler alternative to~\cite[\S 6]{BFGJS20}, which requires the subtraction of divergent quantities and achieves similar results. In this section we first show that our approach is convergent and consistent with the theory developed in previous sections, and subsequently show that it can be used to calibrate a new rough volatility model to market data. Some of the following experiments would benefit from faster solve times; this could be achieved by devising a \say{direct} numerical scheme (like~\cite{multilevel} for RDEs driven by classical Gaussian rough paths) for our type of equations, as this would remove the need to solve ODEs on a finer mesh than that on which the noise is generated. 

\subsection{Numerical tests}
We numerically validate the method above with the following instance of~\eqref{eq:Model}:
\begin{equation}\label{eq:numericsRDE}
\left\{
\begin{array}{rl}
    \displaystyle \frac{\dif S_t}{S_t} & =  \displaystyle \sqrt{a(Z_t - b)^2 + c} \, \d \bfW_t 
 - \frac{1}{2} \left( \frac{\sigma_1 a(Z_t - b)Z_t^{\gamma_1}}{\sqrt{a(Z_t - b)^2 + c}} + \left(a(Z_t - b)^2 + c\right)\right)\d t, \\
    \dif Z_t & = \sigma_0 Z_0^{\gamma_0} \d \bfX_t + \sigma_1 Z_t^{\gamma_1} \dif \bfW_t + (\alpha + \beta Z_t)\dif t.
\end{array}
\right.
\end{equation}

with~$W$ one-dimensional, $X$ and $\rho = \rho_1$ as in~\eqref{eq:fBm}
\[
\sigma_0 = \sigma_1 = a = b = c = \alpha = \beta = 0.1, \quad \gamma_0 = 1, \quad \gamma_1 = 1.5, \quad \rho = 0.8.
\]
This choice of equation and parameters has been made so that the solution is well behaved but general enough to exhibit various interesting behaviours predicted by the theory, which we proceed to replicate numerically. In particular, the choice of the expression under the square root (borrowed from the quadratic rough Heston model, see \autoref{subsec:calib} below) has the benefit of never being zero, key for local existence and to avoid a vanishing denominator in the drift, which has the precise form \autoref{prop:block} making $S$ a local martingale. The form of the volatility (or more precisely the term feeding into the volatility) $Z$ is loosely inspired by the model~\cite[(4)]{Jon03}: this is a \say{rough} version of that, in which the second factor is a correlated fBm instead of an independent Brownian motion.
The fact that $\rho \neq 0$ makes the theory of \autoref{sec:ItoLift} relevant (namely we are not restricted to the case of \autoref{rem:XindW}) and that $\rho \neq \pm 1$ means that this is an RDE driven by a genuinely multidimensional rough path.

\begin{figure}[h!]
    \centering
    \begin{minipage}{0.50\textwidth}
        \centering
        \includegraphics[width=\linewidth]{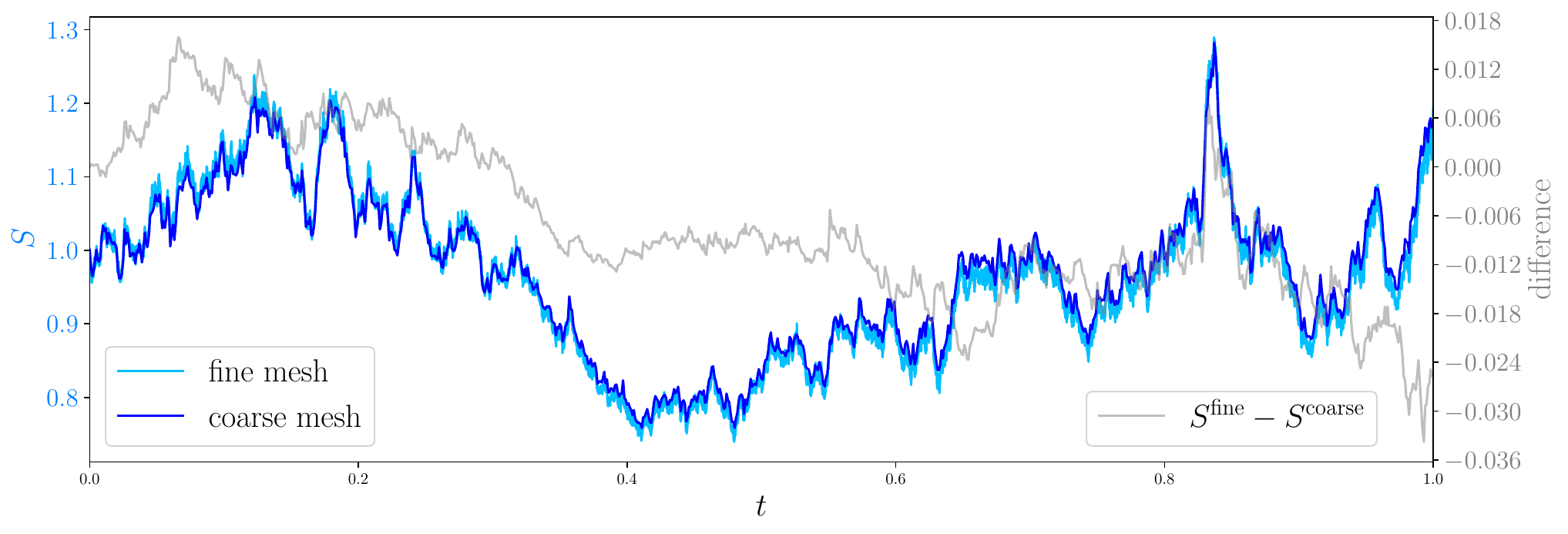}
    \end{minipage}
    \hspace{0pt}
    \begin{minipage}{0.475\textwidth}
        \centering
        \includegraphics[width=\linewidth]{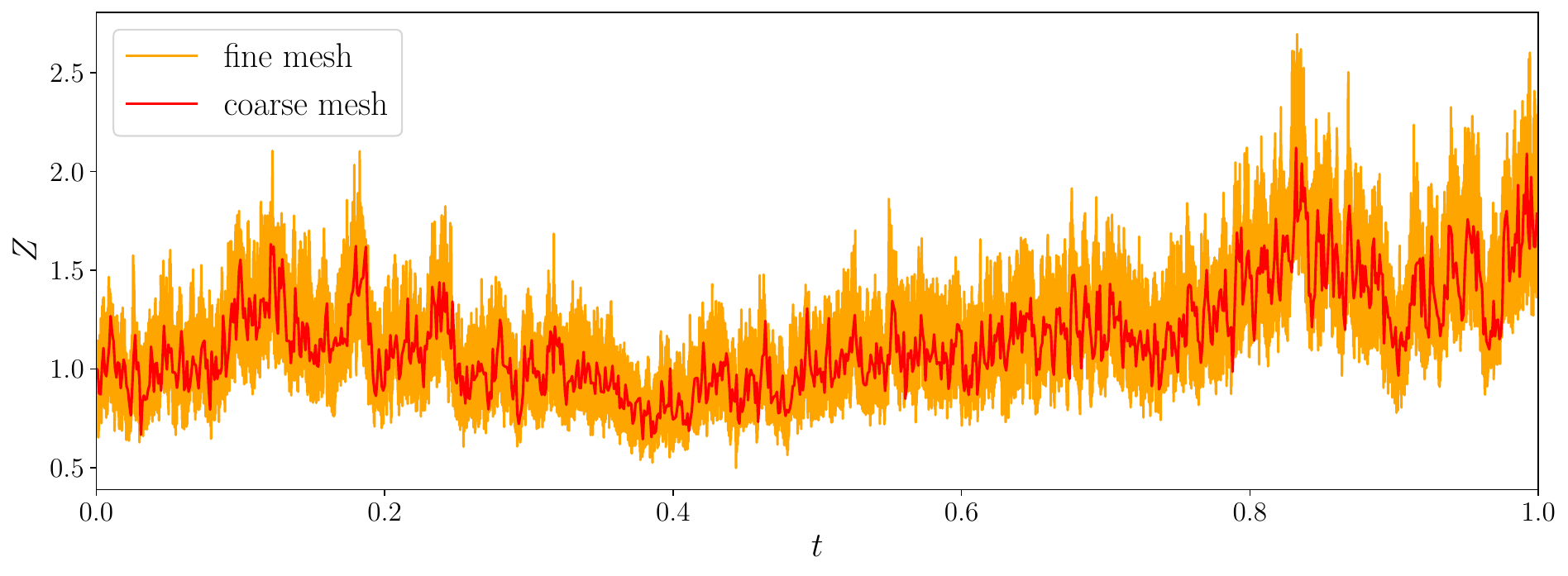}
    \end{minipage}
    \caption{Convergence along mesh refinement}\label{fig:convergence}
\end{figure}

We begin by sampling a single sample path of $(X,W)$ at a very high resolution, on a mesh of size $10^{-7}$ grid points on the interval $[0,1]$. 
We refer the reader to the file \texttt{dynamics\_rough\_vol.ipynb} in our code repository~\cite{code}. Next, we subsample it at a high resolution, mesh size $\eps = 5 \cdot 10^{-7}$, and low resolution, $\eps = 10^{-3}$. In both cases we form the lead-lag control of the interpolated paths $(X^\eps,W^\eps)$ on these meshes; in testing, we found a lag of $1.2\eps$ to be more reliable than simply the mesh size (theoretically there is no difference). We solve the ODEs approximating~\eqref{eq:numericsRDE} with a standard ODE solver operating on a mesh of size $0.1\eps$. Note that having a finer mesh for the solver is crucial, since the ODEs must be solved at a scale at which the noise can be considered smooth; taking the discretisation of the solver to be equal to or greater than that of the noise would incur It\^o-Stratonovich-type corrections. In \autoref{fig:convergence} we have plotted the results of solving~\eqref{eq:numericsRDE} with the noise discretised at the two different choices of the mesh size $\eps$, and observe convergence of both $S$ and $Z$; in the case of the price, we additionally plot the difference on the secondary $y$-axis.

\begin{figure}[h!]
    \centering
    \begin{minipage}{0.49\textwidth}
        \centering
        \includegraphics[width=\linewidth]{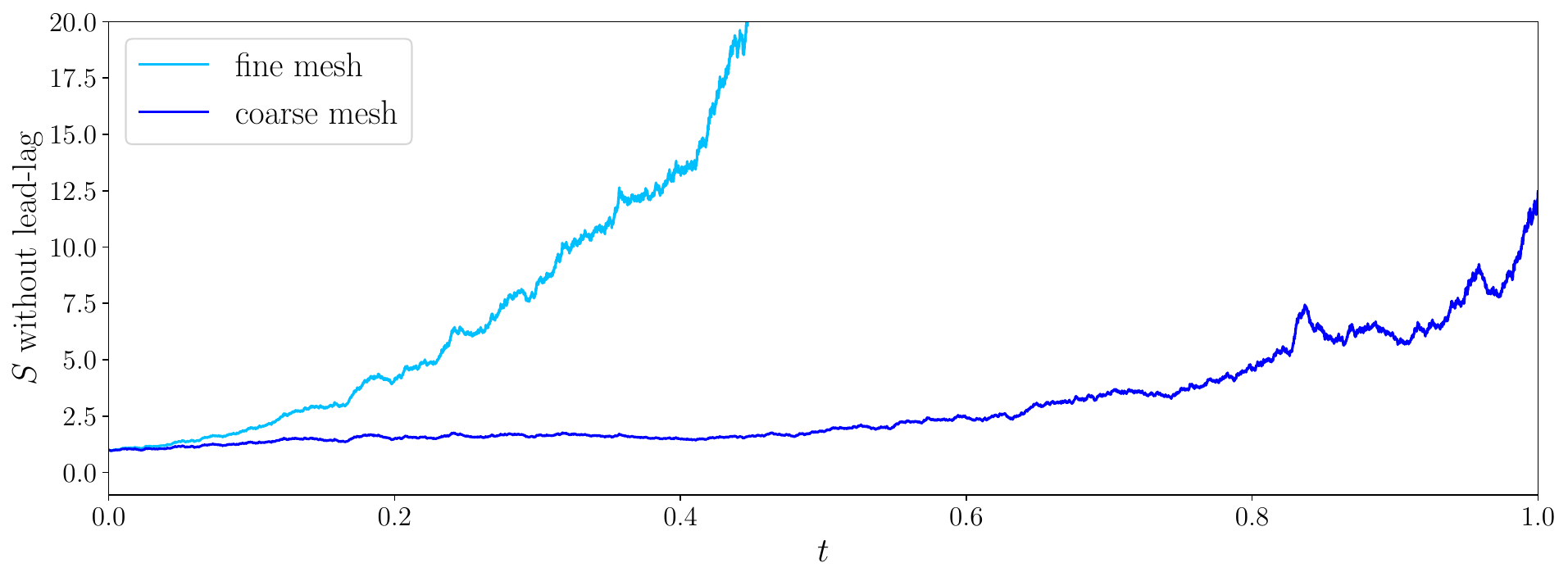}
    \end{minipage}
    \hspace{0pt}
    \begin{minipage}{0.49\textwidth}
        \centering
        \includegraphics[width=\linewidth]{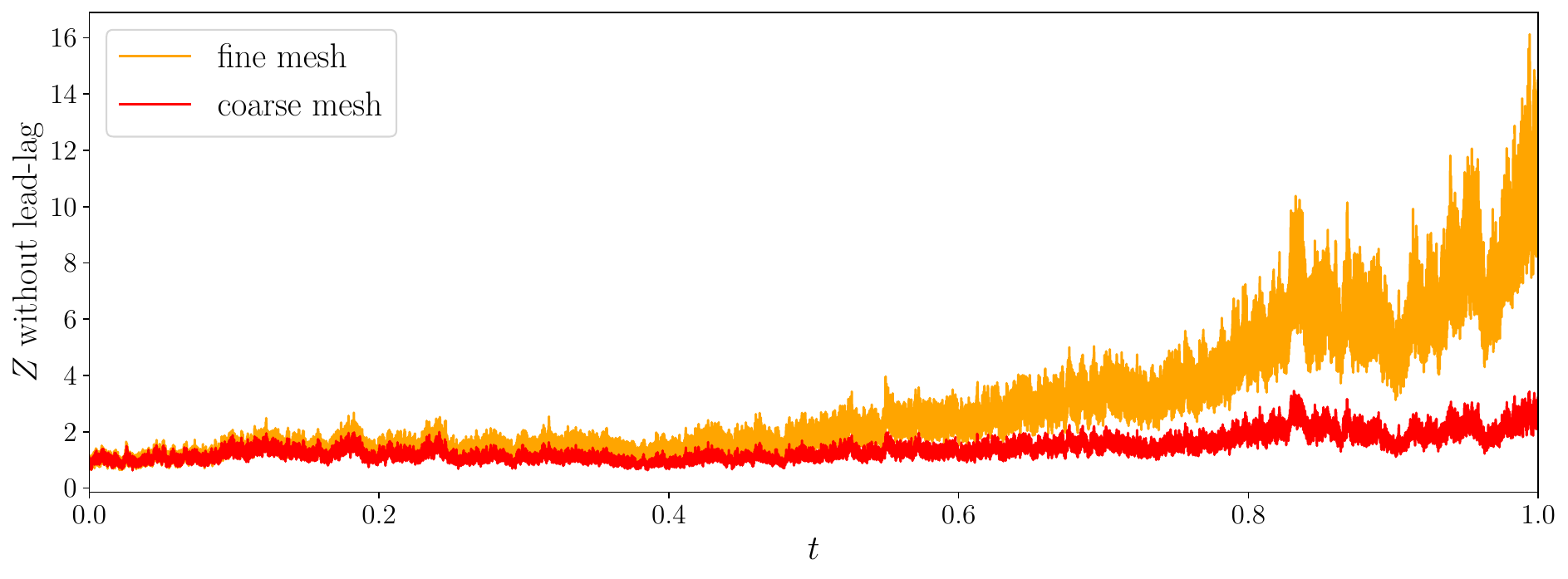}
    \end{minipage}
    \caption{No lead-lag $\implies$ explosion}\label{fig:divergence}
\end{figure}

Next, we carry out the same experiment but omitting the lag between $X$ and~$W$, and plot the resulting solutions for mesh sizes $\eps = 10^{-5}, 5 \cdot 10^{-7}$: we observe divergence of both $Z$ and $S$ in \autoref{fig:divergence}. The divergence of $Z$ is as predicted by~\cite{BFGJS20}, due to the infinite It\^o-Stratonovich correction. The divergence of $Z$ is similarly due to the infinite quadratic covariation $[X,W]$, but is only be observed if the term contracting with it---the Lie bracket $\sigma_0\sigma_1 (\gamma_1 - \gamma_0) Z^{\gamma_0 +\gamma_1 - 1}$---is non-zero. Indeed, a similar experiment with $\gamma_0 = 1 = \gamma_1$ still yields a divergent price but no divergence in~$Z$.

\begin{figure}[h!]
    \centering
    \begin{minipage}{0.49\textwidth}
        \centering
        \includegraphics[width=\linewidth]{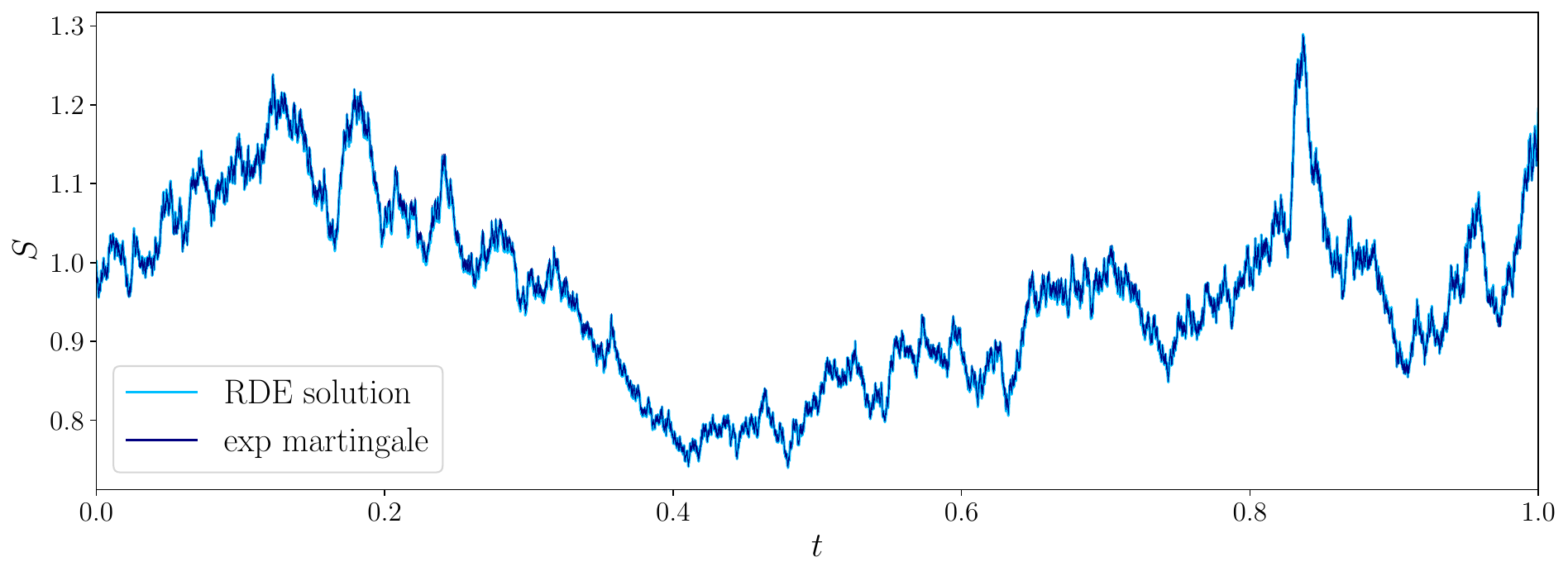}
    \end{minipage}
    \hspace{0pt}
    \begin{minipage}{0.49\textwidth}
        \centering
        \includegraphics[width=\linewidth]{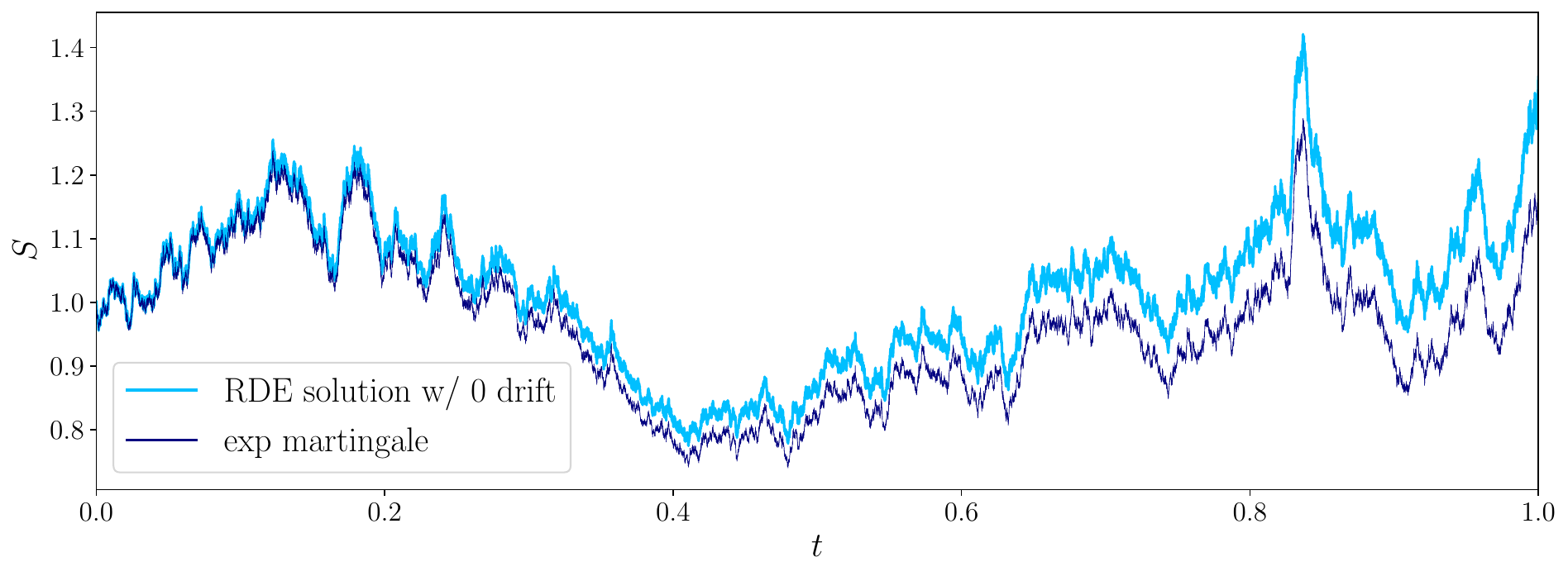}
    \end{minipage}
    \caption{Correct drift $\iff$ equality with exponential martingale}\label{fig:exp}
\end{figure}

In our last plot \autoref{fig:exp}, we check the equality~\eqref{eq:SExpMart} with the exponential martingale in; this is done by taking the solution~$Z$ as in~\eqref{eq:numericsRDE} (which does not involve~$S$), computing the It\^o integral with drift and taking the exponential. We also carry out this test by solving the RDE with zero drift, in which case~$S$ is no longer equal to the It\^o integral $\int_0^t \sqrt{a(Z_s - b)^2 + c} \, S_s \dif W_s$. We observe that the exponential martingale is practically identical to the lead-lag RDE solution for $S$ when solved with correct drift, but not when the drift is set to $0$; moreover, individually removing each of the two additive contributions to the RDE drift similarly results in a discrepancy. This is an additional corroboration of the fact that the solution to the RDE for $S$ is correct and a local martingale (independently of $Z$), a fundamental assumption in asset pricing; moreover, this identity is contingent on the precise form of the drift being correct.

Next, referring to the file \texttt{dynamics\_rough\_vol\_many.ipynb}, we compute the following empirical, relative $L^2$-errors over $1000$ samples. Here, $\overline\varepsilon = 10^{-6}$ and $S^\mathrm{exp}$ is given by~\eqref{eq:SExpMart} (with $Z$ as in the second line of~\eqref{eq:numericsRDE}). These tests, performed on a laptop and heavily constrained by time and memory, corroborate our theoretical results: convergence is slow but occurs. In certain cases (not reported), removing outliers was beneficial. A further interesting point, which we state informally, is that, modulo the error in the volatility process, the error in the price process converges faster than the Euler scheme for an It\^o equation. 
This is a significant point 
when accuracy of the price process is more important than that of~$V$.

\begin{table}[h!]\label{table:L2error}
\centering
\begin{tabular}{|c|c|c|c|}
\hline
\diagbox[width=4em]{$\eps$}{error} & $\phantom{\Bigg{[}}\displaystyle\frac{\lVert S^{\mathrm{exp},\eps}_1 - S^{\eps}_1 \rVert_{L^2}}{\lVert S^{ \eps}_1 - S^{ \eps}_0\rVert_{L^2}}$ & $\phantom{\Bigg{[}}\displaystyle\frac{\lVert S^{\overline \eps}_1 - S^{\eps}_1 \rVert_{L^2}}{\lVert S^{\overline \eps}_1 - S^{\overline \eps}_0\rVert_{L^2}}$  & $\phantom{\Bigg{[}}\displaystyle\frac{\lVert Z^{\overline \eps}_1 - Z^{\eps}_1 \rVert_{L^2}}{\lVert Z^{\overline \eps}_1 - Z^{\overline \eps}_0\rVert_{L^2}}$  \\ \hline
$\phantom{\Big[}10^{-3}$     &   $0.0811$    & $0.3268$      & $0.3742$     \\ \hline
$\phantom{\Big[}10^{-4}$      &   $0.0990$   & $0.1637$       & $0.2893$      \\ \hline
$\phantom{\Big[}10^{-5}$       &  $0.0308$  & $0.1079$      & $0.2286$   \\ \hline
\end{tabular}
\caption{empirical relative $L^2$ errors}
\label{tab:example}
\end{table}
A convenient feature of our model is that solving~$S$ and~$V$ can be done in parallel; in fact, solving for~$V$ first and evaluating~$S$ as the exponential martingale is marginally slower than solving for $(S,V)$ jointly, thanks to \texttt{JAX} vectorisation. Finally, while all error analysis in this and the previous section has been done in the strong~$L^2$ sense, the more relevant metric for option pricing would involve weak rates. These could, for instance, be studied using signature kernel-MMD~\cite{cass2024lecturenotesroughpaths} specified to the signature of the rough path in question.

\subsection{Pricing in the quadratic RDE Heston model}

We now test our framework in the context of calibrating stochastic volatility models to option prices, in particular considering the RDE-driven version of the \emph{quadratic rough Heston model}, proposed in~\cite{GJR20} and further investigated in~\cite{RZ21}.
With a fixed time horizon $T$, and a standard one-dimensional Brownian motion $(W_t)_{t\in[0,T]}$, 
denote by~$S$ the stock price and~$Z$ the auxiliary volatility process, which behave as follows:
\begin{equation}\label{{eq:SExpMart}}
\left\{
    \begin{array}{rll}
    \displaystyle\frac{\dif S_t}{S_t} & = \displaystyle \sqrt{a(Z_t - b)^2 + c} \circ \dif W_t - \half \left(a(Z_t - b)^2 + c\right) \dif t,
    & S_0 = s_0>0, \\ 
    \dif Z_t & = \lambda \theta(t)\dif t + \lambda \eta \sqrt{a(Z_t - b)^2 + c} \circ \dif  W^H_t,
    & Z_0 = z_0>0,
    \end{array}
    \right.
\end{equation}
with $a, b, c, \lambda, \eta$ positive parameters and $\theta:\bbR^+\rightarrow\bbR$ a suitably chosen deterministic function.
The notations~$\circ \d W_t$ and~$\circ \d W^H_t$ are used to indicate that integrals are interpreted via rough integration with respect to the rough path~$\overline{\bfX}$ that extends
the adapted $H$-integrable rough path $\bfX=\mathbf{W}^H$ above the one-dimensional fBm~$W^H$ (given by~\eqref{eq:CanocicalLift1dim}) and~$W$.
Given~$Z$, the stock price is then the exponential martingale:
\begin{equation}\label{eq:SExpMart}
    S_t =S_0 \exp\bigg\{\int_0^t\sqrt{a(Z_u - b)^2 + c} \dif W_u - \frac 12 \int_0^t \left(a(Z_u - b)^2 + c\right) \dif u \bigg\}.
\end{equation}

\begin{remark}
To reduce the computational cost, one can pre-calibrate the Hurst exponent~$H$ using asymptotic approximation, as in~\cite{BFG16,jacquier2025rough,lacombe2021asymptotics}.
This has the advantage that all the Brownian paths can then be pre-generated and stored offline, thus drastically reducing computation and calibration time.
For example, for a time grid with~$40$ points and using $50000$ paths, 
the initialisation time (simulating the Brownian paths) takes~$9.64$ seconds\footnote{The simulations were run on a MacBook Pro model MPHK3B/A with an Apple M2 Max chip, 12 cores and 32 GB memory.}.
The paths for the SDEs solution can then be generated in~$2.86$ seconds. 
The computation of option prices and implied volatilities is then more or less instantaneous from the stored stock price paths.
\end{remark}

\textbf{Algorithm (Simulation of the quadratic RDE Heston model).}

Consider the time grid $\mathbb{T} := \{t_i\}_{i=0,\dots,n_T}$.
\begin{itemize}
    \item The trajectories of~$W^H$ and~$W$ are generated via the hybrid scheme in Section~\ref{subsec:HybridLeadLag} (with $\kappa=1$) and then lagged and stored.
    \item The trajectories of~$Z$ are generated solving the lagged-RDE and stored.
    \item The trajectories of~$S$ are computed by simple left-end point discretisation of~\eqref{{eq:SExpMart}}.
    \item The option prices are then obtained by averaging the payoff over all terminal values of the stock price.
\end{itemize}

\subsection{Calibration of SPX options}\label{subsec:calib}
\begin{figure}[ht]
    \centering
\begin{minipage}{0.49\textwidth}
        \centering
        \includegraphics[width=\linewidth]{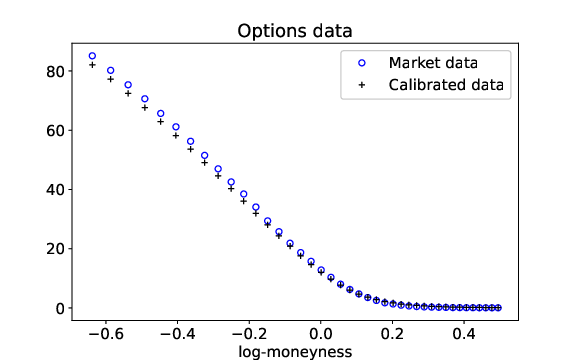}
    \end{minipage}
    \hspace{0pt}
    \begin{minipage}{0.49\textwidth}
        \centering
        \includegraphics[width=\linewidth]{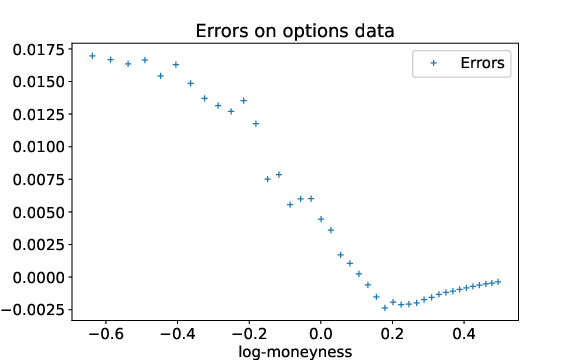}
    \end{minipage}
\caption{Call option prices and errors (differences divided by $S_0$) in the RDE quadratic rough Heston model for maturity $T = 0.548$ years.}
\label{fig:CFP}
\end{figure}

We calibrate the RDE-driven  quadratic rough Heston model to SPX option data 
extracted from the CBOE/OptionMetrics website on 21/11/2013.
We fix (with an educated guess from the short-maturity SPX skew) the Hurst parameter to $H = 0.2$  and minimise, over $\pf = (a, b, c, \theta, \eta, z_0)$ the objective function
\begin{align*}
    \mathcal{L}(\pf) :=\sum_{j=1}^{L_{T}}\left(\mathrm{C}_{T,K_j}(\pf) - \mathrm{C}^{\mathrm{obs}}_{T,K_{j}} \right)^2,
\end{align*}
where, $i\in\{1, \dots,L_T\}$,  $\mathrm{C}_{T,K_{i}}(\pf)$ denotes the computed Call price with maturity~$T$ and strike~$K_{j}$ using the parameters~$\pf$, 
and $\mathrm{C}^{\mathrm{obs}}_{T,K_{j}}$ the observed one. 
We also set the parameter~$\lambda$ to~$1$ as suggested in~\cite{GJR20,RZ21}.
The calibration itself is run with $50$ time steps and $10^5$ paths.
For $T=0.548$, we obtain the optimal vector of parameters 
$\pf^* = (0.3152, 0.3044, 0.0316, 0.2468, 0.9102, 0.1154)$,
and we display in Figure~\ref{fig:CFP} the calibrated vs real market option prices as well as the relative errors.

\appendix\normalsize
\section{Appendix}

One of the main ingredients that we need is Kolmogorov's continuity criterion (\cite[Theorem~2.1, page~25]{RY91} and~\cite[Theorem~3.1]{FH20}), which we state as follows:
\begin{theorem}[Kolmogorov's continuity criterion for two-parameter processes]\label{thm:Kolm}\ \\
Let $Z \colon [0,T]^2 \times \Omega \to \bbR$ be a stochastic process taking two time variables such that $Z_{ss} = 0$ for $s \in [0,T]$. Assume there exist $a,b,K > 0$ such that for all $s,t \in [0,T]$,
	\begin{equation}\label{eq:kolm}
		\bbE[|Z_{st}|^a] \leq K|t-s|^{2+b}.
	\end{equation}
Then for any $c \in(0,\frac{b}{a})$, there exists a random variable $J_{K,b,c} \in L^a(\Omega)$ such that
\begin{equation*}
\sup_{0 \leq s < t \leq T} |Z_{st}| \leq J_{K,b,c} (t-s)^c
\qquad\text{almost surely}.
\end{equation*}
\end{theorem}
Since the two-parameter processes for which we are computing the regularity are  given by It\^o integrals, we need to combine the above result with the following simple consequence of the BDG inequalities.

\begin{proposition}[Moments of It\^o integrals]\label{prop:BDG}
Let $p \in [2,+\infty)$ and $Y$ an $\mathcal{F}_{\bullet}$-adapted process bounded in $L^p(\Omega)$. Then there exists a constant $C_p$ (independent of $Y$ and $T$) such that, for all $s,t \in [0,T]$, 
    \[
    \bbE \bigg[\bigg| \int_s^t Y_u \dif W_u \bigg|^p \bigg] \leq C_p \sup_{u \in [s,t]} |Y_u|_{L^p} (t-s)^{\frac{p}{2}}
    \]
\end{proposition}

\begin{proof} 
Though standard, we include the proof for completeness. 
The process $Y^{s,t}_u \coloneqq \mathbf{1}_{[s,t]}(u) Y_u$ is adapted and almost surely with sample paths in $L^2[0,T]$, so that its It\^o integral exists and is a local martingale, and
\begin{align*}
    \bbE \bigg[\bigg| \int_s^t Y_u \dif W_u \bigg|^p \bigg] &\leq \bbE \bigg[\sup_{0 \leq v \leq T}\bigg|\int_0^v Y^{s,t}_u \dif W_u\bigg|^p \bigg]  \\
    &\leq C_p\bbE \bigg[\bigg[\int Y^{s,t}_u \dif W_u \bigg]_T^{\frac{p}{2}}\bigg] \\
    &= C_p \bbE \bigg[ \bigg( \int_s^t Y^2_u \dif u \bigg)^{\frac{p}{2}} \bigg] \\
    &\leq C_p(t-s)^{\frac{p}{2} - 1} \bbE \bigg[ \int_s^t |Y_u|^p \dif u \bigg] \\ 
    &= C_p (t-s)^{\frac{p}{2} - 1} \int_s^t\bbE[|Y_u|^p] \dif u \\
    &= C_p \sup_{u \in [s,t]} |Y_u|_{L^p} (t-s)^{\frac{p}{2}}.
\end{align*}
by the Burkholder-Davis-Gundy inequality \cite[Theorem~42.1]{RY91}, the expression for the quadratic variation of an It\^o integral, H\"older's inequality, Fubini's theorem, and boundedness in $L^p$.
\end{proof}

For $\nu\in\N$, a finite-dimensional vector space~$\Vv$ and a centered Gaussian process~$X$ taking values in~$\Vv$,
denote by~$\mathscr{C}_{\nu
}(\Vv)$ the $\Vv$-valued homogeneous Wiener chaos of degree~$\nu$ with respect to~$X$.
By hypercontractivity of the Ornstein-Uhlenbeck semigroup,
the following integrability lemma, 
proved and explained in~\cite[Chapters 1.1, 1.4.3, 1.5]{Nua06}, holds for random variables in $\mathscr{C}_{\nu}(\Vv)$, 
and is used frequently in Section~\ref{sec:lead-lag}:

\begin{lemma}\label{lem:Hypercontractivity} Let $\nu\in\N$ and $X_\nu$ a random variable in $\mathscr{C}_{\nu}(\Vv)$.
For any $0<p<q<\infty$,
$$|X_\nu|_{L^q(\Omega)}\leq\bigg(\frac{q-1}{p-1}\bigg)^{\frac{\nu}{2}} |X_\nu|_{L^p(\Omega)}.      $$   
\end{lemma}

\begin{acknowledgements}
For the purpose of open access, the authors have applied a Creative Commons Attribution (CC BY) licence (where permitted by UKRI, ‘Open Government Licence’ or ‘Creative Commons Attribution No-derivatives (CC BY-ND) licence’ may be stated instead) to any Author Accepted Manuscript version arising’.
\end{acknowledgements}

\bibliographystyle{siam}
\bibliography{refs}
\end{document}